\documentclass[12pt,oneside,english]{amsart}
\usepackage[T1]{fontenc}
\usepackage[latin9]{inputenc}
\usepackage{geometry}
\usepackage{amstext}
\usepackage{amsthm}
\usepackage{amssymb}
\usepackage{stmaryrd}
\usepackage{graphicx}
\usepackage{mathtools}
\usepackage{wasysym}
\usepackage{mathrsfs}
\usepackage{bbm}
\usepackage{esint}
\usepackage{color}
\usepackage{subfigure}
\usepackage{url}
\usepackage[percent]{overpic}

\makeatletter
\numberwithin{equation}{section}
\numberwithin{figure}{section}
\theoremstyle{plain}
\newtheorem{thm}{\protect\theoremname}
\theoremstyle{plain}
\newtheorem*{thm*}{\protect\theoremname}
\theoremstyle{plain}
\newtheorem{prop}[thm]{\protect\propositionname}
\theoremstyle{plain}
\newtheorem{lem}[thm]{\protect\lemmaname}
\theoremstyle{remark}
\newtheorem{rem}[thm]{\protect\remarkname}
\theoremstyle{definition}
\newtheorem{defn}[thm]{\protect\definitionname}
\theoremstyle{plain}
\newtheorem{cor}[thm]{\protect\corollaryname}
\theoremstyle{plain}
\newtheorem*{prop*}{\protect\propositionname}
\numberwithin{thm}{section}

\definecolor{kallecol}{rgb}{.99,.1,.5}

\newcommand{\ii}{\mathbbm{i}}
\newcommand{\cconj}[1]{\overline{#1}}

\newcommand{\projtoline}[1]{\mathrm{pr}_{#1}}
\newcommand{\unitcnum}{\xi}

\newcommand{\const}{\mathrm{const.}}

\newcommand{\corori}[1]{\boldsymbol{\nu}(#1)}
\newcommand{\cororibar}[1]{\overline{\boldsymbol{\nu}(#1)}}

\newcommand{\id}{\mathsf{id}}
\newcommand{\idof}[1]{\id_{#1}}
\newcommand{\indicator}{\mathbf{1}}

\newcommand{\mdl}{\mathfrak{m}}
\newcommand{\dmd}{\diamond}
\newcommand{\corn}{\mathfrak{c}}
\newcommand{\corndir}{{\nu}}

\newcommand{\C}{\mathbb{C}} 
\newcommand{\R}{\mathbb{R}} 
\newcommand{\Z}{\mathbb{Z}} 

\newcommand{\dcint}[1]{\sqint_{{#1}}}
\newcommand{\dd}[1]{[ \ud {#1} ]_{\meshsz} }
\newcommand{\ud}{\mathrm{d}} 
\newcommand{\contour}{\gamma}
\newcommand{\contourbis}{\alpha}
\newcommand{\contouralt}{\tilde{\gamma}}
\newcommand{\contourbisalt}{{\tilde{\alpha}}}
\newcommand{\contoursmall}{{\gamma}_-}
\newcommand{\contourlarge}{{\gamma}_+}

\newcommand{\interior}[1]{\mathrm{Int}[{#1}]}

\newcommand{\meshsz}{{\delta}}

\newcommand{\deebar}{{{\bar{\partial}}}}
\newcommand{\ddeebar}{\deebar_{\meshsz}}
\newcommand{\dee}{{\partial}}
\newcommand{\ddee}{\dee_{\meshsz}}

\newcommand{\Lapl}{{\Delta}}
\newcommand{\dLapl}{\Lapl_{\meshsz}}
\newcommand{\GreenF}{\mathbf{G}}
\newcommand{\CauchyKer}{\mathsf{k}}
\newcommand{\potKer}{\mathsf{a}}

\newcommand{\dmon}[2]{{#1^{[{#2}]}}} 
\newcommand{\dmonavg}[2]{{#1^{\{{#2}\}}}} 
\newcommand{\dhimon}[2]{{#1^{[{#2}]}}} 
\newcommand{\dhimonavg}[2]{{#1^{\{{#2}\}}}} 

\newcommand{\domain}{{\Omega}}
\newcommand{\ddomain}{{\domain}_{\meshsz}}
\newcommand{\ddomaindual}{{\domain}_{\meshsz}^*}
\newcommand{\ddomainmdl}{{\domain}_{\meshsz}^{\mdl}}
\newcommand{\ddomaindmd}{{\domain}_{\meshsz}^{\dmd}}
\newcommand{\ddomaincor}{{\domain}_{\meshsz}^{\corn}}
\newcommand{\dC}{{\C_{\meshsz}}}
\newcommand{\dCdual}{{\C_{\meshsz}^*}}
\newcommand{\dCmdl}{{\C_{\meshsz}^{\mdl}}}
\newcommand{\dCdmd}{{\C_{\meshsz}^{\dmd}}}
\newcommand{\dCcor}{{\C_{\meshsz}^{\corn}}}
\newcommand{\dCcormdl}{{\C_{\meshsz}^{\corn\mdl}}}
\newcommand{\dCmdldmd}{{\C_{\meshsz}^{\mdl\dmd}}}

\newcommand{\dE}{{\mathcal{E}_{\meshsz}}}
\newcommand{\dEdual}{{\mathcal{E}_{\meshsz}^*}}
\newcommand{\dEmdl}{{\mathcal{E}_{\meshsz}^{\mdl}}}
\newcommand{\dEdmd}{{\mathcal{E}_{\meshsz}^{\dmd}}}
\newcommand{\dEcor}{{\mathcal{E}_{\meshsz}^{\corn}}}
\newcommand{\edgeendpt}[1]{#1_+}
\newcommand{\edgebegpt}[1]{#1_-}

\newcommand{\dblcov}[1]{{\left[ #1 \right]}}
\newcommand{\dblcovproj}[1]{\underline{#1}}

\newcommand{\DGFF}{{\varphi}}
\newcommand{\curr}{{J}}
\newcommand{\spin}{{\sigma}}
\newcommand{\spinalt}{\widetilde{\spin}}
\newcommand{\spinfield}{{\varsigma}}

\newcommand{\energyfield}{{\varepsilon}}
\newcommand{\disenergy}[1]{{\mathcal{E}}_{#1}}
\newcommand{\disline}{{\varrho}}
\newcommand{\disord}{{\mu}}
\newcommand{\dispair}[2]{(\disord_{#1}\disord_{#2})}
\newcommand{\cordefline}{{\lambda}}
\newcommand{\mdldefline}{{\lambda}}

\newcommand{\ferm}{{\psi}}
\newcommand{\fermpair}[2]{(\ferm_{#1}\ferm_{#2})}

\newcommand{\gaugetr}{\mathsf{S}}

\newcommand{\LTEelem}{{\vartheta}}
\newcommand{\LTEishelem}{{\gamma}}
\newcommand{\LTEset}{{\mathscr{C}}}
\newcommand{\LTEsetin}[1]{{\LTEset}^{{#1}}}
\newcommand{\LTEsetptsin}[2]{{\LTEset}_{{#1}}^{{#2}}}

\newcommand{\genfield}{{\phi}}
\newcommand{\genoper}{{\mathcal{O}}}
\newcommand{\opernbrhood}{{\mathcal{D}}}
\newcommand{\opernbrhoodof}[1]{\opernbrhood_{#1}}
\newcommand{\actn}{{\mathcal{S}}} 
\newcommand{\dpath}[1]{{\mathcal{D} #1}} 

\newcommand{\abbrIsing}{\mathcal{I}} 
\newcommand{\abbrDGFF}{\mathcal{G}} 
\newcommand{\abbrModel}{\mathcal{M}} 
\newcommand{\abbrEven}{\mathrm{even}}
\newcommand{\abbrOdd}{\mathrm{odd}}
\newcommand{\locf}{\mathfrak{F}^\mathrm{loc}}
\newcommand{\locfields}{\locf_{\abbrModel}}
\newcommand{\nullf}{\mathfrak{F}^\mathrm{null}}
\newcommand{\nullfields}{\nullf_{\abbrModel}}
\newcommand{\correqf}{\mathfrak{F}}

\newcommand{\locfieldsDGFF}{\locf_{\abbrDGFF}}
\newcommand{\locfieldsIsing}{\locf_{\abbrIsing}}
\newcommand{\locfieldsIsingEven}{\locf_{\abbrIsing; \abbrEven}}
\newcommand{\locfieldsIsingOdd}{\locf_{\abbrIsing; \abbrOdd}}
\newcommand{\nullfieldsDGFF}{\nullf_{\abbrDGFF}}
\newcommand{\nullfieldsIsing}{\nullf_{\abbrIsing}}
\newcommand{\nullfieldsIsingEven}{\nullf_{\abbrIsing; \abbrEven}}

\newcommand{\correqfieldsDGFF}{\correqf_{\abbrDGFF}}
\newcommand{\correqfieldsIsing}{\correqf_{\abbrIsing}}
\newcommand{\correqfieldsIsingEven}{\correqf_{\abbrIsing; \abbrEven}}
\newcommand{\correqfieldsIsingOdd}{\correqf_{\abbrIsing; \abbrOdd}}

\newcommand{\currmode}[1]{\mathsf{J}_{{#1}}}
\newcommand{\fermmode}{{\Psi}}

\newcommand{\ourorder}[3]{%
	{(\overset{\underleftrightarrow{\phantom{x} #3 \phantom{x}}}%
	{{{{#1}}{{#2}}}}})}
\newcommand{\ourforder}[3]{%
	{(\pm\overset{\underleftrightarrow{\phantom{x} {#3} 
		\phantom{x}}}%
	{{{{#1}}{{#2}}}}})}

\newcommand{\virL}[1]{\mathsf{L}_{#1}}

\newcommand{\crit}{\mathrm{cr.}}
\newcommand{\invtemp}{{{\beta}}}
\newcommand{\invtempcrit}{\invtemp_{\crit}}
\newcommand{\Hamiltonian}{\mathcal{H}}

\newcommand{\EX}{\mathbb{E}}
\newcommand{\PR}{\mathsf{P}}
\newcommand{\correlsym}{\EX}
\newcommand{\correl}[1]{\correlsym \! \left[ {#1} \right]}
\newcommand{\correlin}[2]{\correlsym_{#1} \left[ {#2} \right]}
\newcommand{\QFTcorrel}[1]{\big\langle\, {#1} \,
			  \big\rangle}

\newcommand{\set}[1]{\left\{ #1 \right\}}

\usepackage{babel}

  \providecommand{\corollaryname}{Corollary}
  \providecommand{\definitionname}{Definition}
  \providecommand{\lemmaname}{Lemma}
  \providecommand{\propositionname}{Proposition}
  \providecommand{\remarkname}{Remark}
\providecommand{\theoremname}{Theorem}

\usepackage{babel}

  \providecommand{\corollaryname}{Corollary}
  \providecommand{\definitionname}{Definition}
  \providecommand{\lemmaname}{Lemma}
  \providecommand{\propositionname}{Proposition}
  \providecommand{\remarkname}{Remark}
\providecommand{\theoremname}{Theorem}

\usepackage{babel}
\providecommand{\corollaryname}{Corollary}
  \providecommand{\definitionname}{Definition}
  \providecommand{\lemmaname}{Lemma}
  \providecommand{\propositionname}{Proposition}
  \providecommand{\remarkname}{Remark}
\providecommand{\theoremname}{Theorem}

\makeatother

\usepackage{babel}
\providecommand{\corollaryname}{Corollary}
\providecommand{\definitionname}{Definition}
\providecommand{\lemmaname}{Lemma}
\providecommand{\propositionname}{Proposition}
\providecommand{\remarkname}{Remark}
\providecommand{\theoremname}{Theorem}

\begin{document}

\title[Discrete Complex Analysis and Virasoro Structure]%
{Conformal Field Theory at the Lattice Level: \\
 Discrete Complex Analysis and Virasoro Structure}

\author[Hongler \& Kyt\"ol\"a \& Viklund]%
{Cl\'ement Hongler, Kalle Kyt\"ol\"a, and Fredrik Viklund}

\address{Chair of Statistical Field Theory, Institute of Mathematics, EPFL
Station 8 CH-1015 Lausanne Switzerland }

\email{clement.hongler@epfl.ch}

\address{Department of Mathematics and Systems Analysis, Aalto University,
P.O. Box 11100, FI-00076 Aalto, Finland}

\email{kalle.kytola@aalto.fi}

\address{Department of Mathematics, KTH Royal Institute of Technology, 100
44 Stockholm, Sweden}

\email{fredrik.viklund@math.kth.se}

\begin{abstract}
Critical statistical mechanics and Conformal Field Theory (CFT) are
conjecturally connected since the seminal work of Beliavin, Polyakov,
and Zamolodchikov \cite{belavin-polyakov-zamolodchikov-i}. Both exhibit
exactly solvable structures in two dimensions. A long-standing question
\cite{itoyama-thacker} concerns whether there is a direct link between
these structures, that is, whether the Virasoro algebra representations
of CFT, the distinctive feature of CFT in two dimensions, 
can be found within lattice models of statistical mechanics.
We give a positive answer to this question for
the discrete Gaussian free field and for the Ising model, by connecting
the structures of discrete complex analysis in the lattice models with
the Virasoro symmetry that is expected to describe their scaling limits.

This allows for a tight connection of a number of objects from the
lattice model world and the field theory one. In particular, our results
link the CFT local fields with lattice local fields introduced in
\cite{gheissari-hongler-park} and the probabilistic formulation of
the lattice model with the continuum correlation functions. Our construction
is a decisive step towards establishing the conjectured correspondence
between the correlation functions of the CFT fields and those of the
lattice local fields.

In particular, together with the upcoming \cite{chelkak-hongler-izyurov-ii},
our construction will complete the picture initiated in 
\cite{hongler-smirnov,hongler-ii,chelkak-hongler-izyurov},
where a number of conjectures relating specific Ising lattice fields
and CFT correlations were proven. 

\end{abstract}

\maketitle

\newpage

\tableofcontents

\newpage

\section{Introduction}

\subsection{\label{subsec:stat-mech-and-cft}Statistical Mechanics and Conformal
Field Theory }

Physical arguments suggest that 2D lattice models at continuous phase
transitions have conformally invariant scaling limits that can be
described by Conformal Field Theories (CFTs). The 2D CFTs are exactly
solvable in the sense that they can be studied in terms of representations
of the Virasoro algebra. This has led to exact formulae for the (conjectural)
scaling limits of correlations, partition functions, and critical
exponents of such lattice models. See, e.g., 
\cite{belavin-polyakov-zamolodchikov-i,belavin-polyakov-zamolodchikov-ii,
di-francesco-mathieu-senechal,mussardo}
and the references in the latter.

Despite the success of the application of CFT to lattice models,
it usually constitutes a non-rigorous approach to statistical mechanics.
Indeed, one needs to assume that the fields of the models have conformally
invariant scaling limits and that they can be described within the
framework of certain quantum field theories.

In the special case of the (discrete) Gaussian free field, the CFT
approach can carried out rigorously and Virasoro representations can
be found in the continuum in terms of
insertions~\cite{gawedzki,kang-makarov}.
For the Ising model, significant progress towards connecting its correlations
with the correlation of the relevant CFTs has been made
recently~\cite{chelkak-hongler-izyurov,hongler,hongler-smirnov}.

Schramm's SLEs provide a different route towards a rigorous understanding
of conformally invariant scaling limits~\cite{schramm-i,schramm-iii}.
These random curves describe the scaling limits of cluster interfaces
in the lattice models. Moreover, SLE processes are amenable to calculations
and SLE techniques have been successfully applied to produce interpretations
and rigorous proofs of a number of the conjectures of CFT, see, e.g.,
the references in~\cite{kang-makarov}. Studies of more direct and
systematic connections between SLE and CFT have been carried out,
leading to beautiful results. See, e.g., 
\cite{bauer-bernard-i,doyon-riva-cardy,friedrich-werner,kang-makarov,kytola,
dubedat-iv,dubedat-v,flores-kleban}.

Still, the connection between lattice models and CFT remains far from
well-understood and, it seems fair to say, rather mysterious from
a mathematical perspective. Thus it appears of significant interest
to advance the mathematical understanding of CFT applied to lattice
models. There are many fundamental difficulties, including in particular
the correspondence between discrete and continuous quantities, the
proofs of conformal invariance, the locality of the limits, and the
positivity of the underlying representations, to name a few.

\subsection{\label{subsec:exact-sovability}Exact Solvability}

In two dimensions, a number of lattice models 
(see Section~\ref{subsec:lattice-models}
below) are considered exactly solvable in a different sense than CFTs.
Typically, exact solvability in this context means that certain lattice-level
relations such as the Yang-Baxter equations are present and this often
yields exact formulae for a number of interesting quantities of the
models. In fact, this is how most of the exact results about lattice
models were derived in the 20th century. See, e.g., \cite{baxter}
and references therein.

Recently, discrete solvability has been formulated in terms of \emph{discrete
holomorphicity}. This has enabled the use of discrete complex analysis
techniques leading to rigorous proofs of conformal invariance for
a number of exactly solvable lattice models and to exact formulae
for their limiting correlations, matching the predictions of the relevant
CFTs. See, e.g., 
\cite{kenyon-i,kenyon-ii,smirnov-ii,smirnov-iii,chelkak-hongler-izyurov}.

Once the lattice solvability has been used to establish conformal
invariance, the latter should in principle help reveal the algebraic
structures of CFT. Thus it is natural to expect lattice and CFT solvability
to be (indirectly) connected via continuous conformally invariant
objects such as SLE or the GFF or by the identification of lattice
precursors of key CFT objects, see, e.g., 
\cite{benoist-hongler,dubedat-ii,dubedat-iii,dubedat-v,
chelkak-glazman-smirnov}
for recent progress in this direction.

This leads to the following natural question formulated already in
the late 1980s \cite{itoyama-thacker}: 
\begin{itemize}
\item Is there a \emph{direct} connection between exact solvability
of lattice models and of conformal field theories?
\end{itemize}
In this paper we resolve this longstanding problem in the
positive for the discrete Gaussian free field and the critical
Ising model: the central algebraic structures of the CFTs describing
their scaling limits are in fact already present at the lattice level.

This question has been investigated in the physics literature in the
case of the 8-vertex model and the Ising model, 
see \cite{itoyama-thacker,koo-saleur}
and the references in the latter, as well as the recent 
development~\cite{gainutdinov-et-at}. However, our results are the first
where the relevant lattice and continuous structures are directly
and exactly connected (without deformation). Moreover, the action
of the operators takes place on a space of lattice ancestors of the
local fields, thus giving a transparent, probabilistic interpretation
of the situation, and is formulated in terms of discrete complex analysis,
enabling control of the limits.

\subsection{\label{subsec:lattice-models}Lattice Models}

A lattice model is informally a probabilistic or quantum model which
``lives'' on a graph or lattice such as $ \Z^2$: there
are (random or quantum) degrees of freedom associated with each vertex,
edge, or face of the lattice. Of particular interest are the large-scale
features of such models, particularly when they result in randomness
at large scale suggesting the existence of scaling limits, i.e., macroscopic
random objects which describe the models as one ``looks at them from
far away.''

This paper focuses on two classical probabilistic lattice models:
the discrete Gaussian free field
(defined precisely in Section~\ref{subsec:discrete-gaussian-free-field})
and the Ising model
(defined precisely in Section~\ref{subsec:ising-model}),
both with one
degree of freedom associated with each vertex. Their scaling limits
are the most fundamental examples of CFTs.

\subsubsection{\label{subsec:dgff}Discrete Gaussian free field}

The discrete Gaussian free field (dGFF) on a graph~$G$ is a random
Gaussian vector $\left(\DGFF_{x}\right)$ with entries indexed by
the vertices~$x$ of~$G$ with density proportional to 
$\exp \big( -\const \sum_{x\sim y}
\left(\DGFF_{x}-\DGFF_{y}\right)^{2} \big)$
where the sum is over all pairs of adjacent vertices;
see Section~\ref{subsec:discrete-gaussian-free-field} for the precise
conventions that we will use.

By taking an appropriate scaling limit of the dGFF, one recovers the
Gaussian free field (also known as the massless free boson field)
which plays a central r\^ole in Quantum Field Theory. The continuum
GFF is well-understood mathematically, see, e.g., 
\cite{gawedzki,glimm-jaffe,kang-makarov,sheffield}.

\subsubsection{\label{subsec:ising}Ising Model }

The Ising model is perhaps the most well-studied model of equilibrium
statistical mechanics. It consists of random $\pm1$-valued spins
$\left(\spin_{x}\right)$ living on the vertices~$x$ of a graph~$G$.
The probability of a spin configuration is proportional to 
$\exp \big( -\invtemp \, \Hamiltonian [\spin] \big)$,
where the energy
$\Hamiltonian [\spin] =
   - \sum_{x\sim y} \spin_{x} \spin_{y}$
is obtained by summing over all pairs of adjacent vertices and~$\invtemp>0$
is called the inverse temperature. The large-scale behavior of the
model depends strongly on~$\invtemp$: if we consider the Ising model
on a large subset of $ \Z^2$, a long range alignment will
occur if and only
if~$\invtemp > \invtempcrit := \frac{1}{2}\ln\left(\sqrt{2}+1\right)$,
while the system will look disordered at large scale
for~$\invtemp < \invtempcrit$.
The \emph{critical regime} $\invtemp=\invtempcrit$ has been the object
of much attention in the last decades: in particular, one of the motivations
for the study of CFT is to describe the scaling limit of this model.

\subsection{\label{subsec:cft-and-virasoro-algebra} Conformal Field Theory and
Virasoro Algebra}

In this subsection we briefly outline a number key ideas of CFT,
in particular the Virasoro algebra.

\subsubsection{\label{subsec:statistical-field-theory}Statistical Field Theory}

A (physical) theory aimed at describing a random system with infinitely
many degrees of freedom is often called a \emph{statistical field
theory}. Of particular interest are the (conjectural) scaling limits of lattice
models, i.e., the limits of lattice models living on $\meshsz$-meshed
discretizations~$\ddomain$ of continuous 
domains~$\domain\subset \R^{n}$,
as~$\meshsz \to 0$. It is expected, but unproven except in a few cases,
that most lattice models with an infinite correlation length
converge
to non-trivial scaling limits that can be described by statistical
field theories.

A statistical field theory on $\domain\subset \R^{n}$ is usually
\emph{thought of }as a random process $\genfield$ with a 
measure~$\PR\left\{ \genfield\right\} \propto 
\exp \big( -\actn\left[\genfield\right] \big)$,
where $\actn$ is a functional called the \emph{action}.
The main objects of interest are the correlations of \emph{local fields}
\emph{$\genoper_{j}$}, roughly speaking functions of~$\genfield$
in an infinitesimal neighborhood of their point of insertion,
for example derivatives of $\genfield$.
The correlations 
$\QFTcorrel{ \genoper_{1}(z_{1})
    \cdots \genoper_{n}(z_{n}) }_{\domain}$
are thought of as functional integrals 
\begin{equation}
\label{eq:functional-integrals}
\frac{\int\genoper_{1}(z_{1})
  \left[\genfield\right]\cdots\genoper_{n}(z_{n})
  \left[\genfield\right]e^{-\actn\left[\genfield\right]}
   \; \dpath{\genfield}}{%
  \int e^{-\actn\left[\genfield\right]} \; \dpath{\genfield}},
\end{equation}
over all the possible realizations of the field $\genfield$ defined
on $\domain$. This is natural, e.g., by analogy with the definition
of lattice models such as the dGFF and the Ising model.

Unfortunately, all of the above is difficult to make precise. Instead,
a common approach is to \emph{define} local fields as being \emph{objects
one can take (abstract) correlations of}. One hence considers a space
$\mathcal{F}$ of local fields, equipped with multilinear operations
\[
\mathcal{F}^{n} \ni \left( \genoper_{1}, \ldots, \genoper_{n} \right)
\mapsto \QFTcorrel{ \genoper_{1}(z_{1}) \cdots \genoper_{n}(z_{n}) 
        }_{\domain} \in \C ,
\]
defined for distinct points $z_{1}, \ldots, z_{n} \in \domain$. A number
of axioms are then added corresponding to
what the abstract correlations
are expected to satisfy (positivity, etc.), were they to arise from
functional integrals as in \eqref{eq:functional-integrals}. This
is one of the standard approaches to CFT~\cite{segal,segal-ii}. 
One of the
eventual outcomes of this paper is an alternative route to understanding
(at least) certain field theories, which restores part of this probabilistic
picture, and brings the original spirit of functional integrals much
closer.

\subsubsection{Conformal Field Theory}

A Euclidean \emph{Conformal Field Theory }on $\domain \subset \C $
is informally a statistical field theory with conformal symmetry.
Conformal symmetry is thought of as a symmetry of the action functional
$\actn$. Conformal symmetry can then be \emph{defined} by postulating
the existence of a special local field $T$, called the \emph{holomorphic
Stress-Energy Tensor}\footnote{In the functional integral picture,
$T$ is defined via the variation of
the action with respect to a change of metric; informally the insertion
of $T$ hence represents a change of measure (à la Radon-Nikodym
derivatives) corresponding to an infinitesimal change of metric. }.
Its correlations 
\[
\domain \setminus \set{ z_{j} } \ni z
    \mapsto \QFTcorrel{ T(z) \prod_{j} \genoper_{j}(z_{j}) }
\]
are holomorphic and have prescribed poles as $z \to  z_{j}$.

The poles of $T(z) T(w) $ as $z \to  w$ are in
particular given by the so-called 
\emph{Conformal Ward Identity}\footnote{Again, this should
be interpreted within 
correlations only.}
\begin{equation}
T(z) 
T(w) =\frac{c/2}{\left(z-w\right)^{4}}+\frac{2T(w) }{
\left(z-w\right)^{2}}+\frac{\partial 
T(w) }{z-w}+\mathrm{reg},\label{eq:t-t-ope}
\end{equation}
where the number $c\in \R$ is an important parameter, characteristic
of the CFT in question, and is called its \emph{central charge.}

\subsubsection{Virasoro Algebra}

A key insight of 2D field theory is that the modes of the stress-energy
tensor $T$ can act as operators on other local fields: for each $n\in\Z $
and $\genoper\in\mathcal{F}$, one defines a field\footnote{Since a field is 
merely an object one can take correlations of, this
indeed defines a field. Also, since the correlations of $T(z) $
are holomorphic, one can just take $\epsilon$ small enough. (But
how small $\epsilon$ needs to be depends on the locations of the
other inserted fields in the correlation.) } 
$\virL{n} \genoper \in \mathcal{F}$ by 
\begin{equation}
\virL{n}\genoper(z) :=
  \lim_{\epsilon \to 0_{+}} \;
  \oint_{|\zeta-z|=\epsilon}
    T(\zeta) \genoper(z) \left(\zeta-z\right)^{n+1} \ud\zeta .
\label{eq:ln-def}
\end{equation}
From \eqref{eq:t-t-ope}, the operators $\left(\virL{n}\right)_{n\in\Z }$
can then be shown to form a representation of the Virasoro algebra
of central charge $c$, i.e., their commutation relations are 
\begin{equation}
\left[ \virL{n} , \virL{m} \right]
= \left(n-m\right) \virL{n+m}
 + \frac{c}{12}\left(n^{3}-n\right) \delta_{n,-m} .
\label{eq:virasoro-comm-rel}
\end{equation}
The Virasoro algebra is the cornerstone for the algebraic exact solution
of CFT: the $\virL{n}$ operators can be studied in terms of Virasoro
representation theory; further down this road, by their definition
in terms of $T$, the $\virL{n}$'s yield precise geometric information
about the correlations which, e.g., can be cast as linear partial differential 
equations for correlation functions.

\subsubsection{Vertex Operator Algebras and Sugawara Construction}

The key feature that enables the construction of linear operators
on the space of local fields is the holomorphicity of the stress tensor
(and it is the geometric importance of the latter which then allows
one to derive results about correlations). Large classes of CFTs possess
a number of holomorphic fields besides the stress energy tensor and
its descendants (the Virasoro subrepresentation generated by $T$). These
CFTs, whose algebraic axiomatization is \emph{Vertex Operator Algebras}
(VOAs), are the object of many beautiful insights of mathematical
physics, representation theory, and string theory 
(see \cite{borcherds,frenkel-lepwosky-meurman,kac}).

The CFTs of the GFF and the Ising model both possess a VOA structure,
based on the current field (for the GFF) and the fermion field (for
the Ising CFT) respectively. The modes of these holomorphic fields
and their commutation relations can be studied in a fashion that is
similar to the way the modes of the stress-tensor are studied. Furthermore,
the stress tensor of these theories can be constructed in terms of
the currents and fermions, and as result, the modes of the stress
tensor can be constructed in terms of the modes of the current and
fermion, through what is known as the Sugawara-Sommerfield construction
\cite{sugawara,sommerfield}. The main result of this paper relies
crucially on this construction.

\subsection{\label{subsec:strategy}Strategy}

The approach suggested in \cite{gheissari-hongler-park} and in the
present paper is quite different from the usual axiomatic approach
to CFT: we look at lattice models as precursors of the field theories.
The (often ill-defined) process $\genfield \colon \domain  \to  \R$
is replaced by
a random function $\genfield_{\meshsz} \colon \ddomain \to  \R$,
where $\ddomain$ is a discretization of $\domain $. The functional
integral formalism is then perfectly well defined. Indeed, the lattice
analogues of local fields $\genoper_{\meshsz}$ are defined as functions
of values of the process $\genfield_{\meshsz}$ on finitely many neighbors
of the insertions, and correlation functions are just expected values.

The obvious drawback is that lattice models have no conformal symmetry:
indeed, such models are not invariant under scaling or (most)
rotations, let alone more general conformal mappings. Nevertheless,
some lattice models, such as the dGFF and the Ising model, possess
a number of discrete holomorphic fields, i.e., fields whose correlations
satisfy lattice analogues of the Cauchy-Riemann equations. If we could
find a suitable lattice ancestor of $T$~--- a discrete
holomorphic lattice local field satisfying a discrete version of the
Conformal Ward Identity~\eqref{eq:t-t-ope}~--- we might
hope to be able to realize the Virasoro algebra at the lattice level.

However, the discrete holomorphic fields of the dGFF and the Ising
Model are not lattice ancestors of the stress tensor, but of (in some
sense) more primitive objects; the current and the fermion, respectively.
Since discrete holomorphicity is a rather fragile property (for instance,
it is not even preserved by squaring), it is not obvious how to construct
more sophisticated discrete holomorphic fields from them.

The approach of this paper relies on revealing, at the lattice level,
the extended Vertex Operator Algebra structure that both models carry,
which involves the current and fermion modes. The Virasoro generators
can then be constructed on relevant lattice local field spaces as
bilinear products of these modes, through the Sugawara construction.
Remarkably, the whole construction can be carried out at the lattice
level, yielding the same exact commutation relations as in the scaling
limit, 
while acting on probabilistically transparent objects.

\subsection{Main Result and Applications}

As explained in Section \ref{subsec:cft-and-virasoro-algebra}, the
Virasoro algebra in CFT acts on a space of local fields. In this paper,
we consider the lattice analogue of local fields proposed in 
\cite{gheissari-hongler-park},
and we define relevant operators on that space, allowing for a construction
of the full Virasoro symmetry on it.

\subsubsection{Lattice Local Fields}

A lattice local field is a natural generalization of fields of the
form $x\mapsto\DGFF_{x}$, $x\mapsto\DGFF_{x}^{2}$, 
$x\mapsto\spin_{x}\spin_{x+\meshsz}$,
$x\mapsto\spin_{x+\meshsz}-\spin_{x}$, etc., namely a 
translation-invariant
functional that depends on a finite number of variables applied to
the dGFF and Ising basic fields~$\DGFF_{x}$ and~$\spin_{x}$.
See Definition \ref{def:null-lattice-local-field} in 
Section~\ref{subsec:lattice-models-and-field-theory}
for a more precise definition. We call a lattice local field \emph{null}
if its correlations against other lattice local fields (taken at large
enough distance) are zero. Note that null fields are not necessarily
zero in a given realization, for example the discrete Laplacian of
the dGFF has vanishing correlations. The connection between lattice
local fields and CFT local fields (which can serve as a probabilistic
definition of the latter, since none has been given) has been conjectured
in, e.g., \cite{gheissari-hongler-park}.

\subsubsection{Main Result}

Let $\correqfieldsDGFF$ denote the space of lattice local
fields of the dGFF, modulo its null fields, and let $\correqfieldsIsing$
denote the space of lattice local fields of the critical Ising model,
modulo its null fields.

Our main theorem can then be informally phrased as follows. 
\begin{thm*}
The Sugawara constructions of the Virasoro modes
of the dGFF and of the critical Ising model can be naturally and exactly
realized at the lattice level on the space $\correqfieldsDGFF$
and $\correqfieldsIsing$ respectively, by considering discrete
complex Laurent modes of the lattice current and lattice fermion,
respectively. 
\end{thm*}
The precise form of the theorem, as well as its proof, is given in
Section \ref{sec:virasoro-algebra-at-lattice-level}, in the form
of Theorem \ref{thm:gff} (dGFF case) and Theorem \ref{thm:ising}
(Ising case).

By the link between the discrete and continuous structures that it
establishes, 
our main theorem yields an improved understanding
of both: 
\begin{itemize}
\item On the one hand, the construction demonstrates how the lattice solvability
can be directly expressed in terms of the algebraic structures of
CFT. This gives a convincing answer to the classical question of their
connection \cite{itoyama-thacker} and opens the possibility to understand
more structures related to Vertex Operator Algebras in a similar manner. 
\item On the other hand, by giving a natural lattice construction of the
objects of CFT, it gives the possibility of understanding CFTs
in probabilistic terms. As mentioned in Section~\ref{subsec:strategy},
this can be achieved by finding the proper (manifestly probabilistic)
discrete analogues of the field-theoretic concepts, and (later) establishing
their convergence in the scaling limit. 
\end{itemize}
These two directions, and some applications, are detailed in the next
two paragraphs.

\subsubsection{Application: Algebraic Structures}

The study of the CFTs in terms of their Vertex Operator Algebra structures
is a major branch of CFT \cite{frenkel-lepwosky-meurman}, which has
ramifications in string theory, condensed matter physics, and representation
theory. Most of the mathematical works on such theories rely on a
formal and abstract axiomatic construction of field theories, and
as such often appear daunting. It now appears that a significant part
of the relevant structures can be constructed very concretely at the
lattice level, using the techniques introduced in this paper, thus
significantly facilitating the understanding of these structures.
While these developments will be studied in a subsequent paper, we
briefly outline two particular (related) constructions of CFT, which
appear to be amenable to lattice constructions such as the ones proposed
in this paper: the Coulomb gas formalism and the Affine Kac-Moody
algebra CFTs. The main idea that emerges is: \emph{many of the important
algebraic structures of CFT can emerge from lattice solvability phrased
as discrete holomorphicity.}

The so-called Dotsenko-Fateev Coulomb gas construction is a fundamental
idea of CFT 
\cite{dotsenko-fateev,Felder-BRST_approach,di-francesco-mathieu-senechal},
which informally relies on considering complex exponentials of the
GFF. Within this framework it is possible to vary the central charge
of the theory, and thus to construct explicitly a large number of
CFTs modeled on a Gaussian structure. The constructions of the Coulomb
gas theory can be phrased in terms of deformations of the Sugawara
construction (see, e.g.,~\cite{Felder-BRST_approach,kang-makarov,mickelsson}). 
Using a modified
version of our dGFF construction (corresponding to other central charges)
it is possible to reveal them exactly at the lattice level. This will
allow for constructions of lattice precursors of a number of objects
of central importance in CFT, which moreover appear connected to other
lattice models, such as dimers.

A number of important examples of CFTs are those endowed with extensions
of the Virasoro algebra symmetry called Affine Kac-Moody (AKM) algebras.
These have found applications in condensed matter physics and in string
theory, and are at the heart of coset field theories, which are among
the most general classes of CFTs (in particular, coset CFTs include
all the minimal models). The most basic example of AKM CFT is the
Gaussian free field, which is endowed with its Heisenberg current
algebra, which is precisely the structure that we reveal at the discrete
level. Thanks to the so-called Wakimoto construction, AKM CFTs can
be constructed by taking several independent copies of the GFF, through
a scheme similar to that of the Coulomb Gas construction. Further
down this road, a number of constructions involving several copies
of the Ising fermions, in particular the theory of the framed Vertex
Operator Algebras, have recently yielded important results in representation
theory of finite groups \cite{miyamoto}, and it appears that realizing
them on the lattice level would give new insights and allow for a
number of simplifications.

\subsubsection{Application: Probabilistic Field Theories}

The other promising line of research emerging from this paper is the
possibility to give clear and precise probabilistic meaning to CFT
objects, thus enabling one to restore the ``original'' point of
view of such theories in terms of functional integrals. While discretizing
quantum field theories as a way to regularize them is an old idea,
promoted in particular by Kenneth Wilson \cite{cardy-ii,wilson-kogut}
and now viewed as the best way to mathematically understand them,
the following is new: the possibility to identify transparentely the
operator content of CFTs such as the one describing the Ising model,
at the lattice level. Relatedly, a dual point of view, looking at
the correlations as linear functionals on the space of fields, allows
one to bridge the classical thermodynamical point of view of statistical
mechanics, in terms of Gibbs measures, with the point of view of CFT
correlations, hence allowing one to view the Virasoro algebra as an
action on space of measures. The main idea that emerges is: \emph{the
whole operator content and algebraic structure of certain CFTs can
be explicitly constructed at the lattice level, and hence given a
probabilistic meaning}.

The dual point of view of realizing the Virasoro algebra consists
in looking at correlation functionals $\mu$ defined by 
$\genoper \mapsto \QFTcorrel{
\genoper(0) \prod\genoper_{j}(z_{j}) }_{\domain }$,
defined for any data of a domain $\domain $, boundary conditions, with
insertions $\genoper_{j}(z_{j}) $ at $z_{j}\neq0$,
and in defining an adjoint (contragredient)
action on such functionals,
by defining 
$\virL{n}^{\dagger} \mu \left(\genoper\right)
:= \mu \left(\virL{-n}\genoper\right)$.
Our main result yields a lattice analogue of this, as follows.
Consider a sequence of
discrete domains $\left(\domain_{\meshsz_{k}}\right)_{k}$ with mesh
sizes~$\meshsz_{k}=1/k$, together with boundary conditions and insertions of
lattice local fields~$\genoper_{j}^{\meshsz_{k}}(z_{j})$
at points~$z_{j}\neq0$. We can then form the \emph{sequence} of correlation
functionals $\left(\mu_{k}\right)_{k}$ defined by 
$\genoper^{\meshsz} \mapsto 
\correlin{\domain_{\meshsz_{k}}}{ \genoper^{\meshsz_{k}}(0)
    \prod \genoper_{j}^{\meshsz_{k}} (z_{j}) }$
(for each $\genoper^{\meshsz}$, $\mu_{k}\genoper^{\meshsz}$
is defined for large enough $k$). The dual 
action gives rise to the sequence
$\left(\virL{n}^{\dagger}\mu_{k}\right)_{k}$ by 
$\virL{n}^{\dagger}\mu_{k}(\genoper^{\meshsz})
:= \mu_{k}\left(\virL{-n}\genoper^{\meshsz}\right)$,
where $\virL{-n}\genoper^{\meshsz}$ is the Virasoro action on the lattice
local field $\genoper^{\meshsz}$ (for each $\genoper^{\meshsz}$
and $n\in\Z $, this is defined for large enough~$k$). This
is a natural generalization of the Gibbs measure, where instead of
just looking at the (unnormalized) limits of $\mu_{k}$ as $k \to \infty$
(which is the definition of a Gibbs measure), one looks at the entire
sequence itself (or more precisely, its tail)\footnote{The idea
is that various limits will be of interest, which might involve
renormalizing by certain powers of $\meshsz_{k}$ (depending on 
$\genoper^{\meshsz}$
in particular) as $k \to \infty$.}. This point of view naturally bridges the 
Gibbs measure picture with
the one of CFT.\footnote{This is also the point of view which was chosen as the 
primary point
of view in an earlier version of the present paper.}

In \cite{gheissari-hongler-park}, a conjecture linked the local field
picture of the Ising model with the operator content of the corresponding
CFT: each lattice local field is conjectured to converge, with some
proper normalization, to a CFT local field, and all CFT local fields
can be obtained as such limits. The second part of this conjecture
is particularly interesting as it allows one to give a probabilistic
meaning to the operator content of the CFTs. This seems in particular
to make sense of the operator content of the massive field theories
emerging from perturbed CFTs (e.g. the one describing the critical
Ising model with an infinitesimal magnetic field), where the axiomatic
formalism breaks down. Our main result is a key step for establishing
this second part: the operator content of the Ising CFT consists of
descendants of three primary fields (the identity, the spin and the
energy). Since the correlations of these fields have been established
to converge to their continuous counterparts (the identity is trivial,
see \cite{chelkak-hongler-izyurov} for the spin and \cite{hongler}
for the energy), it remains to prove that the lattice descendants,
as constructed by our main result, indeed converge.

In a subsequent paper, this will be proven, by combining the results
of the present paper with the upcoming paper \cite{chelkak-hongler-izyurov-ii},
where it is proven that multipoint correlations of spins, energies
and fermions, taken at far apart points, converge. A deformation of
the discrete contour integrals appearing in the present paper will
indeed allow one to reduce the correlations of any lattice descendant
field, to those of the spin, energy and fermions and hence yield the
result.

\subsection{Organization of the Paper}

In Section \ref{sec:discrete-complex-analysis}, we introduce the
discrete complex analytic tools we need, in particular discrete holomorphic
functions, discrete contours integrals, lattice integer and half-integer
monomials. The proofs of the statements of this section are postponed
to Section \ref{sec:discrete-complex-analysis-proofs}.

In Section \ref{sec:gaussian-free-field-and-ising-model}, we introduce
the relevant objects for the lattice models that we consider, in particular
the lattice local fields, the discrete Gaussian free field current
and the Ising fermion. The technical proofs of this section, pertaining
to the Ising fermion, are postponed to Section \ref{sec:lattice-fermion-proofs}.

In Section \ref{sec:virasoro-algebra-at-lattice-level}, we combine
the objects and results of Sections \ref{sec:discrete-complex-analysis}
and \ref{sec:gaussian-free-field-and-ising-model} to define the Virasoro
algebra actions, and hence prove the main theorem.

\subsection{Acknowledgements}

CH acknowledges support from the ERC SG CONSTAMIS grant, the NCCR
SwissMAP grant, the NSF DMS-1106588 grant, the Minerva Foundation,
the Blavatnik Family Foundation. FJV acknowledges support from the
Knut and Alice Wallenberg Foundation, the Swedish Research Council,
the Gustafsson
Foundation, the Simons foundation, and the NSF grant DMS-1308476.
KK is supported by the Academy of Finland.

We would like to thank G\'erard Ben Arous, St\'ephane Benoist, Richard
Borcherds, Paul Bourgade, John Cardy, Dmitry Chelkak, Gesualdo Delfino,
Julien Dub\'edat, Hugo Duminil-Copin, Christophe Garban, Krzysztof Gaw\c{e}dzki,
Jan de Gier, Alexander Glazman, Reza Gheissari, Alessandro Giuliani,
Konstantin Izyurov, Igor Krichever, Antti Kupiainen, Vieiri Mastropietro,
Michael McBreene, Jouko Mickelsson, Bertrand Nienhuis, Andrei Okounkov,
Eveliina Peltola, Sungchul Park, Duong H. Phong, Stanislav Smirnov,
Thomas Spencer, Valerio Toledano Laredo, Fabio Toninelli, and Wendelin
Werner for interesting discussions.

CH would also like to thank the late Pierluigi Falco, for many good
and inspiring discussions.

\section{\label{sec:discrete-complex-analysis}Discrete Complex Analysis}

In two dimensions, conformal symmetry is deeply linked to complex
analysis. On the lattice level, the combinatorial structures of the
models we consider in this paper are linked with \emph{discrete} complex
analysis and this is what has allowed for proofs of conformal invariance
of their scaling limits.

\subsection{\label{subsec:lattices-and-discrete-domains}Lattices and Discrete
Domains}

\begin{figure}[tb]
\centering
\subfigure[Square lattice~$\dC$ and its dual~$\dCdual$.] 
{
    \includegraphics[scale=1.25]{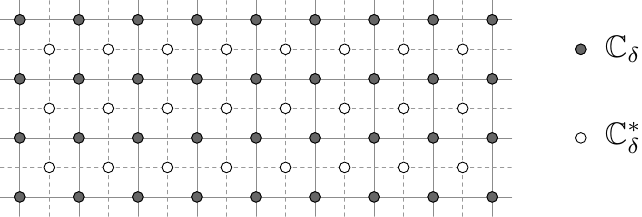}%
	\label{sfig: discrete dee}
} \\
\vspace{0.2cm}
\subfigure[Medial lattice~$\dCmdl$.] 
{
    \includegraphics[scale=1.25]{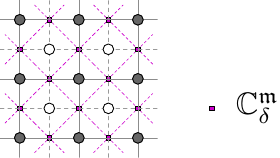}%
	\label{sfig: discrete Laplacian}
}
\hspace{1.5cm}
\subfigure[Corner lattice~$\dCcor$.] 
{
    \includegraphics[scale=1.25]{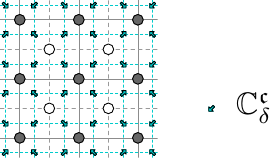}%
	\label{sfig: discrete deebar}
}
\caption{\label{fig:lattices}%
Illustrations of the lattices.
}
\end{figure}

We will work with a number of lattices associated with the square
lattice $\meshsz \Z^2$ of mesh size $\meshsz$, and in particular
use the following (see Figure \ref{fig:lattices}): 
\begin{itemize}
\item {Let $(\dC ,\dE)$ denote the \emph{discrete
complex plane}, i.e., the graph $\meshsz \Z^2$. } 
\item {Let $(\dCdual, \dEdual)$ be the
\emph{dual} of $\meshsz \Z^2$.} 
\item {Let $(\dCdmd, \dEdmd)$
be the \emph{diamond graph} whose vertices are
$\dC \cup \dCdual$
and with an edge connecting each pair of vertices at distance $ \meshsz 
/\sqrt{2}$.} 
\item {Let $(\dCmdl, \dEmdl )$ be the
\emph{medial lattice} with respect to $\meshsz \Z^2$, with
a vertex for each edge of~$\dE$; two medial vertices
are adjacent if the corresponding edges share an endpoint.} 
\item {Let $(\dCcor, \dEcor )$ denote
the \emph{bi-medial lattice (corner lattice)}: each vertex of the bi-medial 
lattice is called a corner. A corner lies between a vertex and a dual vertex,
and two corners are adjacent if they are at distance $\frac{\meshsz }{2}$
from each other. An edge~$e \in \dEcor $ of the bi-medial
lattice lies between a vertex of the diamond lattice and a vertex
of the medial lattice, denoted $e_{\dmd}$ and~$e_{\mdl}$.} 
\end{itemize}
Adjacency is denoted by~$\sim$ on any of the above graphs:
we denote $v \sim w$ if vertices~$v$ and~$w$ are the two endpoints of an edge.
Moreover, for two points of different lattices,
we still use the symbol~$\sim$ to denote that the pair of points are nearest 
neighbors~--- e.g., for $z \in \dCdmd$, $\zeta \in \dCmdl$ we have
$z \sim \zeta$ if and only if $|z-\zeta| = \frac{\meshsz}{2}$.

%

Occasionally we work simultaneously with two of the above lattices, and 
for this purpose we use the shorthand notations
$\dCmdldmd := \dCmdl \cup \dCdmd$
and~$\dCcormdl := \dCcor \cup \dCmdl$.

\subsection{\label{subsec:discrete-differential-operators}Discrete Differential
Operators}

\begin{figure}[tb]
\centering
\subfigure[Discrete Wirtinger derivative~$\ddee$.] 
{
    \includegraphics[scale=1.25]{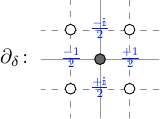}%
	\label{sfig: discrete dee}
}
\hspace{0.1cm}
\subfigure[Discrete Wirtinger derivative~$\ddeebar$.] 
{
    \includegraphics[scale=1.25]{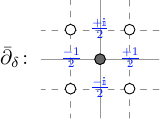}%
	\label{sfig: discrete deebar}
}
\subfigure[Discrete Laplacian~$\dLapl$.] 
{
    \includegraphics[scale=1.25]{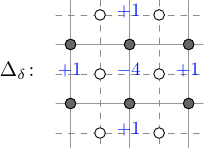}%
	\label{sfig: discrete Laplacian}
}
\caption{\label{fig: discrete differential operators}
Discrete differential operators.
}
\end{figure}

Below, we introduce the
lattice analogues of differential
operators that we use. The coefficients of these finite difference operators 
$\ddee, \ddeebar, \dLapl$ are illustrated also in 
Figure~\ref{fig: discrete differential operators}.
Throughout the paper, whenever needed, we will extend functions
$f$ defined on subsets (subgraphs) $\ddomaindmd$
of $\dCdmd$ by setting 
$f|_{\dCdmd\setminus\ddomaindmd }\equiv0$,
and similarly for functions defined on subgraphs $\ddomainmdl$
of $\dCmdl$.
\begin{itemize}
\item For $f \colon \dCdmd \to  \C $, we define
discrete \emph{Wirtinger derivatives} 
$\ddee  f, \ddeebar f \colon \dCmdl \to  \C $
by 
\begin{align*}
\ddee f(z)  & 
= \frac{1}{2} \left(
      f \big( z+\frac{\meshsz}{2} \big)
      - f\big( z-\frac{\meshsz}{2} \big) \right)
  - \frac{\ii}{2} \left(
      f \big( z+\frac{\ii \meshsz}{2} \big)
      - f \big( z-\frac {\ii \meshsz}{2} \big) \right) \\
 \ddeebar f(z)  & 
= \frac{1}{2} \left(
      f \big( z+\frac{\meshsz}{2} \big)
      - f\big( z-\frac{\meshsz}{2} \big) \right)
  + \frac{\ii}{2} \left(
      f \big( z+\frac{\ii \meshsz}{2} \big)
      - f \big( z-\frac {\ii \meshsz}{2} \big) \right) 
\end{align*}
and for $f \colon \dCmdl \to  \C $, we define 
$\ddee f,\ddeebar f \colon \dCdmd  \to  \C $
by the same formulae.
\item We define
the discrete \emph{Laplacian} as
$\dLapl = 4 \, \ddee \ddeebar = 4 \, \ddeebar \ddee$, so that
for $f \colon \dCdmd \to \C$
we have ${ \dLapl f \colon \dCdmd \to \C}$ given by
\begin{align*}
\dLapl f (z) =
    \sum_{x \in \set{ \pm \meshsz , \pm \ii \meshsz}} f \big( z+x \big)
    -4 \, f \big( z \big)
\end{align*}
and similarly for $f \colon \dCmdl \to  \C $.
\item {A function $f$ from $\dCdmd$ or 
$\dCmdl$
to $ \C $ is said to be \emph{discrete holomorphic} (on a region
of $\dC$) if $\ddeebar f=0$ (on that
region of $\dC$). If $f$ is discrete holomorphic,
then it can be locally integrated, i.e., there exists $F$ (at least
locally defined) such that $\ddee F=f$. } 
\end{itemize}
Note that we are not scaling the right-hand sides,
so the continuum differential operators are approximated
as~$\meshsz \to 0$ by, e.g.,
$\meshsz^{-1} \ddee \to \dee$
and $\meshsz^{-2} \dLapl \to \Lapl$.

\subsection{\label{subsec:discrete-integration}Discrete Integration}

\begin{figure}
\includegraphics[scale=0.9]{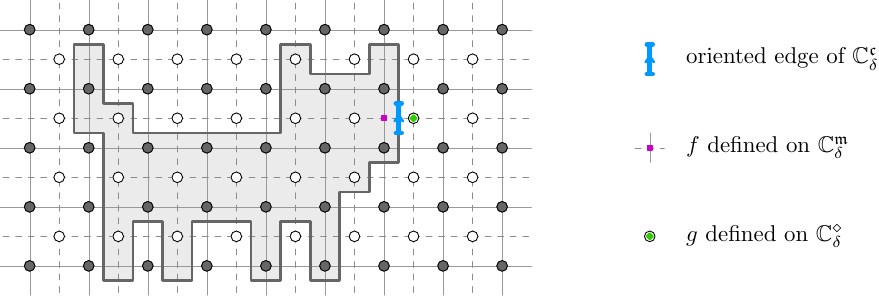}

\caption{\label{fig:discrete contour}
Discrete contour~$\contour$ on the corner lattice~$\dCcor$.
Each (oriented) edge $\vec{e} \in \contour$ of the contour separates a 
vertex $e_{\dmd} \in \dCdmd$  of the diamond lattice
from a vertex $e_{\mdl} \in \dCmdl$ of the medial lattice.}
\end{figure}

In the lattice setting, we will need to integrate products of two 
functions over discrete contours.
By a \emph{discrete contour}, we mean be an oriented
path $\contour$ of edges on the corner lattice $\dCcor$,
see Figure~\ref{fig:discrete contour}.
For two functions $f \colon \dCmdl \to  \C $ and 
$g \colon \dCdmd \to  \C $,
we then define the \emph{discrete contour integral} of $f$ times~$g$
along~$\contour$ by 
\[
\dcint{\contour}f(z_{\mdl}) g(z_{\dmd}) \, \dd{z}
:=
\frac{1}{2} \sum_{\vec{e}\in\contour} f(e_{\mdl}) g(e_{\dmd})
\frac{\edgeendpt{\vec{e}} - \edgebegpt{\vec{e}}}%
     {| \edgeendpt{\vec{e}} - \edgebegpt{\vec{e}} |}
    ,
\]%
where the sum is over all oriented edges 
$\vec{e} = (\edgebegpt{\vec{e}} , \edgeendpt{\vec{e}}) $ 
of~$\contour$, and where $e_{\mdl}$ and $e_{\dmd}$ denote the medial
and diamond vertices separated by~$\vec{e}$.
Note that the continuum contour integral approximation
as $\meshsz \to 0$ again requires a scaling,
$\meshsz \dcint{\contour} f(z_{\mdl}) g(z_{\dmd}) \, \dd{z}
\to \oint f(z) g(z) \, \ud z$.

If the discrete contour $\contour$ is closed,
we denote by $\interior{\contour}$
the \emph{interior of~$\contour$}, i.e.,
the set of points surrounded by~$\contour$.
For closed counterclockwise discrete contours~$\contour$, 
we have the following discrete Stokes-like formula 
\begin{equation}
\dcint{\contour}f(z_{\mdl}) g(z_{\dmd}) \, \dd{z}
=\ii \sum_{w_{\mdl} \in \dCmdl \cap \interior{\contour}}
  f(w_{\mdl}) \ddeebar g(w_{\mdl}) 
+ \ii \sum_{w_{\dmd} \in \dCdmd\cap \interior{\contour}}
  \ddeebar f(w_{\dmd}) g(w_{\dmd}) 
. \label{eq:discrete-stokes-formula}
\end{equation}

In particular, if both $f$ and $g$ are discrete holomorphic in the 
symmetric
difference $\interior{\contour} \oplus \interior{\contouralt}$ of two
closed counterclockwise contours 
$\contour, \contouralt$,
then we have the contour deformation property
\begin{align}\label{eq:contour-deformation-property}
\dcint{\contour}f(z_{\mdl}) g(z_{\dmd}) \, \dd{z}
= \dcint{\contouralt} f(z_{\mdl}) g(z_{\dmd}) \, \dd{z} .
\end{align}

Moreover, if $f,g \colon \dCmdl \to  \C $ are discrete
holomorphic in a lattice neighborhood of a closed integration contour~$\contour$
(i.e. 
$\ddeebar f(e_{\dmd})
= \ddeebar g(e_{\dmd}) = 0$
for any $e\in\contour$), it is elementary to check (using Abel's resummation)
that we have the integration by parts formula 
\begin{equation}
\dcint{\contour}\left(\ddee f(z_{\dmd}\right))
     g(z_{\mdl}) \, \dd{z}
= - \dcint{\contour}f\left(z_{m}\right)
    \left(\ddee g(z_{\dmd}) \right) \dd{z} . 
\label{eq:discrete-contour-integration-by-parts}
\end{equation}

\subsection{\label{subsec:discrete-integer-monomials}Discrete Integer Monomials}

We now summarize basic facts about the discrete analogues $z\mapsto \dmon{z}{k}$
of the monomial functions $z\mapsto z^{k}$ (for $k\in\Z $),
leaving proofs for 
Section~\ref{subsec:proof-of-propositions-discrete-monomials}.
These functions will later be used to construct the lattice counterparts
of the holomorphic modes of discrete holomorphic fields both on~$\dCdmd$
and~$\dCmdl$. In the following statement, we use the notation 
\[ \indicator_w(z) = \begin{cases}
                     1 & \text{ if } z = w \\
                     0 & \text{ if } z \neq w
                     \end{cases}
\]
for the Kronecker delta function.
\begin{prop}
\label{prop:integer-monomials}There exists a unique family of functions
$(\dmon{z}{k})_{k \in\Z }$, defined on 
$\dCdmd\cup\dCmdl$,
for which the following properties hold:
\begin{enumerate}
\item For all $k \in \Z$,
the function $\dmon{z}{k}$ has the same
90~degree rotational symmetry around~$0$
as the continuous function $z \mapsto z^{k}$
does. 
\item We have $\dmon{z}{0} \equiv 1$ on 
$\dCdmd\cup\dCmdl$. 
\item For all $k\geq0$,
we have $\ddeebar \dmon{z}{k} \equiv 0$. For all $k<0$, there 
exists $R>0$ such that for any 
$z\in\dCdmd\cup\dCmdl$
with $\left|z\right|\geq R$, we have $\ddeebar \dmon{z}{k} \equiv 0$. 
\item For all $k \in \Z$, we have $\ddee 
\dmon{z}{k} = k \, \dmon{z}{k-1}$. 
\item We have $\ddeebar \dmon{z}{-1} = 2\pi \indicator_{0}$
on $\dCdmd$ and 
$\ddeebar \dmon{z}{-1}  =
    2\pi \frac{1}{4}
    \sum_{x=\pm\frac{\meshsz }{2},\pm\frac{\ii \meshsz }{2}}
    \indicator_{x}$
on $\dCmdl$. 
\item For all $k\leq-1$, we have $\dmon{z}{k}  \to 0$ as $z \to \infty$. 
\item For any fixed $z\in\dCdmd\cup\dCmdl$
there exists $N\geq0$ such that $\dmon{z}{k} =0$ for all $k\geq N$. 
\item As $\meshsz  \to 0$, we have that
$\meshsz^{k} \dmon{z}{k} $ (extended,
e.g., by linear interpolation) converges to
the function $z \mapsto z^{k}$
uniformly on compact sets for $k\geq0$ and uniformly away from the
origin for $k<0$. 
\item For any $k,\ell\in\Z $, we have
\[ \frac{1}{2\pi \ii } \dcint{\contour}
    \dmon{z_{\mdl}}{k}
    \dmon{z_{\dmd}}{\ell} \, \dd{z}
= \delta_{k+\ell,-1} \]
if $\contour$ is a sufficiently large closed counterclockwise contour
surrounding the origin.
\item Setting \[ \dmonavg{z_{\mdl}}{k}
:= \frac{1}{4}\sum_{x\in\set{ \pm 1, \pm \ii } }
    \dmon{ \Big( z-\frac{\meshsz x}{2} \Big)_{\dmd}}{k} \]
for each $k \in \Z$, we have for all $k,\ell \in \Z$
\[ \frac{1}{2\pi \ii } \dcint{\contour}
    \dmonavg{z_{\mdl}}{k}
    \dmon{z_{\dmd}}{\ell} \, \dd{z}
= \delta_{k+\ell,-1} \]
if $\contour$ is a sufficiently large closed counterclockwise contour
surrounding the origin.
\end{enumerate}
\end{prop}
\begin{proof}
See Section \ref{subsec:proof-of-propositions-discrete-monomials}. 
\end{proof}

\subsection{\label{subsec:discrete-half-integer-monomials}Discrete Half-Integer
Monomials}

In this subsection we discuss discrete analogues $z\mapsto \dhimon{z}{p}$
to the functions $z\mapsto z^{p}$ for \emph{half-integer} exponents
$p \in \Z + \frac{1}{2}$. As is the case for their continuous analogues, these 
functions
are not naturally defined on $\dC$, but rather on
the \emph{double cover} of $\dC$, ramified at $0$,
denoted $\dblcov{\dC,0}$. Above each vertex of
$\dC\setminus\left\{ 0\right\} $, there are now two
vertices $v_{1},v_{2} \in \dblcov{\dC , 0}$, each
one with a well-defined square root $\sqrt{v}_{1}=-\sqrt{v_{2}}$.
For a vertex of $\dblcov{\dC , 0}$, there is a
unique vertex of $\dC\setminus\left\{ 0\right\} $
called the \emph{base point}, and another vertex of $\dblcov{ \dC , 0 }$
called \emph{the point on the opposite sheet}. Two vertices 
$v,w \in \dblcov{ \dC , 0 }$
are \emph{adjacent} if their respective base points are adjacent
and they are on the same sheet
(i.e. $\Re\mathfrak{e}\left(\sqrt{v}/\sqrt{w}\right)>0$):
the graph $\dblcov{ \dC , 0 }$ is hence connected.
We define analogously dual vertices, medial vertices, and diamond
vertices, and denote the relevant sets $\dblcov{ \dCdual , 0 }$,
$\dblcov{\dCmdl, 0}$ and 
$\dblcov{\dCdmd, 0}$,
respectively, and we continue to denote adjacency by~$\sim$.
For a simple path
$\lambda \subset \dC \setminus \set{0} $,
we say that two points $a,b \in \dblcov{\dC, 0}$
with base points $a_{0},b_{0}\in\lambda$ are \emph{on the same sheet}
\emph{of} $\lambda$ if following the square root branch along $\lambda$
one gets from $\sqrt{a}$ to $\sqrt{b}$.

For a function $f \colon  \dblcov{\dCdmd, 0} \to  \C $,
we will always set $f(0) := 0$ and define 
$\ddee f, \ddeebar f
 \colon  \dblcov{\dCmdl, 0} \to  \C $
in the natural manner (i.e. taking
$z\pm\frac{ \meshsz }{2},z\pm \ii \frac{ \meshsz }{2}$
on the same sheet as $z$). We say that such a function has monodromy
$-1$ around $0$ if its values at points on opposite sheets are opposite
(e.g. the square root function has $-1$ monodromy), and we say that
it is single-valued if the values are equal. 

The following lemma is easily verified.
\begin{lem}
\label{lem:product-two-double-valued-functions}Let 
$f \colon  \dblcov{ \dCmdl, 0} \to \C$
and $g \colon  \dblcov{\dCdmd, 0} \to \C$
be two functions with monodromy $-1$ around $0$. Then the function
on the bi-medial edges $e\mapsto 
f(e_{\mdl})g(e_{\dmd})$
is single-valued.
\end{lem}
The existence, uniqueness and basic properties of the discrete half-integer
monomials are summarized in the following proposition. 
\begin{prop}
\label{prop:half-integer-monomials}There exists a unique family of
functions $\left(\dhimon{z}{p} \right)_{p\in\Z +\frac{1}{2}}$
defined on the double cover of 
$\dCdmd\cup\dCmdl$
ramified at $0$ for which the following statements hold: 
\begin{enumerate}
\item For all $p$, the function $\dhimon{z}{p} $ has the same 90~degree 
rotational symmetry around~$0$ as the continuous function
$z\mapsto z^{p}$ does. 
\item $\dhimon{z_{\mdl}}{-\frac{1}{2}}$ is given by Definition 
\ref{def:discrete-inverse-square-root}. 
\item $\dhimon{z_{\mdl}}{\frac{1}{2}}$ is given by Definition 
\ref{def:discrete-square-root}. 
\item
For all $p\geq\frac{1}{2}$ and
$z \in \dblcov{ \dC^{\mdl\dmd}, 0}$,
we have
$\ddeebar \dhimon{z}{p} =0$. For each $p<0$, there 
exists $K>0$ such that $\ddeebar \dhimon{z}{p} = 0$ for all 
$z \in \dblcov{\dC^{\mdl\dmd}, 0}$
with $\left|z\right|\geq K$. 
\item
For all $p \in \Z + \frac{1}{2}$, we have
$\ddee \dhimon{z}{p} = p \dhimon{z}{p-1}$. 
\item For all $p\leq-\frac{1}{2}$, we have
$\dhimon{z}{p} \to 0$ as $z \to \infty$. 
\item For any fixed $z\in\dCdmd\cup \dCmdl $ there
exists $N\geq0$ such that $\dhimon{z}{p} =0$ for all $p\geq N$. 
\item As $ \meshsz  \to 0$, we have that $ \meshsz^{p} \dhimon{z}{p} $ 
converges to the function
$z\mapsto z^{p}$ uniformly on compact sets for $p\geq0$
and uniformly away from the origin for $p<0$. 
\item For any $p,q\in\Z + \frac{1}{2}$, we have
\[ \frac{1}{2\pi \ii } \dcint{\contour}
    \dmon{z_{\mdl}}{p}
    \dmon{z_{\dmd}}{q} \, \dd{z}
= \delta_{p+q,-1} \]
if $\contour$ is a sufficiently large closed counterclockwise contour
surrounding the origin.
\item Setting
\[ \dhimonavg{z_{\mdl}}{p} := \frac{1}{4}\sum_{x\in\left\{ 
\pm1,\pm \ii \right\} }\left(z-\frac{ \meshsz 
x}{2}\right)_{\dmd}^{[p]} , \]
we have for all $p,q \in \Z + \frac{1}{2}$
\[ \frac{1}{2\pi \ii }\dcint{\contour}
    \dhimonavg{z_{\mdl}}{p} \dhimon{z_{\dmd}}{q} \, \dd{z}
= \delta_{p+q,-1} \]
if $\contour$ is a sufficiently large closed counterclockwise contour
surrounding the origin.
\end{enumerate}
\end{prop}

\begin{proof}
See Section \ref{subsec:proof-of-propositions-discrete-monomials}. 
\end{proof}
\begin{rem}
It is possible to prove that the properties 1, 4, 5, together with
$\dhimon{z}{-1/2} \dhimon{z}{1/2} \to 1$ as $z \to \infty$
imply the other ones. 
\end{rem}

\section{\label{sec:gaussian-free-field-and-ising-model}Gaussian free field
and Ising Model}

\subsection{\label{subsec:lattice-models-and-field-theory}Lattice Models and
Field Theory}


As discussed in the introduction, a lattice model~$\abbrModel$
associates to a 
discretization~$\ddomain \subset \dC$ of a domain~$\domain \subset \C$
a random field $\genfield_{\meshsz} \colon \ddomain \to \C$
living on the discrete domain,
i.e., a collection of (complex valued) random variables
$\genfield_{\meshsz}(z)$ indexed by the 
vertices $z$ of the discrete domain.
We now introduce (as in \cite{gheissari-hongler-park}) a
lattice model analogue 
to the fundamental notion of local field
in Conformal Field Theory. 
Intuitively, the value of a lattice local field
at a point~$z$ is the result
of a translation invariant rule applied to the values that $\genfield_{\meshsz}$
takes on a fixed finite neighborhood of $z$. For this, we assume furthermore 
that the field 
$\genfield_{\meshsz}$ is extended to the complement 
$\dC \setminus \ddomain$ of the domain in some prescribed way.
\begin{defn}[Lattice local field]
\label{def:lattice-local-field}
Fix a lattice model~$\abbrModel$.
For $V \subset \Z^2$ 
a finite subset and $F \colon  \C^{V} \to  \C $ a polynomial
function, the random fields given by
\[ \genoper_{ \meshsz}(z)
= F \left[ \Big(\genfield_{ \meshsz}
         (z + x \meshsz) \Big)_{x\in V}\right]
\]
(for all possible choices of the discrete domain~$\ddomain \subset \dC$ and 
of boundary conditions)
constitute a (polynomial) lattice local field for the 
model~$\abbrModel$.
We denote by~$\locfields$ the $\C$-vector space of such lattice local fields.
\end{defn}


\begin{rem}
The condition that $F$ is polynomial 
does not entail any loss of generality in the case of the
Ising model. For the GFF, on the other hand, more general fields (e.g. $L^{2}$, 
such as exponentials) could be handled by density. 
\end{rem}

Examples of local fields are the field $\genfield_{ \meshsz}$ itself,
its square $\genfield_{ \meshsz}^{2}$, its lattice derivative 
$\genfield_{ \meshsz}\left(\cdot- \meshsz\right)-\genfield_{ \meshsz}$,
the product $\genfield_{ 
\meshsz}( \cdot ) \genfield_{ \meshsz}\left(\cdot+ \meshsz\right)$, etc.
The correlations of lattice
local fields are simply defined by taking the expectation with respect
to the measure of the model.

For critical lattice models such as the Gaussian free field and the
Ising model, it is natural to:
\begin{itemize}
\item expect that every lattice local field converges to some CFT 
local field;
\item expect that every CFT local field can be recovered 
as a limit of a suitably chosen lattice local field. 
\end{itemize}

This convergence should hold in the sense that the (suitably renormalized)
correlations of the lattice local fields converge to those of the
CFT local fields, when taken at far apart points: fields in QFT are
\emph{defined} by their correlations. As a result, fields with the
same correlations should be identified. This motivates the following: 
\begin{defn}[Null field]
\label{def:null-lattice-local-field}
A lattice local field $ \genoper_{ \meshsz}$
is called \emph{null} (for a given model~$\abbrModel$) if its correlations 
against
any other lattice local fields vanish (for that model) as soon as
the domain is large enough and the other insertions are far enough
from $z$, i.e.,
there exists $R>0$ such that if $z$ is at distance at least $\meshsz 
R$ from $z_{1}, \ldots, z_{n}$ and from $\dC \setminus \ddomain$, then we have
\[
\EX_{\ddomain}\left[ \genoper_{ \meshsz}(z) 
  \genfield_{ \meshsz}(z_{1}) \cdots 
  \genfield_{ \meshsz}(z_{n}) \right]
= 0 .
\]
Two lattice local fields are said to be \emph{(correlation-)equivalent}
if their difference is null.
The subspace of null fields within the space 
of all local fields 
of a model is denoted by~$\nullfields \subset \locfields$.
\end{defn}

A more precise formulation of the conjectural correspondence of 
lattice local fields to CFT local fields is:
\begin{itemize}
\item we expect that for any local fields
$\genoper_{1}, \ldots, \genoper_{n}$ of the CFT describing the scaling limit of 
the lattice model in question, 
there exist lattice local fields
$ \genoper_{1}^{ \meshsz}, \ldots,  \genoper_{n}^{ \meshsz} \in \locfields$ 
and scaling dimensions
$D_{1}, \ldots, D_{n} \in [0,\infty)$
(with each $D_{i}$ and $ \genoper_{i}^{ \meshsz}$ depending on $ \genoper_{i}$ 
only)
such that if $z_{j}^{ \meshsz} \to  z_{j}$ as $ \meshsz \to 0$
(with $z_1 , \ldots, z_n$ distinct), we have
\[
\frac{1}{ 
\meshsz^{\sum_{i=1}^{n}D_{i}}}  \EX_{\ddomain}\left[ \genoper_{1}^{ \meshsz}
\left(z_{1}^{\meshsz}\right) \cdots 
     \genoper_{n}^{\meshsz} \left(z_{n}^{\meshsz}\right)\right]
\underset{\meshsz \to 0}{\longrightarrow}
\QFTcorrel{  \genoper_{1}(z_{1}) \cdots 
        \genoper_{n}(z_{n}) }_{\domain}.
\]
\end{itemize}
The construction of the present paper is a decisive tool to establish this 
conjecture for the discrete Gaussian free field and
the Ising model. In particular, it gives an explicit way to construct
the lattice precursors of all the Ising CFT descendant fields 
(\emph{a fortiori},
since the algebraic structure of the descendant fields is already
present at the lattice level).

\subsection{\label{subsec:discrete-gaussian-free-field}
Discrete Gaussian Free Field}

The discrete Gaussian free field (dGFF) on a (finite) discrete domain
$\ddomain\subset\dC$ (with Dirichlet boundary
conditions) is a random field $\DGFF \colon \dC \to  \R$
with $\DGFF\big|_{\dC\setminus\ddomain}\equiv0$
and density proportional
to $\exp \big( -\frac{1}{16\pi} E\left[\DGFF\right] \big)$, where the
discrete Dirichlet energy $E$ is defined by $E\left[\DGFF\right]
:=\sum_{x\sim y}\left(\DGFF\left(x\right)-\DGFF\left(y\right)\right)^{2}$.
Equivalently, the dGFF $\DGFF$ is a centered Gaussian field with
covariance given by a multiple of the discrete Laplacian Green's 
function
of $\ddomain$ with $0$ boundary conditions:
\begin{align*}
\correl{ \DGFF(z) \DGFF(w) }
= 8 \pi \; \GreenF_{\ddomain}(z,w) ,
\end{align*}
where the Green's function is determined by
$\dLapl \GreenF_{\ddomain}(\cdot,w) = - \indicator_w(\cdot)$
and $\GreenF_{\ddomain}(z,w) = 0$ unless $z,w \in \ddomain$.
The dGFF is a
natural discretization of the continuous Gaussian free field on $\domain $
(which is a random generalized function $\domain  \to  \R$). Like
any centered Gaussian field, the dGFF satisfies the bosonic Wick's
formula: 
\[
\correl{ \DGFF (x_{1}) \cdots \DGFF (x_{2n}) }
= \sum_{\left\{ \ell_{j},r_{j}\right\}}
    \prod_{j=1}^{n} \correl{ \DGFF (x_{\ell_{j}} )
        \DGFF (x_{r_{j}}) } ,
\]
where the sum is over all pairings
$\left\{ \ell_{1},r_{1}\right\} , \ldots , \left\{ \ell_{n},r_{n}\right\} $
of $\set{ 1, \ldots ,2n } $. The dGFF is a discrete harmonic
field in the following sense.
\begin{lem}\label{lem: Laplacian of dGFF is null}
We have 
\begin{align*} 
\correl{ (\dLapl \DGFF) (x)
    \prod_{j=1}^{n} \DGFF(x_{j}) }
= \sum_{i=1}^n (-8 \pi \, \indicator_{x_i}(x) ) \times 
    \correl{ \prod_{j\neq i} \DGFF(x_{j}) }
\end{align*}
and in particular $\correl{ (\dLapl \DGFF) (x)
    \prod_{j=1}^{n} \DGFF(x_{j}) } = 0$
for $x \in \ddomain \setminus \set{ x_{1}, \ldots, x_{n}}$.
\end{lem}
\begin{proof}
This follows directly from Wick's formula and the covariance being
$8\pi$ times the discrete Laplacian Green's function.
\end{proof}
In particular, it follows from 
Lemma~\ref{lem: Laplacian of dGFF is null} that the discrete Laplacian of 
the discrete Gaussian free field is a null local field in the sense of 
Definition~\ref{def:null-lattice-local-field}:
$\dLapl \DGFF \in \nullfieldsDGFF$.

One of the most natural lattice local fields associated with the dGFF
is the following
\emph{lattice holomorphic current} $\curr \colon \dCmdl \to  \C $.
We first extend $\DGFF$ to $\dCdmd$ by setting it to
zero outside of~$\ddomain$ and on the dual lattice.
Then we may define the current as
\begin{align*}
\curr \left( z \right) = \ii \, \ddee \DGFF (z) .
\end{align*}

As defined, the current is not exactly the discrete analogue of the
continuous current, as it is 
purely real on midpoints of vertical
edges and purely imaginary on midpoints of horizontal edges.
This is
however not important for our approach, as the objects we will build
out of the current $\curr$ are contour integrals, which do approximate
the continuous integrals. 

We have that the current $\curr$ is discrete holomorphic in the sense
of correlations. 
\begin{lem}
\label{lem:current-is-discrete-hol}Let $\ddomain\subset\dC$
be a discrete domain, let $V\subset\ddomain$ be a finite set
and $F \colon  \R^{V} \to  \C $ a polynomial function. Then we
have that the function $G \colon \ddomain \to  \C $ defined by
$G(z) := \correl{
F\big(\DGFF\big|_{V}\big) \curr (z) } $
is discrete holomorphic for $z$ away
from~$\dC \setminus \ddomain$
and from~$V$. 
\end{lem}
\begin{proof}
This directly follows from the harmonicity of the dGFF since 
$\ddeebar \ddee = \frac{1}{4} \dLapl$.
\end{proof}
The assumption that~$F$ is polynomial is chosen as it is general
enough for our purposes, and specific enough so that the integrals
exist. 

At coinciding points, the lattice current has singularities, yielding
nonzero contour integrals. The following elementary lemma will in
particular be most useful: 
\begin{lem}
\label{lem:two-current-contour-integral}Let $\ddomain\subset\dC$
be a discrete domain, let $V\subset\ddomain$ be a finite set
and $F \colon  \R^{V} \to  \C $ be a polynomial function. Let
$w\in\dCmdl$ be a point away from $V$. Consider
the dGFF on a domain $\ddomain$ that includes a neighborhood
of $w$. Then for any
closed counterclockwise contour~$\contour$ such that
$z\in \interior{\contour}$,
$\interior{\contour}\subset\ddomain$ and
$\interior{\contour}\cap V = \emptyset$,
and any function $f \colon \dCdmd \to \C$
that is discrete holomorphic on $\interior{\contour}$ we have
\[
\frac{1}{2\pi \ii}\dcint{\contour} 
  \correl{ \curr (w) \curr (z_{\mdl})
      F \left( \DGFF\big|_{V} \right) }
      f(z_{\dmd}) \, \dd{z}
\, = \, \ddee f(w) \; \correl{ F\left(\DGFF\big|_{V}\right) } .
\]
\end{lem}
\begin{proof}
Observe first that for any $x\in\dCdmd$ that
is adjacent to $w$, Stokes' formula~\eqref{eq:discrete-stokes-formula}
combined with Wick's formula for the dGFF~$\DGFF$ 
with the explicit covariance $8 \pi \, \GreenF_{\ddomain}$ yield
\begin{align*}
& 
\dcint{\contour} 
  \correl{ \DGFF (x) \, \ddee \DGFF (z_{\mdl})  \,
      F \left(\DGFF\big|_{V} \right) }
  f (z_{\dmd}) \, \dd{z} \\
= \; & 
  \ii \sum_{z_\dmd \in \interior\contour}
  \correl{ \DGFF (x) \, \ddeebar \ddee \DGFF (z_{\dmd}) \,
      F \left(\DGFF\big|_{V} \right) }
  f (z_{\dmd}) \\
= \; & - 2 \pi \ii \, f(x) \,
    \correl{ F\left(\DGFF\big|_{V}\right) } .
\end{align*}
By taking a linear combination of the above over the
four $x\in\dCdmd$ adjacent to $w$,
the assertion of the lemma follows.
\end{proof}

\subsection{\label{subsec:ising-model}Ising Model }

We consider the Ising model on finite square grid 
domains~$\ddomain\subset\dC$, and we allow for general boundary 
conditions. The boundary conditions are implemented by a choice of a fixed
configuration 
$\overline{\spin} \colon \dC \setminus \ddomain \to \set{-1,0,+1}$
outside the domain, and the sample space of allowed 
configurations of the model is then
\begin{align*}
\set{ \spin \colon \dC \to \set{-1,0,1} \; \Big| \;
 \spin_x \in \set{-1,+1} \text{ for $x \in \ddomain$, and }
 \spin_x = \overline{\spin}_x \text{ for $x \notin \ddomain$}
 } .
\end{align*}
The constant function~$\overline{\spin} \equiv +1$ is known as
\emph{plus boundary conditions}, the constant function~$\overline{\spin} 
\equiv -1$ as \emph{minus boundary conditions}, and the constant 
function~$\overline{\spin} \equiv 0$ as \emph{free boundary conditions}. Our 
general boundary conditions can thus be combinations of these.
The energy of a configuration~$\spin$ is defined as
\begin{align*}
\Hamiltonian[\spin] = - \sum_{x\sim y}\spin_{x}\spin_{y} ,
\end{align*}
where the sum is over nearest neighbor pairs such that at least one of the 
vertices $x,y$ belongs to the (finite) domain~$\ddomain$. Given an inverse 
temperature parameter~$\beta>0$, the probability measure of the model assigns
probability proportional
to~$e^{-\invtemp \Hamiltonian[\spin] }$ to each allowed configuration~$\spin$.
The critical value for $\invtemp$ that is
of interest for CFT is $\invtempcrit=\frac{1}{2}\ln\left(\sqrt{2}+1\right)$.

Among the most natural local fields of the Ising model are the spin field 
$\spinfield_{\meshsz}\left(x\right):=\spin_{x}$
and the energy field
$\energyfield_{\meshsz}\left(x\right)
:= \spin_{x}\spin_{x+\meshsz}-\frac{\sqrt{2}}{2}$.
A number of results about the convergence and conformal invariance
of the correlations of $\meshsz^{-1/8} \spinfield_{\meshsz}$ and 
$\meshsz^{-1} \energyfield_{\meshsz}$
as $\meshsz \to 0$ with various boundary conditions have been established in
\cite{chelkak-hongler-izyurov, hongler,
hongler-smirnov, chelkak-hongler-izyurov-ii}.

\subsubsection{\label{subsec:disorder-operators}Disorder Operators}

The connection between the Ising model and complex analysis is
more involved than for the GFF: it involves non-local fields, i.e.,
objects which have a point of insertion, which are functions of the spin 
configuration, and which have correlations, but which cannot be represented 
as lattice local
fields. The most basic non-local fields are the \emph{disorder }operators. 
\begin{defn}
By a \emph{disorder line between p and $q$} we mean a simple path
$\disline$ on the dual lattice~$\dCdual$ with endpoints $p,q \in 
\dCdual$.
We denote this $\disline : p \leftrightarrow q$.
For an Ising configuration 
$\left(\spin_{x}\right)_{x\in\dC}$
define the \emph{disorder energy} of $\disline$ by
$\disenergy{\disline}[\spin]
= \sum_{x\sim y: \langle xy \rangle^{*} \in \disline}
	\spin_{x}\spin_{y}$.
For a disorder line $\disline$ between $p$ and $q$ we define the 
\emph{disorder pair}~$\dispair{p}{q}_{\disline}$ as the random variable
\begin{align}\label{eq:def-of-dispair}
\dispair{p}{q}_{\disline}
 = \exp \Big(-2\invtempcrit \disenergy{\disline}[\spin] \Big) .
\end{align}
\end{defn}
Note that for a fixed disorder line~$\disline$,
a disorder pair~$\dispair{p}{q}_{\disline}$ defines a lattice local field of 
the Ising model,
whereas a single disorder operator ``$\disord_p$'' could not be
defined as such.
\begin{figure}
\includegraphics[scale=0.9]{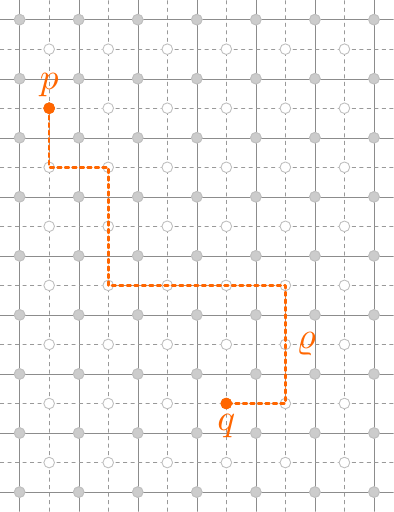}

\caption{\label{fig:disorder line}
A disorder line~$\disline$ is a path 
between two dual vertices $p,q \in \dCdual$.}
\end{figure}

Intuitively, reweighting a correlation by a disorder pair 
$\dispair{p}{q}_{\disline}$,
i.e., considering the reweighted measure 
\[
F \mapsto
\frac{\correl{ \dispair{p}{q}_{\disline} \, F[\spin] }}%
    {\correl{\dispair{p}{q}_{\disline} }} ,
\]
corresponds to an Ising model where the spins `pretend' that their 
neighbors
across $\disline$ are equal to the opposite of their actual values.
The following lemma (due to~\cite{kadanoff-ceva}, see 
also~\cite{dubedat-i,dubedat-iii,chelkak-cimasoni-kassel})
tells us that disorder pair correlations are essentially dependent
on the endpoints of the path only (and hence they are sometimes called
quasi-local fields). 
\begin{lem}
\label{lem:disorder-gauge-invariance}Let $\Gamma_{1},\Gamma_{2}$
be two collections of $k$ disjoint disorder lines such that
the sets of~$2k$ endpoints of both collections are the same. Let 
$\Gamma_{1}\oplus\Gamma_{2}$ denote
the collection of loops made of the symmetric difference of the sets
of dual edges $\cup_{\disline_{1}\in\Gamma_{1}} \disline_{1}$ and 
$\cup_{\disline_{2}\in\Gamma_{2}} \disline_{2}$.
Let $V\subset\dC$ be a finite set. Consider the Ising
model on a large enough domain $\ddomain$, with arbitrary
boundary conditions. We have 
\begin{equation}
\correl{ \left( \prod_{v\in V} \spin_{v} \right)
    \prod_{\disline : p \leftrightarrow q \in \Gamma_{1}}
    \dispair{p}{q}_{\disline} }
= \left(-1\right)^{\mathcal{N}}
  \correl{ \left( \prod_{v \in V}\spin_{v} \right)
      \prod_{\disline:p\leftrightarrow q \in \Gamma_{2}}
      \dispair{p}{q}_{\disline} }
, \label{eq:disorder-gauge-invariance}
\end{equation}
where the $\mathcal{N}$ is the number of pairs $\left(v,\ell\right)$
where $v\in V$ and $\ell\in\Gamma_{1}\oplus\Gamma_{2}$ is a loop
surrounding $v$. 
\end{lem}
\begin{proof}
For each loop $\ell\in\Gamma_{1}\oplus\Gamma_{2}$, let 
$\gaugetr_{\ell} \colon 
    \set{ \pm 1 }^{\ddomain} \to \set{ \pm 1 }^{\ddomain}$
be the involution that flips all the spins contained inside of $\ell$
and leaves the other ones unchanged (see Figures~\ref{fig:fermion-pair}
and~\ref{fig:two-fermion-gauge-transformed} in 
Section~\ref{subsec:corner-lattice-fermions}
for an example of similar gauge transformation). Let $\gaugetr$
denote the gauge transform consisting of composition of all the (commuting)
$\gaugetr_{\ell}$ for $\ell\in\Gamma_{1}\oplus\Gamma_{2}$. For
a configuration~${\spin \in \set{ \pm 1 }^{\ddomain}}$, and $j=1,2$,
consider
\[
\Hamiltonian_{j} :=
  \Hamiltonian[\spin] 
  + 2 \sum_{\disline\in\Gamma_{j}} \disenergy{\disline}[\spin] .
\]
Proving \eqref{eq:disorder-gauge-invariance} hence amounts to showing
that 
\[
\sum_{\spin \in \set{ \pm 1 }^{\ddomain }}
  \left( \prod_{v\in V} \spin_{v}\right)
  e^{-\invtemp \Hamiltonian_{1}[\spin] }
= \left(-1\right)^{\mathcal{N}}
  \sum_{\spinalt \in\set{ \pm 1 }^{\ddomain }}
  \left(\prod_{v\in V}\spinalt_{v}\right)
  e^{-\invtemp \Hamiltonian_{2}\left[\spinalt\right]} ,
\]
which simply follows by observing that if 
$\spinalt=\gaugetr[\spin] $,
then $\prod_{v\in V}\spin_{v} = 
\left(-1\right)^{\mathcal{N}}\prod_{v \in V}\spinalt_{v}$
and $\Hamiltonian_{2}\left[\spinalt\right] = \Hamiltonian_{1}[\spin] $. 
\end{proof}

\subsubsection{\label{subsec:corner-lattice-fermions}Corner Lattice Fermions}

Informally, a fermion operator $\ferm$ consists of a spin (living
on the primal lattice~$\dC$) next to a disorder (living on the dual 
lattice~$\dCdual$);
a natural
location for a fermion is hence at a corner (between a vertex and
and a dual vertex). Again, due to the non-locality of the disorder
operator, we define correlations of pairs of fermions with a defect
path between them; later, we show that only the sign of correlations
is affected by the choice of the path. 
\begin{defn}\label{def:corner-defect-line}
Let $c$ be a corner between $x \in \dC$ and $p \in \dCdual$
and let $d$ be a corner between $y \in \dC$ and $q \in \dCdual$.
We define a 
\emph{corner defect line $\cordefline$ with corner-ends
$c,d\in\dCcor$} as the concatenation
$\left[cp\right] \oplus \disline \oplus \left[qd\right]$
of a disorder line $\disline$ with endpoints $p,q$ with the two \emph{corner
segments} $\left[cp\right]$ and $\left[qd\right]$. We denote
$\cordefline : c \leftrightarrow d$, and we call 
$\disline : p \leftrightarrow q$ the
\emph{main part} of $\cordefline$.
We call $x,y$ the \emph{spin-ends} and $p,q$ the \emph{disorder-ends}
of $\cordefline$.
We denote by $\mathbf{W}\left(\cordefline:c\leadsto d\right)$ the
cumulative angle of turns by~$\cordefline$
(also known as \emph{winding}) traversed from~$c$ to~$d$.
\end{defn}
\begin{figure}[tb]
\centering
\subfigure[A corner defect line~$\cordefline$ associated with the  
corner lattice 
fermion pair $\left(\ferm_{c}\ferm_{d}\right)_{\cordefline}$,
for $c,d \in \dCcor$ (cf. Definitions~\ref{def:corner-defect-line} 
and~\ref{def:lattice-fermion}).
In this figure the winding is $\mathbf{W} \left(\cordefline:c \leadsto d\right) 
= \pi/4$ and the direction of the corner~$c$ is 
$\corndir(c) = e^{-\ii \pi /4}$.]
{
    \includegraphics[scale=0.9]{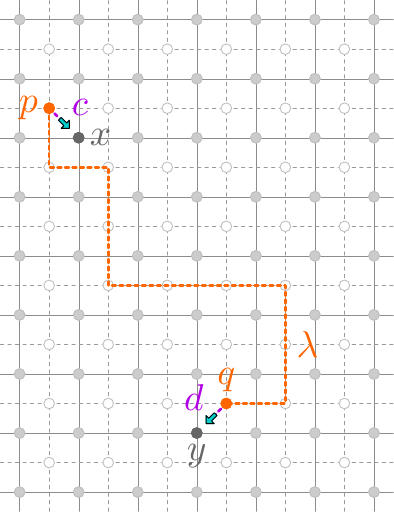}%
	\label{sfig: corner defect line}
}
\hspace{0.2cm}
\subfigure[A 
medial defect line~$\mdldefline$ associated with
fermion pair $\left(\ferm(w)\ferm(z)\right)_{\mdldefline}$
(cf. Definitions~\ref{def:medial-defect-line} 
and~\ref{def:lattice-holomorphic-fermion}).
This fermion pair is defined by summing over corner lattice fermion pairs 
corresponding to the four corners~$z_x \in \dCcor$
adjacent to~$z \in \dCmdl$ and the four corners~$w_y \in \dCcor$
adjacent to $w \in \dCmdl$ (where $x,y \in \set{\pm1\pm\ii}$).
] 
{
    \includegraphics[scale=0.9]{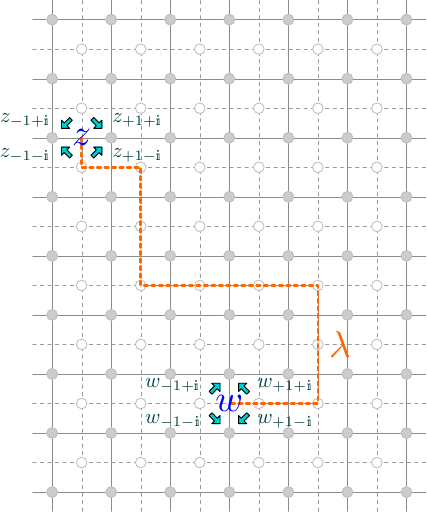}%
	\label{sfig: medial defect line}
}
\caption{\label{fig: defect lines for fermions}
Corner lattice fermions and 
medial lattice fermions.
}
\end{figure}

\begin{rem}
While the definition of $\mathbf{W}$ is the same as that in related
works \cite{chelkak-hongler-izyurov,chelkak-smirnov,hongler,hongler-smirnov},
the defect line is not the same object as the path appearing in the
low-temperature expansion of the fermionic observables of these works.
That path should interpreted as a line of frustration and should be
viewed as a configuration-dependent object, unlike the defect line,
which is fixed. 
\end{rem}

We now introduce the lattice fermion pair that we will work with,
a complexification of that introduced by Kadanoff and Ceva \cite{kadanoff-ceva},
see also \cite{dubedat-i,dubedat-iii}. 
\begin{defn}
\label{def:lattice-fermion}
Let $\cordefline$ be a corner defect line
with corner-ends $c,d$, spin-ends $x,y$ and disorder-ends $p,q$.
Let
\[ \corndir(c) := \frac{x-p}{\left|x-p\right|} \in
    \big\{ e^{\pm\frac{\pi \ii }{4}},e^{\pm\frac{3\pi \ii }{4}} \big\} \]
denote the direction of the corner $c$. We define the fermion pair
$\left(\ferm_{c}\ferm_{d}\right)_{\cordefline}$ as 
\[
\left(\ferm_{c}\ferm_{d}\right)_{\cordefline}
= - \overline{\corndir(c)} \, e^{-\frac{\ii}{2}
  \mathbf{W} \left(\cordefline:c\leadsto d\right)} \,
  \dispair{p}{q}_{\disline} \, \spin_{x}\spin_{y}.
\]
\end{defn}

\begin{rem}
The two-point correlation function of the corner lattice fermion coincides with 
the two-point observable defined in~\cite{gheissari-hongler-park}, in the 
special case when
$+$~boundary conditions are imposed and no other fields are inserted in the 
correlations. However, the
definition of corner lattice fermion given above makes the nature of the 
fermion pair (with fixed defect line) as a
local field (i.e., as a function of a finite number of spins) transparent and 
explicit.

%
\end{rem}

Despite the apparent difference of r\^ole of $c$ and $d$ in the definition,
the fermion pair is antisymmetric. 
\begin{lem}
\label{lem:corner-fermion-antisymmetry}Let $\cordefline$ be a corner
disorder line with corner-ends $c,d$. Then we have 
$\left(\ferm_{c}\ferm_{d}\right)_{\cordefline}
+\left(\ferm_{d}\ferm_{c}\right)_{\cordefline } =0$. 
\end{lem}
\begin{proof}
It is elementary to check that
$ - \overline{\corndir(c)} \, e^{-\frac{\ii}{2}
    \mathbf{W} \left(\cordefline:c\leadsto d\right)}
= + \overline{\corndir(d)} \, e^{-\frac{\ii}{2}
    \mathbf{W} \left(\cordefline:d\leadsto c\right)}$,
and the rest is unchanged.
\end{proof}

\begin{figure}
\includegraphics[scale=0.8]{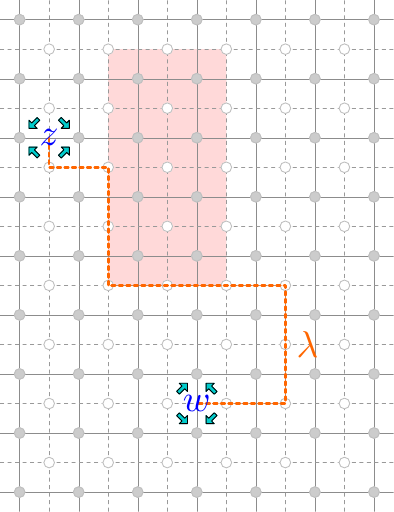}%
\hspace{1.0cm}
\includegraphics[scale=0.8]{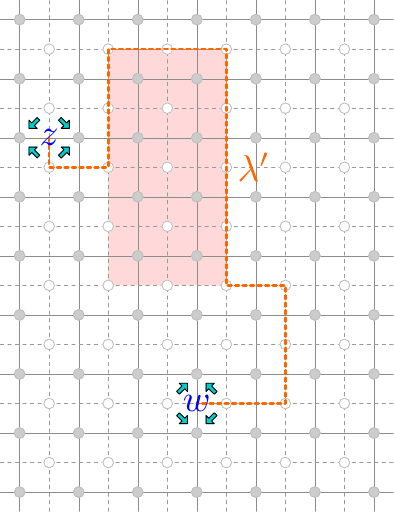}%
\caption{\label{fig:two-fermion-gauge-transformed}%
The defect line~$\mdldefline'$ on the right is obtained from the defect 
line~$\mdldefline$ on the left by 
performing the gauge transformation which flips the spins in the 
shaded area.}
\end{figure}

Fixing the two corners $c,d$, we have that the dependence on $\cordefline$
of $\left(\ferm_{c}\ferm_{d}\right)_{\cordefline}$ is not due to local
factors.
\begin{lem}
\label{lem:two-fermion-defect-line-independence}
Let~$\cordefline,\tilde{\cordefline}$
be two corner defect lines sharing the same corner-ends $c,d$.
Let $\cordefline\oplus\tilde{\cordefline}$ denote the collection of loops
of $\dCdual$ made of the symmetric difference of~$\cordefline$
and~$\tilde{\cordefline}$. Let $V\subset\dC$ be a finite
subset. Consider the Ising model on a large enough $\ddomain$,
with arbitrary boundary conditions. We have 
\[
\correl{ \left( \prod_{v\in V}\spin_{v} \right)
  \left(\ferm_{c}\ferm_{d}\right)_{\cordefline} }
= \left(-1\right)^{\mathcal{N}} \,
  \correl{ \left( \prod_{v\in V} \spin_{v}\right)
      \left(\ferm_{c}\ferm_{d}\right)_{\tilde{\cordefline}} }
,
\]
where $\mathcal{N}$ is the number of pairs $\left(v,\ell\right)$
where $v\in V$ and $\ell\in\cordefline\oplus\tilde{\cordefline}$ surrounds~$v$. 
\end{lem}
\begin{proof}
The proof uses the same bijection as the proof of 
Lemma~\ref{lem:disorder-gauge-invariance}
(see Figure \ref{fig:two-fermion-gauge-transformed}). It is elementary
to check that the $e^{-\frac{\ii}{2}\mathbf{W}\left(\cordefline\right)}$
term in the definition of the fermion compensates for the change of the
spins adjacent to the corners $c,d$. The rest behaves in the same
manner, thus yielding the result. 
\end{proof}

The following lemma, which allows one to exchange defect lines between
four fermions, will be instrumental in our construction:.
\begin{lem}
\label{lem:four-fermion-exchange}Let $V\subset\dC$
be a finite set and $F \colon \set{ \pm 1 }^{V} \to  \C $. Let
$c_{1},c_{2},c_{3},c_{4}$ be distinct corners.
For $i<j$, let 
$\cordefline_{ij}:c_{i}\leftrightarrow c_{j}$
be corner defect lines which are disjoint when the indices have no
overlap, i.e., 
$\cordefline_{12}\cap\cordefline_{34} = \cordefline_{13}\cap\cordefline_{24}
 = \cordefline_{14} \cap\cordefline_{23} = \emptyset$.
If 
$\left(\cordefline_{12}\oplus\cordefline_{34}\right)\oplus\left(\cordefline_{13}
\oplus\cordefline_{24}\right)$
and 
$\left(\cordefline_{12}\oplus\cordefline_{34}\right)\oplus\left(\cordefline_{14}
\oplus\cordefline_{23}\right)$
do not contain loops surrounding any point of~$V$, then for any 
large enough $\ddomain$
(with arbitrary boundary conditions) we have 
\begin{align*}
\phantom{-} \correl{ F\left(\spin|_{V}\right) \,
    \left(\ferm_{c_{1}} \ferm_{c_{2}}\right)_{\cordefline_{12}}
    \left(\ferm_{c_{3}} \ferm_{c_{4}}\right)_{\cordefline_{34}}
  }
\, = \; & 
  - \correl{
    F\left(\spin|_{V}\right) \,
    \left(\ferm_{c_{1}} \ferm_{c_{3}}\right)_{\cordefline_{13}}
    \left( \ferm_{c_{2}} \ferm_{c_{4}}\right)_{\cordefline_{24}}
  } \\
\, = \; & \phantom{-} \correl{ F\left(\spin|_{V}\right) \,
    \left(\ferm_{c_{1}} \ferm_{c_{4}}\right)_{\cordefline_{14}}
    \left(\ferm_{c_{2}} \ferm{c_{3}} \right)_{\cordefline_{23}}
  } .
\end{align*}
\end{lem}
\begin{proof}
Let us only prove the first equality (the other one is symmetric).
For each loop 
$\ell\in\left(\cordefline_{12}\oplus\cordefline_{34}
\right)\oplus\left(\cordefline_{ 13} \oplus\cordefline_{24}\right)$,
which by assumption does not surround points of~$V$,
define the gauge transform $\gaugetr_{\ell}$ that flips the spins
inside it as before. We can assume that $\ell$ is a loop that includes
edges of $\cordefline_{12}$, $\cordefline_{34}$, $\cordefline_{13}$ and 
$\cordefline_{24}$:
otherwise the gauge transform $\gaugetr_{\ell}$ just amounts to
displacing a piece of an individual defect line, without exchanging
the lines endpoints,
and this case
can be handled by an application of 
Lemma~\ref{lem:two-fermion-defect-line-independence}.
By composing enough gauge transforms which do not affect the left-hand
and right-hand side, we can actually assume that $\ell$ is a square
surrounding only one spin, with its horizontal edges belonging to
$\cordefline_{12}$ and $\cordefline_{34}$ and its vertical edges belonging
to $\cordefline_{13}$ and $\cordefline_{24}$. It is then elementary to check
that $\gaugetr_{\ell}$ affects the fermions in the desired way
(see Figure \ref{fig:gauge-transformation-four-fermion}).

\begin{figure}[tb]
\centering
\subfigure[Corner lattice fermion pairs 
$ \left(\ferm_{c_{1}} \ferm_{c_{2}}\right)_{\cordefline_{12}}
  \left(\ferm_{c_{3}} \ferm_{c_{4}}\right)_{\cordefline_{34}}$.]
{
    \includegraphics[scale=0.8]{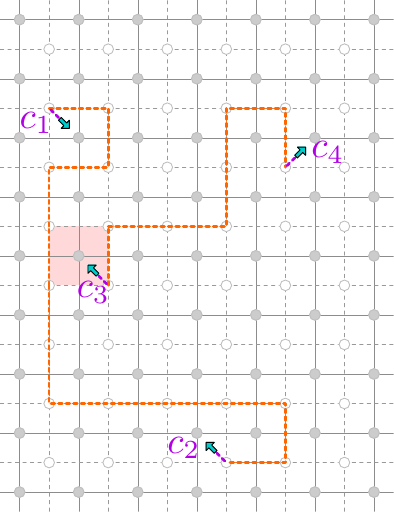}
	\label{sfig: discrete Laplacian}
}
\hspace{1.5cm}
\subfigure[Corner lattice fermion pairs 
$ \left(\ferm_{c_{1}} \ferm_{c_{3}}\right)_{\cordefline_{13}}
  \left(\ferm_{c_{2}} \ferm_{c_{4}}\right)_{\cordefline_{24}}$.]
{
    \includegraphics[scale=0.8]{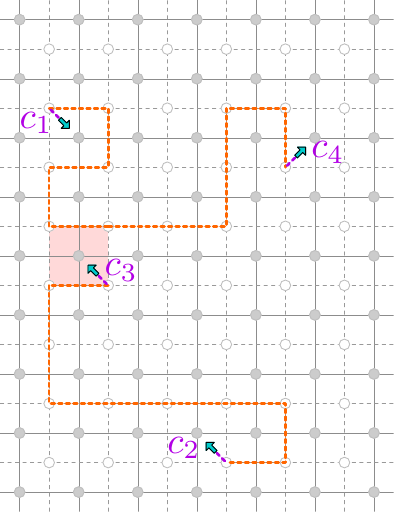}
	\label{sfig: discrete deebar}
}
\caption{\label{fig:gauge-transformation-four-fermion}A gauge transformation
involving two fermion pairs, and consisting of a flip of the spin
at the center of the shaded square.}
\end{figure}
%
\end{proof}
The above lemma generalizes to the following.
\begin{prop}
\label{prop:n-fermion-exchange}Let $V\subset\dC$
be finite. Let $c_{1},\ldots,c_{2k}$ be distinct corners.
Let
\[\Lambda=\left\{ \cordefline_{1}:c_{1}\leftrightarrow 
c_{2},\ldots,\cordefline_{k}:c_{2k-1}\leftrightarrow c_{2k}\right\} \]
be a collection of $k$ disjoint corner defect lines and let $\tilde{\Lambda}$
be a collection of $k$ disjoint corner defect lines 
$\tilde{\cordefline}_{j}: c_{m_{j}}\leftrightarrow c_{n_{j}}$
linking the $c_{1},\ldots,c_{2k}$ pairwise, with $m_{j}<n_{j}$.
Let $\Lambda\oplus\tilde{\Lambda}$ the set of loops made of the symmetric
difference 
${\big(\cup\cordefline_{j}\big) 
\oplus \big(\cup\tilde{\cordefline}_{j}\big)}$,
and let $\mathcal{N}$ denote the number of pairs $\left(v,\ell\right)$
with $v\in V$ and $\ell\in\Lambda\oplus\tilde{\Lambda}$ surrounding~$v$. Let 
$\mathcal{C}$ denote the number of crossings of the pair
partition $\left\{ \left(m_{j},n_{j}\right)\right\} $
of $\set{ 1, \ldots, 2k } $,
i.e. the number of pairs $j<k$ such that $m_{j}<m_{k}<n_{j}<n_{k}$.
Consider the Ising model on a large enough domain~$\ddomain$ with arbitrary
boundary conditions. We have 
\begin{align*}
& \correl{ \left(\prod_{v\in V}\spin_{v}\right)
    \big( \ferm_{c_{1}} \ferm_{c_{2}} \big)_{\cordefline_{1}}
    \cdots
    \big( \ferm_{c_{2k-1}} \ferm_{c_{2k}} \big)_{\cordefline_{k}}
  } \\
= & \left(-1\right)^{\mathcal{N}+\mathcal{C}} \;
  \correl{
    \left( \prod_{v\in V} \spin_{v} \right)
    \big( \ferm_{c_{m_{1}}} \ferm_{c_{n_{1}}} \big)_{\cordefline_{1}}
    \cdots
    \big(\ferm_{c_{m_{k}}} \ferm_{c_{n_{k}}} \big)_{\cordefline_{k}}
  } .
\end{align*}
\end{prop}
\begin{proof}
This follows from Lemmas \ref{lem:two-fermion-defect-line-independence}
and \ref{lem:four-fermion-exchange}, by induction on $k$. 
\end{proof}
This then yields the following important proposition, which allows
one to avoid specifying the defect paths: 
\begin{prop}
\label{prop:fermion-double-cover}Let $V\subset\dC$
be a finite and connected set and let $\ddomain \supset V$ be a large
enough domain. Consider the Ising model on $\ddomain $ with arbitrary
boundary conditions. Let $\dblcov{\ddomain , V}$ denote the double
cover of $\ddomain \setminus V$ ramified around $V$. 
For $c_{1},\ldots,c_{2k} \in \dblcov{\ddomaincor , V}$
and any $F \colon \{ \pm 1 \}^{V} \to  \C $, the correlation
\begin{align*} 
& \correl{ \phantom{\Big| \!\!} F \big( \spin|_{V} \big) \,
  \ferm(c_{1}) \cdots \ferm(c_{ 2k})
    } \\
:= \; & 
\correl{ \phantom{\Big| \!\!} F \big( \spin|_{V} \big)
    \big( \ferm_{c_{1}} \ferm_{c_{2}} \big)_{\cordefline_{1}} \cdots 
    \big( \ferm_{c_{2k-1}} \ferm_{c_{2k}} \big)_{\cordefline_{k}}
  }
\end{align*}
is independent of the choice of $\cordefline_{1}:c_{1}\leftrightarrow 
c_{2},\ldots,\cordefline_{k}:c_{2k-1}\leftrightarrow c_{2k}$,
provided the $\cordefline_{j}$'s stay away from $V$ and that 
$c_{2j-1},c_{2j}$
are on the same sheet of $\dblcov{ \dC, V }$ when
going along~$\cordefline_{j}$. The resulting correlation
$\correl{ \phantom{\big| \!}F\big(\spin|_{V}\big) \,
  \ferm(c_{1}) \cdots \ferm(c_{ 2k}) } $
is totally antisymmetric with respect to permutations of the variables
$c_{1},\ldots,c_{2k}$. It is single-valued as a function of each~$c_j$ if 
$F$ is even and has $-1$ monodromy around~$V$ if $F$ is odd. 
\end{prop}
\begin{proof}
By the Proposition \ref{prop:n-fermion-exchange}, the only dependence
on a path such as $\cordefline_{1}$ is through its lift to the double
cover (if we modify $\cordefline_{1}$ by a symmetric difference of two
loops both surrounding~$V$, it does not change the correlations).
If we modify~$\cordefline_{1}$ by a loop surrounding~$V$, the 
correlation will change
sign if~$F$ is odd and stay constant if~$F$ is even. The antisymmetry
follows from Lemmas~\ref{lem:corner-fermion-antisymmetry} 
and~\ref{lem:two-fermion-defect-line-independence}. 
\end{proof}
\subsubsection{\label{subsec:medial-lattice-fermions}Discrete Holomorphic 
Fermions}

We now introduce the discrete holomorphic fermions, which live on
the medial lattice: informally they are simply the averages of the
corner fermions taken at the four corners surrounding a medial vertex.
At criticality, their correlations are discrete holomorphic (Proposition
\ref{prop:discrete-holomorphicity-fermion-correlations}). 
\begin{defn}\label{def:medial-defect-line}
Let $w,z\in\dCmdl$ be medial vertices. We define a \emph{medial
defect line}~$\mdldefline$ with medial-ends $w,z$ as the concatenation
$\left[wp\right]\oplus\disline\oplus\left[qz\right]$ where $p,q \in \dCdual$
are adjacent to $w,z$ and $\disline$ is a simple path on the dual
lattice, 
called the \emph{main part} of~$\mdldefline$. We say that two (corner, medial) 
defect lines 
\emph{differ only locally}, if their endpoints are either the
same or neighbors and the main parts differ by at most the dual edges
containing the endpoints. 
\end{defn}

Let us now introduce the key object: the discrete holomorphic fermion,
which lives on the medial lattice (see Figure~\ref{sfig: medial defect line}).


\begin{defn}\label{def:lattice-holomorphic-fermion}
For $z \in \dCmdl$ and $x \in \set {\pm1\pm \ii}$, 
let~$z_{x} := z + \frac{\meshsz}{4} x$ denote the 
corner adjacent to~$z$ in direction~$x$;
for $w \in \dCmdl$ and $y \in \set {\pm1\pm \ii}$ write 
$w_{y} := w+ \frac{\meshsz}{4} y$ analogously.
Fix a medial defect line~$\mdldefline$ with medial ends~$z,w$, and let
$\cordefline_{yx}$ denote the corner defect lines with corner ends~$w_{y}$
and~$z_{x}$ such that 
the main parts of $\cordefline_{yx}$ and $\mdldefline$
differ at most by the edges containing~$w,z$.
We define the \emph{discrete holomorphic fermion pair} 
$\left( \ferm(w)  \ferm(z)  \right)_{\mdldefline}$
by 
\[
\left(\ferm(w)  \ferm(z) \right)_{\mdldefline}
= \frac{\pi}{8\sqrt{2}} \sum_{x,y \in \set{\pm1 \pm \ii}}
  \left(\ferm_{w_{y}} \ferm_{z_{x}}\right)_{\cordefline_{yx}},
\]
where if $w_{y}=z_{x}$, we interpret
$-\bar{\corndir}\left(w_{y}\right)
e^{-\frac{\ii}{2}\mathbf{W}\left(\cordefline_{yx}\right)}
:= \frac{\overline{z}-\overline{w}}{|z-w|}$
in Definition \ref{def:lattice-fermion}. 
\end{defn}

\begin{rem}
The correlations of 
$\left( \ferm(w)  \ferm(z)  \right)_{\mdldefline}$
taken in a domain with $+$ boundary conditions and without any other
fields correspond to the observable 
$\sum_{\zeta,\xi} \frac{\sqrt{\zeta}}{\sqrt{\xi}}
    f_{\ddomain} \left( w^{\zeta} , z^{\xi} \right)$
of \cite{hongler}, where the sum is taken over the possible orientations
of the edges $e(w) $ and $e(z) $. 
\end{rem}

The antisymmetry is naturally inherited from the corner-lattice fermion. 
\begin{lem}
\label{lem:medial-fermion-antisymmetry}For a medial defect line
$\mdldefline$ with (distinct) medial-ends $w,z$ have that 
\[ \left( \ferm(w)  \ferm(z)  \right)_{\mdldefline}
    = - \left( \ferm(z)  \ferm(w)  \right)_{\mdldefline} . \]
\end{lem}
\begin{proof}
Straightforward from Lemma \ref{lem:corner-fermion-antisymmetry}. 
\end{proof}

By Lemma~\ref{lem:two-fermion-defect-line-independence} the  
correlations of this fermion pair are independent of the choice of the defect 
line~$\mdldefline$, up to a sign. We thus omit the defect line from the 
notation, and consider the correlations defined on the appropriate double cover.
A fundamental property of the correlations of
$\left( \ferm(w)  \ferm(z)  \right)$
is their discrete holomorphicity apart from singularities when~$w$ and~$z$ 
coincide.
\begin{prop}
\label{prop:discrete-holomorphicity-fermion-correlations}Let $\ddomain 
\subset \dC$
be a discrete domain and let $V \subset \ddomain $ be a connected set.
Consider the Ising model on $\ddomain $ with arbitrary boundary conditions.
Let $\dblcov{\ddomain , V}$ denote the double cover of $\ddomain \setminus V$
ramified around $V$. Let $W \subset V$, let $w \in \dblcov{\ddomainmdl , V}$
and let $w^{*} \in \dblcov{\ddomainmdl , V}$ share the same base
point, on the opposite sheet. Consider the function 
$H \colon \dblcov{\ddomainmdl , V} \to  \C $
defined by 
\[
H(z) :=
  \correl{ \left(\prod_{x\in W}\spin_{x}\right)
  \ferm(w) \ferm(z) } .
\] 
Then for $\zeta \in \dblcov{\ddomaindmd , V}$ away from
$\partial\ddomain , V$, we 
have
\begin{align*}
\ddeebar H(\zeta)
= \; & 0
\qquad\qquad\qquad \text{ if $\zeta\not\sim w,w^{*}$} \\
\text{ and } \qquad
\ddeebar H(\zeta) 
= \; & \frac{\pi}{2} \, \correl{ \prod_{x\in W}\spin_{x} }
\qquad \text{ if $\zeta \sim w$}.
\end{align*}
\end{prop}
\begin{proof}
See Section \ref{subsec:discrete-holomorphicity}. 
\end{proof}
We will mostly use the above result in the following form: 
\begin{cor}
\label{cor:fermion-dance}
Let $W \subset V \subset \dC$, $w \in \dblcov{\ddomainmdl , V}$, and
$H \colon \dblcov{\ddomainmdl, V} \to \C$ be as above.
If $\contour$ is a
contractible closed contour on the double cover~$\dblcov{\ddomaincor, V}$
such that
$\interior{\contour} \supset
    \set{ w \pm \frac{\meshsz}{2},w\pm\frac{\ii \meshsz }{2} } $
and such that $\interior{\contour} \cap V = \emptyset$ and 
$f \colon \dblcov{\ddomaindmd, V} \to  \C $
is discrete holomorphic on $\interior{\contour}$, then 
\[
\frac{1}{2 \pi \ii} 
\dcint{\contour} f(z_{\dmd}) 
  H(z_{\mdl}) \, \dd{z}
= \frac{1}{4} \, \correl{ \prod_{x\in W} \spin_{x} }
  \sum_{x \in \set{\pm\frac{\meshsz}{2},\pm\frac{\ii \meshsz }{2}}} f(w_{x}) .
\]
\end{cor}
\begin{proof}
Using the discrete Stokes' formula~\eqref{eq:discrete-stokes-formula},
the contractible contour~$\contour$ can 
be deformed to the trivial contour plaquette by plaquette, and only the 
plaquettes with non-zero~$\ddeebar H$ contribute to the contour integral.
Proposition~\ref{prop:discrete-holomorphicity-fermion-correlations}
states that these plaquettes are exactly the $\zeta\sim w$ and gives the 
values~$\ddeebar H(\zeta)$.
\end{proof}
\begin{rem}
We can say informally that 
$z \mapsto \ferm(w)  \ferm(z) $
has four discrete poles of residue~$\frac{1}{4}$ at the four 
diamond vertices next to~$w$, as long as we avoid $\partial\ddomain $, $V$ and
$w^{*}$. 
\end{rem}

\section{\label{sec:virasoro-algebra-at-lattice-level}Virasoro Algebra at
the Lattice Level}

In this section, we implement at the lattice level the
Virasoro mode operators $\virL{n}$, $n \in \Z$,
for the discrete Gaussian free field and the Ising model.
The idea is to consider the relevant Laurent modes of the discrete 
holomorphic current and fermion, respectively, to obtain commutation relations
and to construct the Virasoro modes from them.
At all stages of the construction we need to make sure that
we indeed obtain lattice local fields (modulo lattice null fields)
and hence work on probabilistic objects rather than just on an abstract
notion of fields.
It is quite remarkable
that this is possible at all, given how rigid the theory of discrete
complex analysis is (in particular, the fact that the product of two
discrete holomorphic functions is not discrete holomorphic in general),
and given how simple our definition of lattice local field is.

Throughout this section, discrete contour integration is performed over 
(closed counterclockwise) discrete contours~$\contour$ 
on the corner lattice, as in Section~\ref{sec:discrete-complex-analysis},
and we denote by $\interior{\contour}$ the set
of points surrounded by $\contour$.
For a lattice local field $\genoper$,
we will denote by $\opernbrhoodof{\genoper}$ the smallest disk 
$D\left(0,\rho\right)$
centered at the origin such that $\genoper(0) $ does
not depend on field values outside of $D\left(0,\rho\right)$. For
$k\in\Z \cup\left(\Z +\frac{1}{2}\right)$, we will
denote by $D_{k}$ the smallest disk $D\left(0,\rho\right)$ such
that $\dmon{z}{k}$ is discrete holomorphic outside of 
$D\left(0,\rho\right)$.

\subsection{Gaussian free field}

We first address the case of the discrete Gaussian free field.
Let us denote by
$\locfieldsDGFF$ the space of dGFF local fields and
$\nullfieldsDGFF$ the subspace of
null fields, as in Definitions~\ref{def:lattice-local-field}
and~\ref{def:null-lattice-local-field}. Our 
construction of the Laurent mode operators 
will take place in the space
\[ \correqfieldsDGFF = 
\locfieldsDGFF / \nullfieldsDGFF\]
of correlation equivalence classes of local fields of dGFF.

\subsubsection{Current Modes}

Let us recall that we denote by $\curr (z) = \ii \, \partial\DGFF(z)$ 
the discrete holomorphic current.
Below we define its discrete Laurent modes~$\currmode{k}$, $k \in \Z$.
\begin{defn}
Let $\genoper \in \locfieldsDGFF$ be a lattice local field and let $k\in\Z $.
Let $\contour$ be such that $\interior{\contour}\supset 
\opernbrhoodof{\genoper} \cup D_k$.
To define a new local 
field~$\currmode{k}^{\contour} \genoper \in \locfieldsDGFF$, it is enough to 
specify its value~$\currmode{k}^{\contour}\genoper(0)$ at the origin, since 
local fields are given by a 
translation invariant rule.
We define $ \currmode{k}^{\contour}\genoper(0) $ by 
\[ \currmode{k}^{\contour}\genoper(0)
= \frac{1}{2\pi\ii} \dcint{\contour} \genoper(0)
    \curr (z_{\mdl}) \dmon{z_{\dmd}}{k} \, \dd{z} .
\]
\begin{center}
\begin{overpic}[scale=0.6]{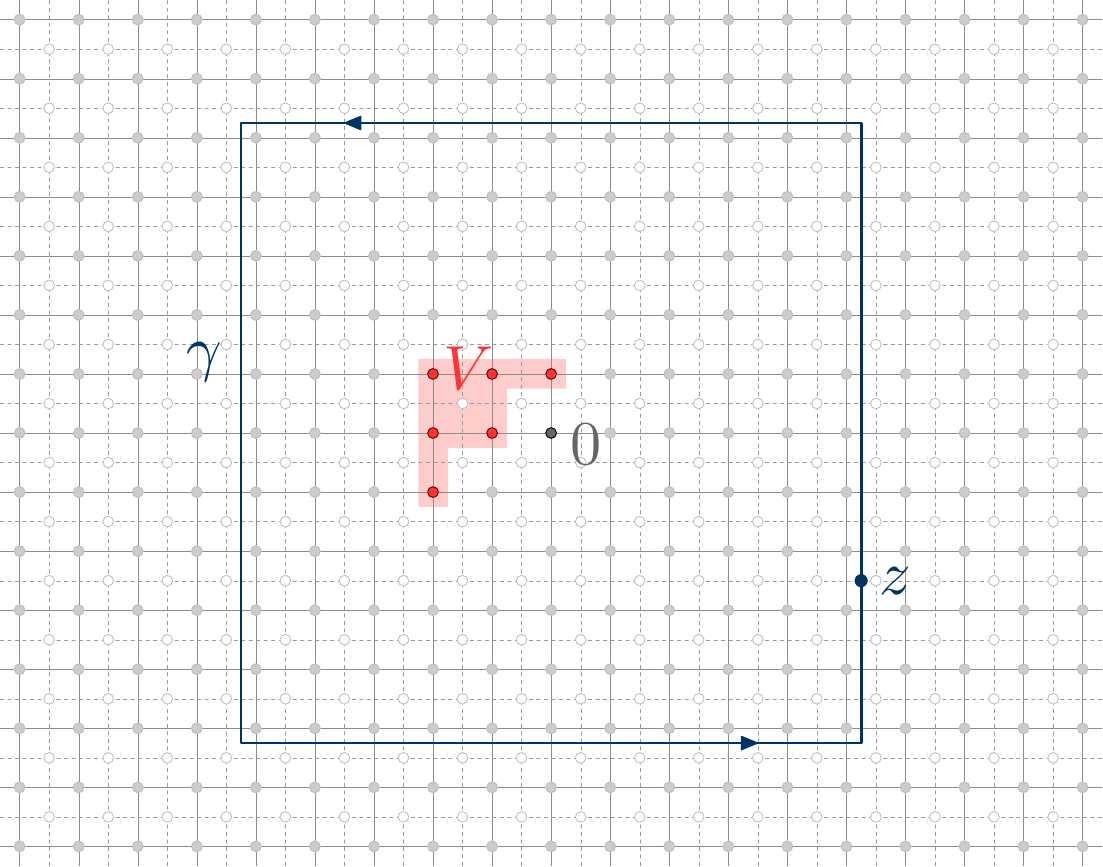}
\end{overpic} \\
\end{center}
\end{defn}

\begin{lem}
\label{lem:equivalence-current-modes}
The following properties hold for $\currmode{k}^{\contour}\genoper$:
\begin{itemize}
\item[(a)] If $\genoper \in \locfieldsDGFF$ is a local field and
$\contour,\contouralt$ are two
large enough 
closed counterclockwise contours, then we have
$ { \currmode{k}^{\contour}\genoper -
\currmode{k}^{\contouralt}\genoper \in \nullfieldsDGFF } $.
\item[(b)]
If $\genoper, \widetilde{\genoper} \in \locfieldsDGFF$ are two local 
fields such that
$\genoper - \widetilde{\genoper} \in \nullfieldsDGFF$ and 
$\contour$ is a large enough closed counterclockwise contour, then we have
$ \currmode{k}^{\contour}\genoper -
\currmode{k}^{\contour}\widetilde{\genoper} \in \nullfieldsDGFF$.
\end{itemize}
\end{lem}
\begin{proof}
independent 
%
Consider two different contours $\contour$ and $\contouralt$.
It follows from discrete holomorphicity of~$\dmon{z_{\dmd}}{k}$
(Proposition~\ref{prop:integer-monomials}(3)), discrete holomorphicity of 
correlations of~$\curr (z_{\mdl})$
(Lemma~\ref{lem:current-is-discrete-hol}), and
the contour deformation
property~\eqref{eq:contour-deformation-property}, that the correlations of
$\currmode{k}^{\contour} \genoper(0)$ and
$\currmode{k}^{\contouralt} \genoper(0)$ with other insertions at large enough 
distance are equal. In other words,
$\currmode{k}^{\contouralt} \genoper - \currmode{k}^{\contour} \genoper$
is null. This proves~(a).

Consider then two different local fields~$\genoper$ and~$\widetilde{\genoper}$
such that $\genoper - \widetilde{\genoper}$ is null. Provided that the contour 
$\contour$ is sufficiently large, then for any $z_{\mdl}$ on
$\contour$, the correlations of
$\big( \genoper(0) - \widetilde{\genoper}(0) \big) \curr (z_{\mdl})$
with other insertions sufficiently far away are vanishing. Therefore, by 
linearity, the correlations
of $\currmode{k}^{\contour} \genoper(0) -
\currmode{k}^{\contour} \widetilde{\genoper}(0)$ with other insertions are 
also vanishing, so indeed
$\currmode{k}^{\contour} \genoper - \currmode{k}^{\contour} 
\widetilde{\genoper}$
is null. This proves~(b).
\end{proof}

This lemma hence allows one to define the current mode action on the space
$\correqfieldsDGFF$ of dGFF local fields modulo null fields. 
\begin{defn}
\label{def:current-modes-gff}
For a lattice local field $\genoper \in \locfieldsDGFF$
and $k \in \Z $, we define
$\currmode{k}\genoper \in \correqfieldsDGFF$ as the correlation
equivalence class
\begin{align*}
\currmode{k}\genoper :=
\currmode{k}^{\contour} \genoper + \nullfieldsDGFF ,
\end{align*}
which is independent of the choice of a large 
enough $\contour$ by Lemma~\ref{lem:equivalence-current-modes}(b).
Moreover, by Lemma~\ref{lem:equivalence-current-modes}(a) we have that
the $\currmode{k}\genoper$ only depends on the correlation equivalence class
of $\genoper$, so this defines operators
\[ \currmode{k} \colon
\correqfieldsDGFF \to \correqfieldsDGFF . \]
\end{defn}

\begin{figure}
\begin{overpic}[scale=0.4]{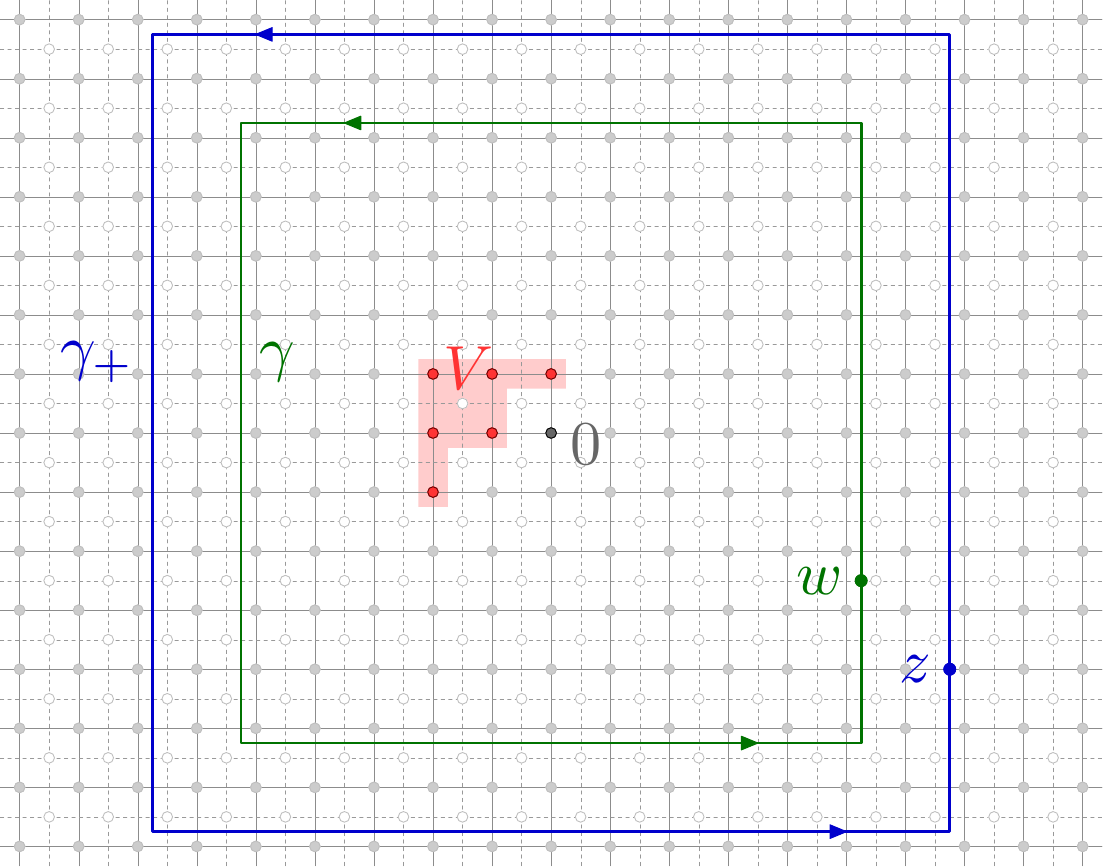}
      \put (103,37) {$-$}
\end{overpic}\qquad
\begin{overpic}[scale=0.4]{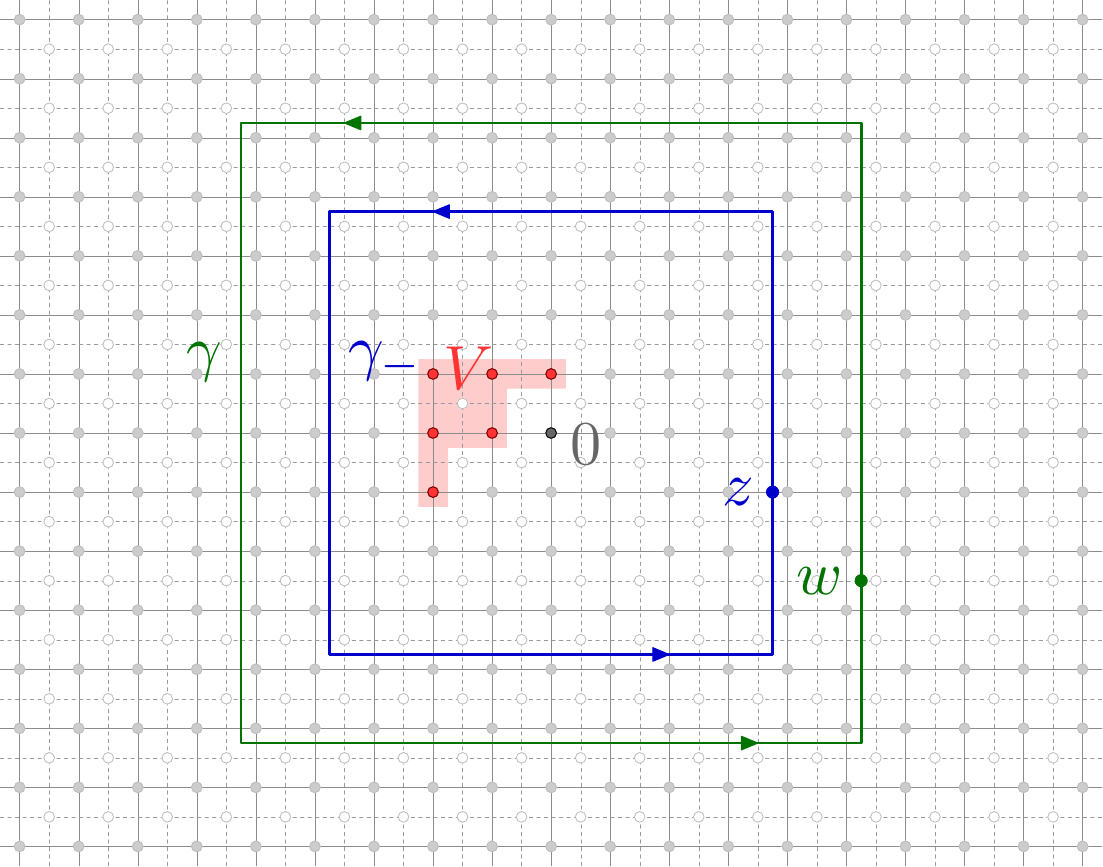}
\end{overpic} \\
\vspace{.5cm}
\begin{overpic}[scale=0.4]{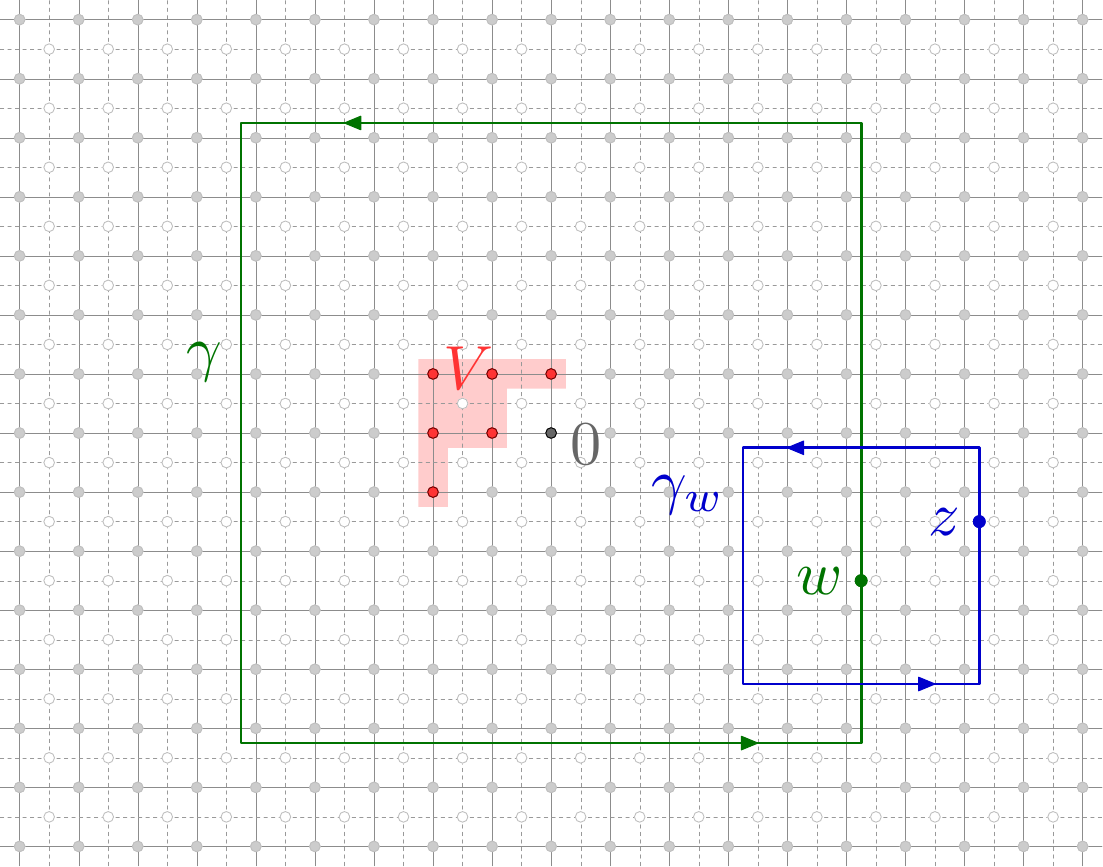}
      \put (-10,37) {$=$}
\end{overpic}

\caption{\label{fig:satellite-integral}The commutator 
$\left[ \currmode{k}, \currmode{\ell}\right]$
is expressed as the difference of two integrals where the order of the~$z$ 
and~$w$ contours are exchanged, yielding a `satellite' contour
integral over $z$ `orbiting' around $w$, which itself orbits around~$0$ and the 
finite set~$V$ of points used to define the local field~$\genoper$ on which 
the modes are acting.
}
\end{figure}

The following annihilation property for the current modes will be
important to define the Virasoro modes below.
\begin{lem}
\label{lem:finite-energy-dgff}
Let $\genoper$ be a lattice local
field. There exists $K>0$ such that for any $k\geq K$, $ \currmode{k}\genoper=0$
modulo $\nullfieldsDGFF$. 
\end{lem}
\begin{proof}
With a large but fixed contour $\contour$,
by Item~7 of Proposition~\ref{prop:integer-monomials}
we can choose $K$ such that for all $z_{\diamond}$ on the contour and $k 
\geq K$ we have $\dmon{z_{\diamond}}{k} = 0$.
It then follows directly from the definition that 
$\currmode{k}^{\contour} \genoper = 0$.
\end{proof}
\begin{prop}
\label{prop:gff-modes-heisenberg-algebra}
The operators
$\left( \currmode{k}\right)_{k\in\Z }$ form a representation of the
Heisenberg algebra on $\correqfieldsDGFF$, i.e.,
for all $k,l \in \Z$ we have the commutation relation
\[ \left[ \currmode{k}, \currmode{\ell} \right]
= \; k \delta_{k,-\ell} \; \, \idof{\correqfieldsDGFF} .
\]
\end{prop}
\begin{proof}
Let $\genoper$ be a lattice local field
and let 
$\contour,\contoursmall,\contourlarge$ be sufficiently large closed 
counterclockwise contours nested around each other so that
$\interior{\contoursmall} \subset \interior{\contour}
\subset \interior{\contourlarge}$.
The local field
${ \currmode{k}^{\contourlarge} \currmode{\ell}^{\contour}\genoper
    - \currmode{\ell}^{\contour} \currmode{k}^{\contoursmall}\genoper }$
is by definition a difference of two double contour integrals.
The discrete holomorphicity of the
current~$\curr$ (Lemma~\ref{lem:current-is-discrete-hol})
allows us to deform these contours within correlations.
More precisely, if $V \subset \dC$ is a finite set located
far enough outside the outermost contour~$\contourlarge$ and
$F(\DGFF|_V)$ is a polynomial of the values of the DGFF in~$V$,
then we can write
\begin{align*}
& \correl{ \left( 
\currmode{k}^{\contourlarge} \currmode{\ell}^{\contour}\genoper(0)
    - \currmode{\ell}^{\contour} \currmode{k}^{\contoursmall} \genoper(0)
  \right) F(\DGFF|_V)} \\
= \; & \correlsym \Bigg[
    \frac{1}{(2 \pi \ii)^2} \bigg(
    \dcint{\contourlarge} \dd{z} \dcint{\contour} \dd{w}
      \; \curr (z_{\mdl}) \curr (w_{\mdl})
      z_{\dmd}^{\left[k\right]} w_{\dmd}^{\left[\ell\right]} \\
& \qquad\qquad - \dcint{\contour} \dd{w} \dcint{\contoursmall} \dd{z}
      \; \curr (w_{\mdl}) \curr (z_{\mdl})
      z_{\dmd}^{\left[k\right]} w_{\dmd}^{\left[\ell\right]}
  \bigg) \, \genoper(0) \, F(\DGFF|_V) \Bigg] \\
= \; & \correl{ \frac{1}{(2 \pi \ii)^2} \left( \dcint{\contour} 
    \Big( \dcint{\contour_{w}} 
    \curr (z_{\mdl}) \curr (w_{\mdl})
    z_{\dmd}^{\left[k\right]} w_{\dmd}^{\left[\ell\right]} \; \dd{z}
    \Big) \dd{w} \right)
  \genoper(0) \, F(\DGFF|_V) } ,
\end{align*}
where the $w$-integration is kept intact while for each fixed $w$ the 
difference of the $z$-integrals have been combined and deformed to
a ``satellite integral'' along a part $\contour_w$ encircling~$w$ clockwise
(see Figure~\ref{fig:satellite-integral}).
We then apply Lemma \ref{lem:two-current-contour-integral}
to evaluate the inner satellite integral, using also
properties of discrete monomials (Proposition~\ref{prop:integer-monomials}), 
and 
we get
\begin{align*}
& \correl{ \left( 
\currmode{k}^{\contourlarge} \currmode{\ell}^{\contour}\genoper(0)
    - \currmode{\ell}^{\contour} \currmode{k}^{\contoursmall} \genoper(0) 
  \right)  F(\DGFF|_V)} \\
= \; & \correl{ \frac{1}{2 \pi \ii} \left( \dcint{\contour} 
    k w_{\mdl}^{\left[k-1\right]}
    w_{\dmd}^{\left[\ell\right]}\, \dd{w} \right)
  \genoper(0) \, F(\DGFF|_V) } \\
= \; & k \delta_{k,-\ell} \; 
  \correl{ \phantom{\big|} \genoper(0) \, F(\DGFF|_V) } .
\end{align*}
This equality shows that up to null fields we have
\begin{align*}
\currmode{k}^{\contourlarge} \currmode{\ell}^{\contour} \genoper(0) 
    - \currmode{\ell}^{\contour} \currmode{k}^{\contoursmall} \genoper(0) \,
\equiv \; k \delta_{k,-\ell} \; \genoper(0) ,
\end{align*}
which proves the asserted commutation relation of $\currmode{k}$
and~$\currmode{\ell}$.
\end{proof}

\subsubsection{\label{subsec:gff-vir-modes}Gaussian free field Virasoro Modes}
\begin{defn}
We define the operator 
$\virL{n}^{\abbrDGFF} 
\colon \correqfieldsDGFF \to \correqfieldsDGFF$
by 
\[
\virL{n}^{\abbrDGFF}
    = \frac{1}{2} \sum_{k\in\Z}
    \ourorder{\currmode{n-k}}{\currmode{k}}{k}
\]
where we set
\begin{align*}
\ourorder{A}{B}{k} =
\begin{cases}
A B & \text{ if $k \geq 0$} \\
B A & \text{ if $k < 0$} .
\end{cases}
\end{align*}
\end{defn}

\begin{rem}
By virtue of Lemma \ref{lem:finite-energy-dgff},
the sum above is well-defined as an operator on the space~$\correqfieldsDGFF$ of
lattice local fields: there are only finitely many non-null terms, when the sum 
acts on
(the correlation equivalence class of) any given
lattice local field.
\end{rem}

\begin{rem}
Our choice of the above definition is guided by the convenience of 
the calculations below, but it can also be easily seen to agree with the common
definition
$\virL{n}^{\abbrDGFF}
= {\frac{1}{2}\sum_{k\in\Z }: \currmode{n-k} 
\currmode{k}:}$
where the normal-ordered product ${: \currmode{j} \currmode{k}:}$ is defined as 
$\currmode{j} \currmode{k}$
if $j\leq k$ and as $\currmode{k} \currmode{j}$ otherwise. 
\end{rem}

\begin{lem}
\label{lem:current-virasoro-commutation}
For any $n,m,k \in \Z$, we have
\begin{align*}
\big[\virL{n}^{\abbrDGFF}, \currmode{m} \big]
& = - m \, \currmode{n+m} \\
\big[ \virL{n}^{\abbrDGFF}, 
\ourorder{\currmode{m-k}}{\currmode{k}}{k} \big]
& = - k \ourorder{\currmode{m-k}}{\currmode{n+k}}{k}
    - \left(m-k\right) \ourorder{\currmode{n+m-k}}{\currmode{k}}{k} .
\end{align*}
\end{lem}
\begin{proof}
The second formula follows from the first by using the commutator identity
$[A,BC] = B[A,C] + [A,B]C$, so it remains to prove the first formula.
Observe first that by the commutator 
identity $[AB,C] = A[B,C] + [A,C]B$
and Proposition~\ref{prop:gff-modes-heisenberg-algebra} we have,
when $k + \ell = n$,
\begin{align*}
[\currmode{\ell} \currmode{k} , \currmode{m}]
= \; & \currmode{\ell} [ \currmode{k} , \currmode{m} ]
    + [\currmode{\ell} , \currmode{m} ] \currmode{k} \\
= \; & \currmode{\ell} \; k \, \delta_{k+m,0}
		+ \ell \, \delta_{\ell+m,0} \; \currmode{k}
= -m \, (\delta_{k,-m} + \delta_{k,n+m}) \, \currmode{n+m} .
\end{align*}
Using this, we calculate
\begin{align*}
\big[ \virL{n}^{\abbrDGFF}, \currmode{m} \big]
= \; & \frac{1}{2} \; \sum_{k \in \Z}
    \big[ \ourorder{\currmode{n-k}}{\currmode{k}}{k} , \currmode{m} \big] \\
= \; & \frac{1}{2} \; \sum_{k \in \Z}
    \Big( -m \, (\delta_{k,-m} + \delta_{k,n+m}) \, \currmode{n+m} \Big)
= -m \,  \currmode{n+m} ,
\end{align*}
where by virtue of Proposition~\ref{lem:finite-energy-dgff}, the sums over~$k$ 
again have only finitely many non-zero terms when acting on 
a given correlation equivalence class of local fields.
\end{proof}
\begin{thm}
\label{thm:gff}The operators $\left(\virL{n}^{G}\right)_{n\in\Z }$
form a representation of the Virasoro algebra with central
charge~${c=1}$, namely 
\[
\left[\virL{n}^{\abbrDGFF},\virL{m}^{\abbrDGFF}\right]
= 
\left(n-m\right) \virL{n+m}^{\abbrDGFF}
+ \frac{1}{12} \delta_{n+m,0} \left(n^{3}-n\right) \idof{\correqfieldsDGFF} .
\]
\end{thm}
\begin{proof}
Let us omit the $\abbrDGFF$ superscript. To compute 
$\left[\virL{n},\virL{m}\right]$,
write 
$\virL{m} 
= \frac{1}{2}\sum_{k\in\Z } \ourorder{\currmode{m-k}}{\currmode{k}}{k}$,
use the second formula of Lemma \ref{lem:current-virasoro-commutation},
and perform a change of variables $\ell = k+n$ in the first part to
obtain that 
\begin{align*}
\left[\virL{n},\virL{m}\right]
= \; & \frac{1}{2} \sum_{k\in\Z} \big[ \virL{n}, 
		\ourorder{\currmode{m-k}}{\currmode{k}}{k} \big]
	= \frac{1}{2} \sum_{k\in\Z} \Big(
		- k \ourorder{\currmode{m-k}}{\currmode{n+k}}{k}
		- (m-k) \ourorder{\currmode{n+m-k}}{\currmode{k}}{k} \Big) \\
= \; & \frac{1}{2} \sum_{\ell\in\Z} 
		(n-\ell) \ourorder{\currmode{n+m-\ell}}{\currmode{\ell}}{\ell-n} 
	+ \frac{1}{2} \sum_{k\in\Z} (k-m) 
		  \ourorder{\currmode{n+m-k}}{\currmode{k}}{k} \\
= \; & (n-m) \, \virL{n+m} + \mathcal{A}_{n,m} ,
\end{align*}
where 
\[
\mathcal{A}_{n,m} 
:= \frac{1}{2}\sum_{\ell\in\Z } (n-\ell) \Big(
	\ourorder{\currmode{n+m-\ell}}{\currmode{\ell}}{\ell-n}
	- \ourorder{\currmode{n+m-\ell}}{\currmode{\ell}}{\ell} \Big) .
\]
Now note that by Proposition~\ref{prop:gff-modes-heisenberg-algebra} we get
\begin{align*}
\ourorder{\currmode{n+m-\ell}}{\currmode{\ell}}{\ell-n}
	- \ourorder{\currmode{n+m-\ell}}{\currmode{\ell}}{\ell}
= \; & \begin{cases}
       \ell \, \delta_{n+m,0} & \text{ if } 0 \leq \ell < n \\
       -\ell \, \delta_{n+m,0} & \text{ if } n \leq \ell < 0 \\
       0 & \text{ otherwise} .
       \end{cases}
\end{align*}
Observing furthermore that
$\sum_{\ell=0}^{n} (n-\ell) \ell = \frac{1}{6}(n^3-n)$ for $n\geq0$, we can 
therefore simplify
\begin{align*}
\mathcal{A}_{n,m} 
= \; & \frac{1}{12} \, (n^3-n) \, \delta_{n+m,0} ,
\end{align*}
and conclude the proof.
\end{proof}

\subsection{Ising Model}


The Ising model involves a number of additional difficulties. First
of all, since the fermion field is not local, its modes cannot act
on local fields, and we must hence consider modes of fermion pairs.
Second, since the fermion is quasi-local with respect to spin-antisymmetric
fields, we must take half-integer power Laurent
modes of it (while taking integer
modes of it when acting on spin-symmetric fields),
see Definition~\ref{def:fermion-bilinear-modes}. Third, since the
fermion pairs involve defect lines, one must be careful while choosing
them and exchanging them to get the right commutation modes.

Recall that we denote by
$\locfieldsIsing$ the space of Ising local fields,
$\nullfieldsIsing$ the subspace of
null fields, and by
\[ \correqfieldsIsing = 
\locfieldsIsing / \nullfieldsIsing \]
the space of correlation equivalence classes of local fields of the Ising model.
Note that the space of local fields splits to a direct sum
$\locfieldsIsing = \locfieldsIsingEven \oplus \locfieldsIsingOdd$
of spin-symmetric fields
(fields~$\genoper_{\meshsz}$ which are even under global spin flip
$\spin \mapsto -\spin$,
i.e., $\genoper_{\meshsz} [-\spin] = \genoper_{\meshsz}[\spin]$)
and spin-antisymmetric fields (fields~$\genoper_{\meshsz}$ which are odd under 
global spin flip,
i.e., $\genoper_{\meshsz} [-\spin] = - \genoper_{\meshsz}[\spin]$).
All considerations below will be done separately for these two sectors.


\subsubsection{Fermion Modes}
We now define fermion Laurent mode pairs in the even and odd sector, separately.
In the even sector we use integer powers, but half-integer indices, and in the 
odd sector we use half-integer powers but integer indices~--- this convention 
of indexing is due to half-integer scaling dimension of the fermion field.
As before, since lattice local fields are translation invariant in
their definition, to define a lattice local field, we only need to
give the definition
at the origin. 

\begin{defn}
Let~$\genoper \in \locfieldsIsingEven$ be a spin-symmetric
lattice local field. 
Let \[ p,q\in\Z +\frac{1}{2} . \]
Set $E := D_{p}\cup D_{q}\cup \opernbrhoodof{\genoper}$. Let 
$\contour, \contourbis$ be counterclockwise closed paths such that 
$\interior{\contour} \supset \interior{\contourbis} \supset E$
and let $\cordefline$
be a choice of medial defect
lines that does not cross $ \R_{-}\cup E$ for all
$z,w \in \dCmdl \setminus E$.
We define
\begin{align*}
\left( \fermmode^{\contour}_{p} \fermmode^{\contourbis}_{q}
    \genoper \right)_{\cordefline} (0) 
= \; & \frac{1}{\left(2\pi \ii \right)^{2}}
  \dcint{\contour} \dcint{\contourbis} \genoper
  (0) \left(\ferm(z_{\mdl})  \ferm(w_{\mdl}) \right)_{
  \cordefline : z_{\mdl}\leftrightarrow w_{\mdl}} 
	\dmon{z_{\dmd}}{p-\frac{1}{2}}
    \dmon{w_{\dmd}}{q-\frac{1}{2}}
  \, \dd{z} \dd{w} .
\end{align*}
\end{defn}
\begin{defn}
Let $\genoper\in \locfieldsIsingOdd$ be a spin-antisymmetric lattice local 
field. Let
\[ k , \ell\in\Z .\]
Set $E := D_{k} \cup D_{\ell} \cup 
\opernbrhoodof{\genoper}$.
Let $\contour,\contourbis$ be counterclockwise closed paths such that 
$\interior{\contour} \supset \interior{\contour} \supset E$
and let $\cordefline$ be a choice of medial defect
lines that does not cross $ \R_{-} \cup E$ for all
$z,w \in \dCmdl \setminus E$.
We define
\begin{align*}
\left( \fermmode_{p}^{\contour}\fermmode_{q}^{\contourbis} \genoper 
\right)_{\cordefline}(0) 
= \; & \frac{1}{\left(2\pi \ii \right)^{2}}
    \dcint{\contour} \dcint{\contourbis} \genoper (0) 
  \left(\ferm(z_{\mdl})  \ferm(w_{\mdl}) \right)_{ 
  \cordefline : z_{\mdl}\leftrightarrow w_{\mdl}}
	\dhimon{z_{\dmd}}{k-\frac{1}{2}}
	\dhimon{w_{\dmd}}{\ell-\frac{1}{2}}
  \, \dd{z} \dd{w},
\end{align*}
where the branches of $\dhimon{z}{k-\frac{1}{2}}$ and 
$\dhimon{w}{\ell-\frac{1}{2}}$
are the principal branches on $ \C \setminus \R_{-}$
(i.e. the real part is nonnegative on the positive real axis).
\end{defn}
\begin{figure}
\includegraphics[scale=0.8]{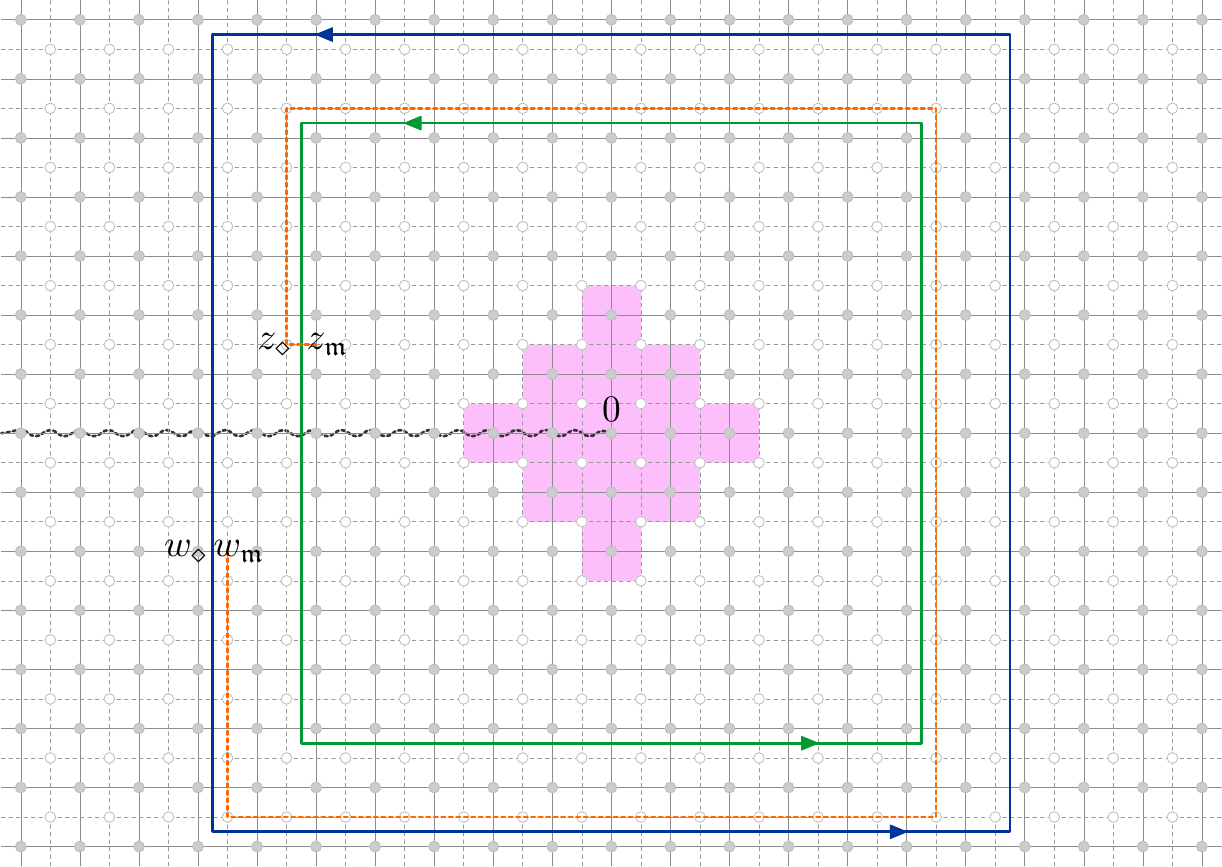}
\caption{
The setup for the integrand of the definition of 
$\left(\fermmode_{p}^{\contour}\fermmode_{q}^{\contourbis}\genoper\right)_{
\cordefline}.$
The set~$E$ is the shaded area. The contours with arrows are the integration
contours $\contourbis$ (inner, green) and $\contour$ (outer, blue) 
respectively.
The marked medial vertices are the $z_{\mdl}$ and $w_{\mdl}$
that appear in the integral (respectively on the inner contour and
on the outer contour), and on which the fermion $\ferm$ is defined.
The marked diamond vertices are the $z_{\dmd}$ and
$w_{\dmd}$ (on the inner and outer contours) on which the monomials
are defined. An example of two pairs $z_{\mdl},z_{\dmd}$ and 
$w_{\mdl},w_{\dmd}$
is drawn here.
The main part of the corresponding medial defect
line is drawn in dashed orange stroke. The defect lines are chosen not to cross 
the branch cut on~$ \R_{-}$, marked by the wiggly line.
}
\end{figure}

The following lemma tells us that the choices of $\contourbis$, $\contour$
and $\cordefline$ are essentially irrelevant, modulo null fields. Similarly
the choice of the branch cut on~$\R_{-}$ is irrelevant as
well.
\begin{lem}
\label{lem:equivalence-fermion-modes-local-fields}
The following properties hold 
for~$\left( \fermmode_{p}^{\contour}\fermmode_{q}^{\contourbis} \genoper 
\right)_{\cordefline}$:
\begin{itemize}
\item[(a)]
If $\genoper \in \locfieldsIsingEven$ is a
spin-symmetric local field, and $\cordefline$ and $\widetilde{\cordefline}$ are 
two choices of the defect lines as above,
then we have for $p,q \in \Z + \frac{1}{2}$
\[ \big( \fermmode^{\contour}_{p} \fermmode^{\contourbis}_{q}
    \genoper \big)_{\cordefline} -
\big( \fermmode^{\contour}_{p} \fermmode^{\contourbis}_{q}
    \genoper \big)_{\widetilde{\cordefline}} \in \nullfieldsIsingEven . \]
The analogous property also holds for
$\big( \fermmode^{\contour}_{k} \fermmode^{\contourbis}_{\ell}
    \genoper \big)_{\cordefline}$
for any spin-antisymmetric field
${\genoper \in \locfieldsIsingOdd}$ and~$k,\ell \in \Z$.
\item[(b)] If $\genoper \in \locfieldsIsingEven$ is a spin-symmetric local 
field and $\contour,\contourbis$ and $\contouralt,\contourbisalt$ are two
pairs of large enough
closed counterclockwise contours, then
for any defect lines $\widetilde{\cordefline}$ chosen for the 
contours $\contouralt,\contourbisalt$, we have
\[ \big( \fermmode^{\contour}_{p} \fermmode^{\contourbis}_{q}
    \genoper \big)_{\cordefline} -
\big( \fermmode^{\contouralt}_{p} \fermmode^{\contourbisalt}_{q}
    \genoper \big)_{\widetilde{\cordefline}} \in \nullfieldsIsingEven . \]
The analogous property also holds for
$\big( \fermmode^{\contour}_{k} \fermmode^{\contourbis}_{\ell}
    \genoper \big)_{\cordefline}$
where ${\genoper \in \locfieldsIsingOdd}$ is any spin-antisymmetric field and 
$k,\ell \in \Z$.
\item[(c)]
If $\genoper, \widetilde{\genoper} \in \locfieldsIsingEven$ are two
spin-symmetric local fields such that
$\genoper - \widetilde{\genoper} \in \nullfieldsIsingEven$ and 
$\contour, \contourbis$ are two large enough closed counterclockwise contours, 
then we have for $p,q \in \Z + \frac{1}{2}$
\[ \big( \fermmode^{\contour}_{p} \fermmode^{\contourbis}_{q}
    \genoper \big)_{\cordefline} -
\big( \fermmode^{\contour}_{p} \fermmode^{\contourbis}_{q}
    \widetilde{\genoper} \big)_{\cordefline} \in \nullfieldsIsingEven . \]
The analogous property also holds for
$\big( \fermmode^{\contour}_{k} \fermmode^{\contourbis}_{\ell}
    \genoper \big)_{\cordefline}$
where ${\genoper, \widetilde{\genoper} \in \locfieldsIsingOdd}$
are any spin-antisymmetric fields and~$k,\ell \in \Z$.
\end{itemize}
%
\end{lem}
%
\begin{proof}
To prove statements modulo null fields,
one argues within correlations as in
Lemma~\ref{lem:equivalence-current-modes}.
The independence on the choice of defect 
lines follows from Lemma~\ref{lem:two-fermion-defect-line-independence}.
The rest of the proof is similar to that of 
Lemma~\ref{lem:equivalence-current-modes}.
Part~(b) relies on the discrete holomorphicity of the correlations
involving two fermion insertions (Proposition 
\ref{prop:discrete-holomorphicity-fermion-correlations}) and
and the discrete holomorphicity
of the discrete monomials (Propositions~\ref{prop:integer-monomials}
and~\ref{prop:half-integer-monomials}). 
For the spin-symmetric case, one furthermore needs to observe the
single-valuedness of the fermion around spin-symmetric fields
Proposition~\ref{prop:fermion-double-cover}, whereas for the spin-antisymmetric 
case one
uses Lemma~\ref{lem:product-two-double-valued-functions}, observing that 
both the half-integer monomial and the fermion have 
monodromy~$-1$ around the set~$E$.
%
%
\end{proof}
This lemma again allows one to define the fermion mode pair action on the spaces
$\correqfieldsIsingEven$ and $\correqfieldsIsingOdd$ of Ising local fields 
modulo null fields. 
\begin{defn}
\label{def:fermion-bilinear-modes}
\label{def:current-modes-gff}
For a spin-symmetric lattice local field $\genoper \in \locfieldsIsingEven$
and $p,q \in \Z + \frac{1}{2}$, we define
$\fermmode_{p} \fermmode_{q} \, \genoper
\in \correqfieldsIsingEven$ as the correlation equivalence class
\begin{align*}
\fermmode_{p} \fermmode_{q} \, \genoper :=
\big( \fermmode^{\contour}_{p} \fermmode^{\contourbis}_{q}
    \genoper \big)_{\cordefline} + \nullfieldsIsingEven ,
\end{align*}
which is independent of the choice of a large 
enough contours $\contour$, $\contourbis$ and defect lines
$\cordefline$ by Lemma~\ref{lem:equivalence-fermion-modes-local-fields}(a,b).
Moreover, by Lemma~\ref{lem:equivalence-fermion-modes-local-fields}(c) we have 
that
the $\fermmode_{p} \fermmode_{q} \, \genoper$ only 
depends on the correlation equivalence class of $\genoper$, so this defines 
operators
\[ \fermmode_{p} \fermmode_{q} \; \colon \;
\correqfieldsIsingEven \to \correqfieldsIsingEven . \]

Similarly, for a spin-antisymmetric lattice local field
$\genoper \in \locfieldsIsingOdd$ and $k,\ell \in \Z$, we define
the correlation equivalence class
$\fermmode_{k} \fermmode_{\ell} \, \genoper
\in \correqfieldsIsingOdd$ and obtain operators
\[ \fermmode_{k} \fermmode_{\ell} \; \colon \;
\correqfieldsIsingOdd \to \correqfieldsIsingOdd . \]
\end{defn}

The following annihilation property for the fermion modes will be
important to define the Virasoro modes below. 
\begin{lem}
\label{lem:bounded-energy-fermion-modes}
Let $\genoper \in \locfieldsIsingEven$
(resp. $\genoper \in \locfieldsIsingOdd$)
be a spin-symmetric (resp. spin-antisymmetric) lattice local field.
Then there exists $K>0$ such that for any
$p,q \in \Z + \frac{1}{2}$ with $q\geq K$
(resp. $k,\ell \in \Z$ with $\ell\geq K$), 
we have $\fermmode_{p} \fermmode_{q} \genoper = 0 \in \correqfieldsIsingEven$
(resp. 
$\fermmode_{k} \fermmode_{\ell} \genoper = 0 \in \correqfieldsIsingOdd$).
\end{lem}
\begin{proof}
With large but fixed contours $\contour$, $\contourbis$,
by Item~7 of Proposition~\ref{prop:integer-monomials}
(respectively Item~7 of Proposition~\ref{prop:half-integer-monomials})
we can choose $K$ such that for all $w_{\diamond}$ on the contour and
$q \geq K$ we have $\dmon{w_{\diamond}}{q} = 0$.
It then follows directly from the definition that 
$\big( \fermmode^{\contour}_{p} \fermmode^{\contourbis}_{q}
    \genoper \big)_{\cordefline} = 0$.
\end{proof}
\begin{prop}
\label{prop:fermion-modes-anticommutation-relations}For $p,q,k,l\in\Z $
or $p,q,k,l\in\Z +\frac{1}{2}$, we have the following anticommutation
relations: 
\begin{align}
\fermmode_{p}\fermmode_{q}+\fermmode_{q}\fermmode_{p} & 
= \delta_{p,-q} \; \id 
\label{eq:first-anticommutation}\\
    \left(\fermmode_{k}\fermmode_{p}\right) 
\left(\fermmode_{q}\fermmode_{\ell}\right)
  + \left(\fermmode_{k}\fermmode_{q}\right) 
\left(\fermmode_{p}\fermmode_{\ell}\right) & 
= \delta_{p,-q} \; \fermmode_{k}\fermmode_{\ell} .
\label{eq:second-anticommutation}
\end{align}
\end{prop}
\begin{proof}
The calculations are to be performed modulo null fields, i.e., it is sufficient 
to prove equalities within correlations as in
Proposition~\ref{prop:gff-modes-heisenberg-algebra}. We will indicate 
equalities up to null fields by~$\equiv$.

Let $\genoper$ be a lattice local field. For an integer or half-integer
$n$, set $n_{-} := n-\frac{1}{2}$. Set $E := D\left(p\right)\cup 
D\left(q\right)\cup 
D\left(\genoper\right)$.
Let us first prove the first identity~(\ref{eq:first-anticommutation}).
For $\interior{\contour}\supset \interior{\contourbis}\supset 
\interior{{\tilde{\contour}}}\supset E$,
we have (omitting the defect lines
$\cordefline:z_{\mdl}\leftrightarrow w_{\mdl}$ to
lighten the notation), 
\begin{align*}
\big( \fermmode^{\contour}_{p} \fermmode^{\contourbis}_{q} + 
    \fermmode^{\contourbis}_{q} \fermmode^{\tilde{\contour}}_{p}
    \big) \genoper(0) 
= \; & \phantom{+} \frac{1}{\left(2\pi \ii \right)^{2}}
  \dcint{\contour} \dcint{\contourbis} 
  \genoper(0) 
  \left(\ferm(z_{\mdl}) \ferm\left(w_{\mdl} \right)\right)
	\dmon{z_{\dmd}}{p_{-}}
    \dmon{w_{\dmd}}{q_{-}}
  \, \dd{z} \dd{w}\\
& + \frac{1}{\left(2\pi \ii \right)^{2}}
  \dcint{\contourbis} \dcint{{\tilde{\contour}}}
  \genoper(0) 
  \left(\ferm(w_{\mdl}) \ferm(z_{\mdl}) \right)
	\dmon{w_{\dmd}}{q_{-}}
	\dmon{z_{\dmd}}{p_{-}}
  \, \dd{w} \dd{z}.
\end{align*}
We can now use 
$\left(\ferm(w)  \ferm(z) \right)
= - \left(\ferm(z)  \ferm(w) \right)$
to rewrite the above (within correlations)
as a satellite integral as above 
\[
\frac{1}{2\pi \ii }\dcint{\contourbis} 
  \genoper(0) 
  \left(\frac{1}{2\pi \ii }
\dcint{\contour(w)}
  \left(\ferm\left(w_{\mdl} \right)
    \ferm(z_{\mdl}) \right)
  \dmon{z_{\dmd}}{p_{-}}
  \, \dd{z} \right)
  \dmon{w_{\dmd}}{q_{-}}
  \, \dd{w},
\]
where $\contour(w) $ is a small contour with
$\interior{\contour} \supset \set{ w \pm \meshsz , w \pm \ii \meshsz }$
and $\interior{\contour}\cap E=\emptyset$. Now, by Corollary 
\ref{cor:fermion-dance},
we have (within correlations) 
\[
\left(\frac{1}{2\pi \ii } \dcint{ \contour(w) }
\left( \ferm(w_{\mdl})  
\ferm(z_{\mdl})  \right)
	\dmon{z_{\dmd}}{p_{-}}
  \, \dd{z} \right)
\equiv \frac{1}{4}\sum_{x=\pm1,\pm \ii }
  \dmon{w_{x}}{p_{-}}
=: \dmonavg{w_{\mdl}}{p_{-}} 
\]
By Property~10 of Proposition~\ref{prop:integer-monomials} and Property~9 of 
Proposition~\ref{prop:half-integer-monomials} (see also 
Lemma~\ref{lem:ising-contour-integral} below), we have 
\[
\frac{1}{2\pi \ii } \dcint{\contourbis}  \left(
	\dmon{w_{\dmd}}{q_{-}}
    \dmonavg{w_{\mdl}}{p_{-}}
    \right)  \dd{w}
= \delta_{p,-q},
\]
which yields 
\[
\big( \fermmode^{\contour}_{p} \fermmode^{\contourbis}_{q} + 
    \fermmode^{\contourbis}_{q} \fermmode^{\tilde{\contour}}_{p}
    \big) \genoper
\equiv \delta_{p,-q} \; \genoper .
\]

The proof of the second identity (\ref{eq:second-anticommutation})
is similar: we have that the integrand in the definitions 
$\fermmode_{k}\fermmode_{p}\fermmode_{q}\fermmode_{\ell}\genoper(0) $
and $\fermmode_{k}\fermmode_{q}\fermmode_{p}\fermmode_{\ell}\genoper(0) $
is 
\[
\genoper(0) 
  \left(\ferm(\zeta_{\mdl})  \ferm(z_{\mdl}) \right)_{
    \iota:\zeta\leftrightarrow z}
\left( \ferm(w_{\mdl})  \ferm\left(\xi_{\mdl}\right)
    \right)_{\kappa:w\leftrightarrow\xi}  
  \dmon{\zeta_{\dmd}}{k_{-}}
  \dmon{z_{\dmd}}{p_{-}}
  \dmon{w_{\dmd}}{q_{-}}
  \dmon{\xi_{\dmd}}{\ell_{-}}
\]
where $\iota$ and $\kappa$ are chosen to avoid $\opernbrhoodof{\genoper}$
and not to cross the negative real axis.

By Lemma \ref{lem:four-fermion-exchange}, we can exchange the medial
defect lines 
\[
\big( \ferm(\zeta_{\mdl})  \ferm(z_{\mdl}) 
    \big)_{\iota:\zeta\leftrightarrow z}  
  \big(\ferm(w_{\mdl})  \ferm(\xi_{\mdl})
    \big)_{\kappa:w\leftrightarrow\xi} 
= \big( \ferm(\zeta_{\mdl})  \ferm(\xi_{\mdl})
    \big)_{ \eta:\zeta\leftrightarrow\xi} 
  \big( \ferm(z_{\mdl}) \ferm(w_{\mdl}) 
    \big)_{\lambda:z\leftrightarrow w},
\]
by choosing medial defect lines $\eta$ and $\lambda$ that avoid
$\opernbrhoodof{\genoper}$ and do not cross the negative real
axis.

Now,
$\fermmode_{k} \fermmode_{p} \fermmode_{q} \fermmode_{\ell}
+ \fermmode_{k} \fermmode_{q} \fermmode_{p} \fermmode_{\ell}$
can be evaluated summing the two quadruple integrals of 
\[
\left(\ferm(\zeta_{\mdl})  \ferm(\xi_{\mdl})
    \right)_{\eta:\zeta\leftrightarrow\xi}
  \left(\ferm(z_{\mdl}) \ferm(w_{\mdl}) 
    \right)_{\lambda:z\leftrightarrow w}
  \dmon{\zeta_{\dmd}}{k_{-}}
  \dmon{z_{\dmd}}{p_{-}}
  \dmon{w_{\dmd}}{q_{-}}
  \dmon{\xi_{\dmd}}{\ell_{-}}
\]
with order of the variables $z$ and $w$ swapped. We obtain that
the result of the second and third integrals can be written as the
same satellite integral as for the first identity, yielding the same
result. What is left is hence easily seen to be equal (up to null fields) to 
\[
\delta_{p,-q} \; \fermmode_{k}\fermmode_{\ell}\genoper(0) ,
\]
thus concluding the proof.
\end{proof}

\subsubsection{\label{subsec:ising-vir-modes}Ising Virasoro Modes}
For the next definition, let us introduce the notation
\[ \ourforder{A}{B}{k}
= \begin{cases}
  \; A B & \text{ if $k \geq 0$} \\
  - B A & \text{ if $k < 0$} .
  \end{cases}
\]
\begin{defn}
\label{def:ln-ising}
For $n \in \Z$, we define 
$\virL{n}^{\abbrEven}  \colon 
\correqfieldsIsingEven \to \correqfieldsIsingEven$
and $\virL{n}^{\abbrOdd}  \colon 
\correqfieldsIsingOdd \to \correqfieldsIsingOdd$
by 
\begin{align}
\virL{n}^{\abbrEven}
:= \; & \frac{1}{2} \sum_{k\in\Z +\frac{1}{2}} 
	k \, \ourforder{\fermmode_{n-k}}{\fermmode_{k}}{k} 
\label{eq:ln-ns-ising} \\
\virL{n}^{\abbrOdd}
:= \; & \frac{1}{2} \sum_{k\in\Z }
  k \, \ourforder{\fermmode_{n-k}}{\fermmode_{k}}{k}
  + \frac{1}{16} \, \delta_{n,0} \; \id .
\label{eq:ln-ram-ising}
\end{align}
\end{defn}
\begin{rem}
By virtue of Lemma~\ref{lem:bounded-energy-fermion-modes},
the sums above are well-defined as operators on the
spaces~$\correqfieldsIsingEven$  and~$\correqfieldsIsingOdd$
of lattice local fields: there are only finitely many non-null terms, when the 
sums act on (the correlation equivalence class of) any given
lattice local field.
\end{rem}
\begin{rem}
Using Lemma \ref{lem:no-fermion-square} below, the above definitions
can easily be checked to equivalently give the following commonly 
used formulas involving normal orderings
\begin{align*}
\virL{n}^{\abbrEven} & =\begin{cases}
\frac{1}{2}\sum_{k\in\Z +\frac{1}{2}}\left(k+\frac{1}{2}\right):\fermmode_{n-k
}\fermmode_{k}: & \text{ if }n\neq0\\
\sum_{k=1}^{\infty}k\fermmode_{-k}\fermmode_{k} & \text{ if }n=0
\end{cases}\\
\virL{n}^{\abbrOdd} & = \begin{cases}
\frac{1}{2}\sum_{k\in\Z}\left(k+\frac{1}{2}\right):\fermmode_{n-k
}\fermmode_{k}: & \text{ if }n\neq0\\
\frac{1}{16} \; \id 
+\sum_{k=1}^{\infty}k\fermmode_{-k}\fermmode_{k} & \text{ if } n=0
\end{cases}
\end{align*}
where $:\fermmode_{j}\fermmode_{k}:$ is defined as $\fermmode_{j}\fermmode_{k}$ 
if $j\leq k$
and as $-\fermmode_{k}\fermmode_{j}$ otherwise. 
\end{rem}
\begin{lem}
\label{lem:no-fermion-square}
For any $s\in\Z $, we have 
\begin{align*}
\sum_{k\in\Z  + \frac{1}{2}} \ourforder{\fermmode_{s-k}}{\fermmode_{k}}{k}
\; = \; 0
\qquad \text{ and } \qquad
\sum_{k\in\Z }\ourforder{\fermmode_{s-k}}{\fermmode_{k}}{k}
\; = \; \frac{1}{2} \delta_{s,0} \; \id .
\end{align*}
\end{lem}
\begin{proof}
Rewriting the sum in terms of two equal pieces and changing variables
to~$k' = s-k$ 
in one of the two, we get
\begin{align*}
\sum_{k} \ourforder{\fermmode_{s-k}}{\fermmode_{k}}{k}
= \frac{1}{2} \sum_{k} \ourforder{\fermmode_{s-k}}{\fermmode_{k}}{k}
    + \frac{1}{2} \sum_{k'} 
            \ourforder{\fermmode_{k'}}{\fermmode_{s-k'}}{s-k'} .
\end{align*}
We conclude the
asserted formulas by the following cancellations among the two pieces
\begin{align*}
\ourforder{\fermmode_{s-k}}{\fermmode_{k}}{k}
    + \ourforder{\fermmode_{k}}{\fermmode_{s-k}}{s-k}
= \begin{cases}
  \id & \text{ if $s=k=0$} \\
  0 & \text{ otherwise}.
  \end{cases}
\end{align*}
The case~$s=k=0$ here is simply the anticommutation relation
$2 \, \fermmode_{0} \fermmode_{0} = \id$ of 
Proposition~\ref{prop:fermion-modes-anticommutation-relations}.
In the cases where $k, s-k$ have the same sign, the cancellation follows from
the anticommutation relation
$\fermmode_{s-k} \fermmode_{k} + \fermmode_{k} \fermmode_{s-k} = 0$.
In the cases where $k, s-k$ have different signs, the cancellation follows
from the definition of $\ourforder{A}{B}{k}$.
\end{proof}
\begin{lem}
\label{lem:commutation-vir-fermion}
For any $n\in\Z $, if
$j,k\in\Z +\frac{1}{2}$ we have 
\[
\Big[ \virL{n}^{\abbrEven}, \ourforder{\fermmode_{j}}{\fermmode_{k}}{k} \Big]
= - \big( j+\frac{n}{2} \big)
      \ourforder{\fermmode_{n+j}}{\fermmode_{k}}{k}
  - \big( k+\frac{n}{2} \big)
      \ourforder{\fermmode_{j}}{\fermmode_{n+k}}{k} .
\]
Also, for any $n,j,k\in\Z $, we have 
\[
\Big[ \virL{n}^{\abbrOdd} , \ourforder{\fermmode_{j}}{\fermmode_{k}}{k} \Big] 
= - \big( j+\frac{n}{2} \big)
      \ourforder{\fermmode_{n+j}}{\fermmode_{k}}{k}
  - \big( k+\frac{n}{2} \big)
      \ourforder{\fermmode_{j}}{\fermmode_{n+k}}{k} .
\]
\end{lem}
\begin{proof}
It is sufficient to prove the statements without the reordering
$\ourforder{\cdot}{\cdot}{k}$, since possible reordering only amounts to 
interchanging $j$ and $k$ and changing signs.
Moreover, the proofs of both statements are completely similar, so we only 
detail the calculation for the first one.
Using Definition~\ref{def:ln-ising} and the anticommutation relations
of Proposition~\ref{prop:fermion-modes-anticommutation-relations}, and the 
identity
\[ AB CD - CD AB = A (BC+CB) D - (AC+CA) B D + C A (BD+DB) - C (AD+DA) B , \]
we get
\begin{align*}
\Big[ \virL{n}^{\abbrEven}, \fermmode_{j} \fermmode_{k} \Big]
= \; & \frac{1}{2} \sum_{\ell\in\Z +\frac{1}{2}} 
	\ell \, \Big( \ourforder{\fermmode_{n-\ell}}{\fermmode_{\ell}}{\ell}
				\fermmode_{j} \fermmode_{k}
	    - \fermmode_{j} \fermmode_{k} 	    
			  \ourforder{\fermmode_{n-\ell}}{\fermmode_{\ell}}{\ell} \Big) \\
= \; & \frac{1}{2} \sum_{\ell\in\Z +\frac{1}{2}} 
	\ell \, \Big( \delta_{\ell+j,0} \, \fermmode_{n - \ell} \fermmode_{k} 
	    - \delta_{n-\ell+j,0} \, \fermmode_{\ell} \fermmode_{k}
	    + \delta_{\ell+k,0} \, \fermmode_{j} \fermmode_{n - \ell} 
	    - \delta_{n-\ell+k,0} \, \fermmode_{j} \fermmode_{\ell} \Big) \\
= \; & \frac{1}{2} \Big(
		(-j) \, \fermmode_{n - \ell} \fermmode_{k} 
	    - (n+j) \, \fermmode_{\ell} \fermmode_{k}
	    + (-k) \, \fermmode_{j} \fermmode_{n - \ell} 
	    - (n+k) \, \fermmode_{j} \fermmode_{\ell} \Big) \\
= \; & - \big( j+\frac{n}{2} \big) \, \fermmode_{n+j}\fermmode_{k}
  - \big( k+\frac{n}{2} \big) \, \fermmode_{j}\fermmode_{n+k} .
\end{align*}
\end{proof}
\begin{thm}
\label{thm:ising-ns} The operators
$\left(\virL{n}^{\abbrEven}\right)_{n\in\Z }$
form a representation of the Virasoro algebra with central 
charge~${c=\frac{1}{2}}$.
\end{thm}
\begin{proof}
Using the definition of the Virasoro mode $\virL{m}^{\abbrEven}$ and 
Lemma~\ref{lem:commutation-vir-fermion}, we calculate
\begin{align*}
[\virL{n}^{\abbrEven} , \virL{m}^{\abbrEven}]
= \; & \frac{1}{2} \sum_{k\in\Z +\frac{1}{2}} 
	k \, \big[ \virL{n}^{\abbrEven} ,
		  \ourforder{\fermmode_{m-k}}{\fermmode_{k}}{k} \big] \\
= \; & \frac{-1}{2} \sum_{k\in\Z +\frac{1}{2}} 
	\Big( k \big( m-k+\frac{n}{2} \big)
	        \ourforder{\fermmode_{n+m-k}}{\fermmode_{k}}{k}
	    + k \big( k+\frac{n}{2} \big)
	        \ourforder{\fermmode_{m-k}}{\fermmode_{n+k}}{k}
	\Big) .
\end{align*}
Performing the change of variables $\ell = k+n$ in the second part of the sum 
and then combining the two parts, and using Lemma~\ref{lem:no-fermion-square}, 
we get
\begin{align*}
[\virL{n}^{\abbrEven} , \virL{m}^{\abbrEven}]
= \; & -\frac{1}{2} \sum_{k\in\Z +\frac{1}{2}} 
	k \big( m-k+\frac{n}{2} \big)
	        \ourforder{\fermmode_{n+m-k}}{\fermmode_{k}}{k}
	- \frac{1}{2} \sum_{\ell\in\Z +\frac{1}{2}} 
	    (\ell - n) \big( \ell-\frac{n}{2} \big)
	        \ourforder{\fermmode_{n+m-\ell}}{\fermmode_{\ell}}{\ell-n}  \\
= \; & -\frac{1}{2} \sum_{k\in\Z +\frac{1}{2}} 
	\Big( km-kn+\frac{n^2}{2} \Big)
	        \ourforder{\fermmode_{n+m-k}}{\fermmode_{k}}{k}
	+ \mathcal{B}_{n,m} \\
= \; & (n-m) \; \frac{1}{2} \sum_{k\in\Z +\frac{1}{2}}
			k \, \ourforder{\fermmode_{n+m-k}}{\fermmode_{k}}{k}
	+ \mathcal{B}_{n,m} \\
= \; & (n-m) \; \virL{n+m}^{\abbrEven} + \mathcal{B}_{n,m} ,
\end{align*}
where
\begin{align*}
\mathcal{B}_{n,m}
= \; & - \frac{1}{2} \sum_{\ell\in\Z +\frac{1}{2}} 
	    (\ell - n) \big( \ell-\frac{n}{2} \big) \Big(
	        \ourforder{\fermmode_{n+m-\ell}}{\fermmode_{\ell}}{\ell-n}
	        - \ourforder{\fermmode_{n+m-\ell}}{\fermmode_{\ell}}{\ell}
	\Big) .
\end{align*}
The theorem will be proven, if we show that
$\mathcal{B}_{n,m} = \frac{n^3-n}{24} \delta_{n+m,0} \, \id$.
Note that in the 
sum over~$\ell$ which defines~$\mathcal{B}_{n,m}$, whenever either
$\ell , \ell -n \geq 0$ or
$\ell , \ell -n < 0$ the two terms directly cancel. The finitely many 
remaining terms are simplified
using Proposition~\ref{prop:fermion-modes-anticommutation-relations} to 
constant multiples of the identity. For example for~$n > 0$, we 
get
\begin{align*}
\mathcal{B}_{n,m}
= \; & - \frac{1}{2} \sum_{\substack{\ell\in\Z +\frac{1}{2} \\ 0 < \ell < n}} 
	    (\ell - n) \big( \ell-\frac{n}{2} \big) \Big( \underbrace{
	        - \fermmode_{n+m-\ell}\fermmode_{\ell}
	        - \fermmode_{n+m-\ell}\fermmode_{\ell} }_{
					= - \delta_{n+m,0} \, \id } \Big) \\
= \; & + \frac{1}{2}
    \Big( \sum_{\substack{\ell\in\Z +\frac{1}{2} \\ 0 < \ell < n}} 
	    (\ell - n) \big( \ell-\frac{n}{2} \big) \Big) \; \delta_{n+m,0} \; \id 
\; = \; \frac{n^3-n}{24} \delta_{n+m,0} \, \id .
\end{align*}
The case~$n<0$ is similar.
\end{proof}
\begin{thm}
\label{thm:ising-r}The operators
$\left(\virL{n}^{\abbrOdd}\right)_{n\in\Z }$
form a representation of the Virasoro algebra with central 
charge~${c=\frac{1}{2}}$. 
\end{thm}
\begin{proof}
As in the proof of Theorem \ref{thm:ising-ns}, we need to show that
\[
\big[ \virL{n}^{\abbrOdd} , \virL{m}^{\abbrOdd} \big]
    - (n-m) \, \virL{n+m}^{\abbrOdd}
\; = \; \frac{n^3-n}{24} \, \delta_{n,-m} \; \id .
\]
If $n,m\neq0$ and $n+m\neq0$, the proof that the left hand side
vanishes is exactly the same as in 
Theorem~\ref{thm:ising-ns}
(the formulas for 
$\virL{n}^{\abbrOdd}, \virL{m}^{\abbrOdd}, \virL{n+m}^{\abbrOdd}$ are the same, 
and moreover the anticommutation
result of Proposition~\ref{prop:fermion-modes-anticommutation-relations},
the commutation result of Lemma~\ref{lem:commutation-vir-fermion} and the
cancellation result of Lemma~\ref{lem:no-fermion-square} apply in exactly the 
same way). If $n=m=0$, the result is trivial. If either~$n$ or~$m$ is zero, 
simply observe that the extra term $\frac{1}{16} \, \id$
commutes with anything, and hence that there is no difference.

The only non-trivial case to check is hence $n>0$ and $m=-n$. We calculate, 
using now the second statement in Lemma~\ref{lem:no-fermion-square},
\begin{align*}
[\virL{n}^{\abbrOdd} , \virL{-n}^{\abbrOdd}]
= \; & \frac{1}{2} \sum_{k\in\Z} 
	k \, \big[ \virL{n}^{\abbrOdd} ,
		  \ourforder{\fermmode_{-n-k}}{\fermmode_{k}}{k} \big] \\
= \; & \frac{-1}{2} \sum_{k\in\Z} 
	\Big( k \big( -k-\frac{n}{2} \big)
	        \ourforder{\fermmode_{-k}}{\fermmode_{k}}{k}
	    + k \big( k+\frac{n}{2} \big)
	        \ourforder{\fermmode_{-n-k}}{\fermmode_{n+k}}{k}
	\Big) \\
= \; & -\frac{1}{2} \sum_{k\in\Z} 
	k \big( -k-\frac{n}{2} \big)
	        \ourforder{\fermmode_{-k}}{\fermmode_{k}}{k}
	- \frac{1}{2} \sum_{\ell\in\Z} 
	    (\ell - n) \big( \ell-\frac{n}{2} \big)
	        \ourforder{\fermmode_{-\ell}}{\fermmode_{\ell}}{\ell-n}  \\
= \; & -\frac{1}{2} \sum_{k\in\Z} 
	\Big( -2 k n + \frac{n^2}{2} \Big)
	        \ourforder{\fermmode_{-k}}{\fermmode_{k}}{k}
	+ \mathcal{B}'_{n} \\
= \; & 2n \; \frac{1}{2} \sum_{k\in\Z} 
		k \, \ourforder{\fermmode_{-k}}{\fermmode_{k}}{k}
	- \frac{n^2}{8} \, \id
	+ \mathcal{B}'_{n} \\
= \; & 2n \; \virL{0}^{\abbrOdd}
	- \frac{n}{8} \, \id
	- \frac{n^2}{8} \, \id
	+ \mathcal{B}'_{n} ,
\end{align*}
where 
\begin{align*}
\mathcal{B}'_{n}
= \; & - \frac{1}{2} \sum_{\ell\in\Z} 
	    (\ell - n) \big( \ell-\frac{n}{2} \big) \Big(
	        \ourforder{\fermmode_{-\ell}}{\fermmode_{\ell}}{\ell-n}
	        - \ourforder{\fermmode_{-\ell}}{\fermmode_{\ell}}{\ell} \Big) \\
= \; & - \frac{1}{2} \sum_{\ell = 0}^{n-1} 
	    (\ell - n) \big( \ell-\frac{n}{2} \big) \Big(
	        - \fermmode_{\ell}\fermmode_{-\ell}
	        - \fermmode_{-\ell}\fermmode_{\ell} \Big) \\
= \; & \Big( \frac{1}{2} \sum_{\ell = 0}^{n-1} 
	    (\ell - n) \big( \ell-\frac{n}{2} \big) \Big) \; \id
\; = \; \frac{1}{24} \big( n^3 + 3 n^2 + 2 n \big) \; \id .
\end{align*}
Now simplifying, we obtain the desired form of the commutator,
\begin{align*}
[\virL{n}^{\abbrOdd} , \virL{-n}^{\abbrOdd}]
\; = \; 2n \, \virL{0}^{\abbrOdd}
	+ \frac{n^3-n}{24} \, \id.
\end{align*}
\end{proof}
We can now construct
the Virasoro representation on the Ising model local 
fields, by putting together the even and odd sectors treated above.
\begin{thm}
\label{thm:ising}Setting $\virL{n}^{\abbrIsing} := 
\virL{n}^{\abbrEven} \oplus \virL{n}^{\abbrOdd}$,
we have that the operators
$\left(\virL{n}^{\abbrIsing}\right)_{n \in \Z}$ form a representation
of the Virasoro algebra with central charge~${c=\frac{1}{2}}$ on 
$\correqfieldsIsing$,
the space of Ising lattice local fields modulo null fields. 
\end{thm}
\begin{proof}
This is immediate from Theorems \ref{thm:ising-ns} and \ref{thm:ising-r}.
\end{proof}

\section{\label{sec:discrete-complex-analysis-proofs}Discrete Complex Analysis
Proofs}

\subsection{Integer Monomial Construction}

The construction of the positive integer monomials is quite straightforward: 
\begin{lem}
\label{lem:pos-int-mon}There exists a unique family of functions
$\dmon{z}{p}  \colon \dCdmd\cup\dCmdl
\to \C $
for $p\geq0$ such that $z^{\left[0\right]}\equiv1$, $z^{[1]}=\meshsz^{-1}z$,
$\ddee \dmon{z}{p}  = p \dmon{z}{p-1} $, $\dmon{0}{p} = 0$
for all $p\geq1$, and 
$\left(\pm\frac{\meshsz}{2}\right)^{[p]}=\left(\pm\frac{\ii \meshsz 
}{2}\right)^{[p]}$
for all $p\geq2$. These functions satisfy the properties 1, 2, 3,
4, 7, 8 of Proposition \ref{prop:integer-monomials}. 
\end{lem}
\begin{proof}
It is not hard to check by induction on $p$, using discrete integration,
that the functions $\dmon{z_{\mdl}}{p}$ and 
$\dmon{z_{\dmd}}{p}$
are both discrete holomorphic and satisfy the relevant symmetry properties,
and are uniquely determined by the values near $0$. The convergence
as $\meshsz \to 0$ follows by discrete integration. The set of medial
points where $\dmon{z}{p+1}$ vanishes is easily shown to be
the set of neighbors of the diamond points where $\dmon{z}{p} $
vanishes (and vice versa), thus showing that the set of points where
$\dmon{z}{p} $ vanishes grows with $p$. 
\end{proof}
The construction of the negative integer monomials is quite simple,
too. 
\begin{lem}
\label{lem:neg-int-mon}
There exists a unique family of functions
$\dmon{z}{p}  \colon \dCdmd \cup \dCmdl \to \C $
for $p<0$ such that $\ddee \dmon{z}{p} = p \dmon{z}{p-1} $,
$\dmon{0}{p} = 0$ for all $p<0$ and with 
$\ddeebar \dmon{z_{\mdl}}{-1}
    = {2\pi} \mathbf{1}_{\set{ 0}} $
and 
$\ddeebar \dmon{z_{\dmd}}{-1}
= \frac{{2\pi}}{4} \sum_{x \in \set{\pm 1 , \pm \ii}}
    \mathbf{1}_{\set{x\meshsz/2} }$,
such that $\dmon{z}{-1}  \to 0$ as $z \to \infty$. These functions
satisfy the Properties 1, 6, and 8 of Proposition \ref{prop:integer-monomials}. 
\end{lem}
\begin{proof}
Recall that the 
discrete Cauchy kernel~$\CauchyKer \colon \dCmdl \to \C$ with
a discrete pole at $0$ is the unique 
function~$\CauchyKer \colon \dCmdl \to \C$ such that
$\ddeebar \CauchyKer = \mathbf{1}_{\set{0}}$ and $\CauchyKer$ decays at infinity
\cite{chelkak-smirnov-i, kenyon-i}~--- it can be constructed as the 
$\ddee$-derivative of the random walk potential 
kernel~$\potKer \colon \dC \to \R$ (see~\cite{lawler-limic}), extended as zero 
on the dual~$\dCdual$.
The function~$\dmon{z_{\mdl}}{-1}$ is then defined as~$2 \pi \CauchyKer$ 
on~$\dCmdl$.
Given this, the function $\dmon{z_{\dmd}}{-1}$ can be constructed
by $\dmon{z}{-1} =\frac{1}{4} \sum \dmon{z_{x}}{-1}$
where the sum is over the four medial neighbors $z_{x} \in \dCmdl$
to $z \in \dCdmd$; it is easy to check that
they satisfy Properties 1 (symmetry) and 6 (decay at infinity), and
Property 8 (convergence) is given by standard discrete complex analysis
techniques \cite{chelkak-smirnov-i}. The uniqueness for $p=-1$ follows
from the maximum principle. The other functions are readily determined
by repeated $\ddee$-differentiation, and it is elementary to check 
their properties. 
\end{proof}

\subsection{\label{subsec:construction-of-disc-sqrts}Construction of Discrete
Square Roots}

In this subsection, we construct the discrete analogues 
$\dhimon{z_{\mdl}}{-\frac{1}{2}},
\dhimon{z_{\mdl}}{\frac{1}{2}}
\colon \dblcov{\dCmdl,0} \to  \C $
of the functions 
$z \mapsto \frac{1}{\sqrt{z}}$ and $z \mapsto \sqrt{z}$
that we will need for our
construction. For this, we rely on constructions introduced in 
\cite{chelkak-hongler-izyurov,gheissari-hongler-park},
and modifications thereof.

Let us denote by~$\dCcormdl := \dCcor \cup \dCmdl$ the lattice formed 
by the corners and midpoints of edges of~$\dC$.
We say that a function $F \colon \dCcormdl \to \C $
is \emph{s-holomorphic} if for any $c\in\dCcor$
adjacent to $z \in \dCmdl , v \in \dC , f \in \dCdual$
we have 
\begin{equation}
F(c) =\frac{1}{2}\left(F(z) 
+\overline{\corndir(c)} \bar{F}(z) \right),\label{eq:s-holomorphicity}
\end{equation}
where $\corndir(c) =\left(v-f\right)/\left|v-f\right|$. 
This requirement will be called a ``projection relation'', since
it can be interpreted as stating that the value of the 
function at a corner~$c$ is the projection of the complex value at an 
adjacent midedge~$z$ to the 
line~$\sqrt{\bar{\corndir}(c)} \, \R \subset \C$ in the complex plane.

We extend the notion of s-holomorphicity
to functions defined on $\dblcov{\dCcormdl , 0}$
by defining adjacent as adjacent and on the same sheet.

\begin{figure}
\includegraphics[scale=0.8]{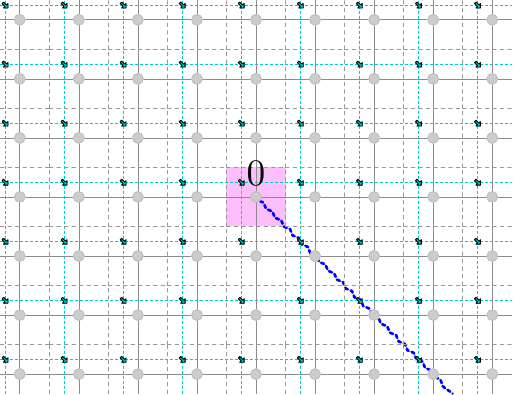}
\caption{The lattice of those corners whose direction is
$\corndir(c) = \varrho$. The wiggly
line denotes the branch cut~$\varrho \R_+$, and the corner immediately on the 
North-West of~$0$ is~$\frac{-1+\ii }{4}\meshsz$. The shaded square is~$Q$. }
\end{figure}

\subsubsection{Discrete Inverse Square Root}

In this paragraph, we introduce the discrete analogue of the function
$z^{-\frac{1}{2}}$ on the medial lattice. 
\begin{lem}
Set $\kappa := e^{\pi \ii /8}$, and $\varrho := e^{-\pi \ii /4}$. There exists
a unique s-holomorphic function 
$K_{\meshsz} \colon  \dblcov{ \dC^{\corn\mdl}, 0 } \to \C $
with $-1$ monodromy around $0$ such that $K_{\meshsz}\left(\frac{-1+\ii 
}{4}\meshsz\right)=\kappa$,
such that $K_{\meshsz}(c) =0$ for all corners with $\corndir(c) =\varrho$
on $\varrho \R_{+}$ and that decays at infinity. 
\end{lem}

\begin{proof}
Set $\Lambda_{\meshsz}
:= \set{ c \in \dblcov{ \dCcor , 0 } : \corndir(c) = \varrho } $
and let $\Lambda_{\meshsz}^{\pm}$ denote the two sheets 
of $\Lambda_{\meshsz}\setminus\varrho \R_{+}$.
We construct $K_{\meshsz}$ on $\Lambda_{\meshsz}^{\pm}$ as $\pm\kappa 
H_{\meshsz}$,
where $H_{\meshsz}(c) $ is the (Beurling-type) probability
that a simple random walk on $\Lambda_{\meshsz}$ starting from $c$
hits the ``tip'' $\frac{-1+\ii }{4}\meshsz$ before the slit 
$\varrho \R_{+}$. The function
$K_{\meshsz}$ is clearly discrete harmonic on $\Lambda_{\meshsz}$,
and also on the cut $\varrho \R_{+}$; indeed, the value on
the cut is zero by definition, as is the average of the values of
$H_{\meshsz}$ on the two sheets, since they are of opposite sign,
by symmetry. That it decays at infinity is standard. The uniqueness
follows from the maximum principle using the decay at infinity. Given
this, we can then use the same strategy as the proof of Lemma 2.14
(Section 3.2.2 and Remark 3.1) of \cite{chelkak-hongler-izyurov},
to uniquely reconstruct $K_{\meshsz}$ from values on $\Lambda_{\meshsz}$,
by enforcing a $-1$ mondromy around $0$. \footnote{Note that compared to our 
paper, the lattice is rotated by 45 degrees
in~\cite{chelkak-hongler-izyurov}, that the 
the lattice square side length is denoted there
by~$\sqrt{2}\,\delta$
that the primal graph
there corresponds to the dual graph.
}

An important difference with the situation of \cite{chelkak-hongler-izyurov}
is that there is no s-holomporphic singularity (failure of s-holomorphicity)
of $K_{\meshsz}$ anywhere. Indeed, the point where the discrete harmonicity
of $H_{\meshsz}$ fails is $\frac{-1+\ii }{4}\meshsz$, which is contained
within the (dual lattice) square $Q$ of sidelength $\meshsz$ centered
at $0$. An s-holomorphic singularity occurs at a corner when two
conflicting values for a given corner are ``suggested'' by the two
adjacent medial vertices through 
the projection relation~\eqref{eq:s-holomorphicity}.
In the case of \cite{chelkak-hongler-izyurov}, the singularity occurs
at a corner which is outside the square $Q$, and the two values suggested
by the adjacent medial vertices are opposite (they come from two sheets);
in our case, thanks to the fact that branching is at the center of
the square $Q$, the s-holomorphicity is preserved: the two values
that come naturally from both sheets do not conflict, since the corners
for which they ``suggest'' a value are actually different (i.e.
on opposite sheets). Note that this situation is exactly the same
as in the proof of Theorem 2.20 in \cite{chelkak-hongler-izyurov}
(see in particular the paragraph before that proof); see also Remark
2.6 in \cite{gheissari-hongler-park}, where it is explained that
discrete holomorphicity may exist despite the failure of harmonicity. 
\end{proof}
Define the constant $c_{*} := \sqrt{\frac{\sqrt{2}}{\pi}}$.
\begin{lem}
\label{lem:convergence-k-function}We have that 
$\frac{1}{c_{*}\sqrt{\meshsz}}K_{\meshsz}(z) \to \ii z^{-\frac{1}{2}}$
as $\meshsz \to 0$, uniformly for $z$ away from $0$. 
\end{lem}
\begin{proof}
The convergence technology of the proof of Lemma 2.14 in 
\cite{chelkak-hongler-izyurov}
leads to the proof of convergence of 
$\frac{1}{\vartheta\left(\meshsz\right)}K_{\meshsz} \to \ii z^{-\frac{1}{2}}$,
for a function $\vartheta\left(\meshsz\right)\asymp\sqrt{\meshsz}$
as $\meshsz \to 0$. But the proof of Lemma 5.15 of~\cite{gheissari-hongler-park}
gives the stronger estimate that 
$\vartheta\left(\meshsz\right)\sim\sqrt{\frac{\sqrt{2}\meshsz}{\pi}}
= c_{*} \sqrt{\meshsz}$
as $\meshsz \to 0$ 
(note that the lattice square side length in those papers is denoted 
by~$\sqrt{2}\,\delta$ whereas lattice square side length in our paper 
is~$\meshsz$). 
\end{proof}
We can now define $z_{\mdl}^{\left[-\frac{1}{2}\right]}$ on 
$\dblcov{\dCmdl, 0}$.
\begin{defn}
\label{def:discrete-inverse-square-root}We define 
$z_{\mdl}^{\left[-\frac{1}{2}\right]}$
on $\dblcov{ \dCmdl , 0 }$ by symmetrizing $K_{\meshsz}$
as follows: 
\[
z_{\mdl}^{\left[-\frac{1}{2}\right]} 
:= \frac{1}{8 \ii c_{*}} \sum_{k=0}^{7}
    e^{\ii k \pi / 4 } \, K_{\meshsz}\left(\ii^{k}z\right).
\]
\end{defn}

\begin{rem}
One can show that the function $z_{\mdl}^{\left[-\frac{1}{2}\right]}$
is the unique function on $\dblcov{\dCmdl, 0}$
that is discrete holomorphic on $\dblcov{\dCmdl, 0}$
except at $0$, that decays at infinity and shares the same 90 degree
rotation symmetries as $z\mapsto1/\sqrt{z}$ and is purely real on
the positive axis. 
\end{rem}

\begin{lem}
We have that the values of $z_{\mdl}^{\left[-\frac{1}{2}\right]}$ at
$e^{\frac{\pi \ii }{2}k}\frac{\meshsz}{2}$ for $k=0,\ldots,7$ are given
by $Ce^{-\frac{\pi \ii }{4}k}$ for some constant $C>0$. 
\end{lem}

\begin{proof}
This follows from a direct calculation and symmetry. 
\end{proof}

\subsubsection{Discrete Square Root}
For the construction of the discrete square root, we rely on a discrete 
analogue of the real part of the square root function (on the 45 degree-rotated 
and factor~$\sqrt{2}$ scaled square lattice
$(\delta + \ii \delta) \Z^2 = \sqrt{2} \varrho \, \dC$), denoted by 
$G_{\dblcov{\dC, 0}}$, defined 
in~\cite{gheissari-hongler-park}, following~\cite{chelkak-hongler-izyurov}.
It is constructed by essentially integrating the harmonic measure
used to define the discrete version of $z^{-1/2}$. In this section,
we first extend $G_{\dblcov{\dC, 0}}$ to an s-holomorphic
function (on the medial and corner lattices of $(\delta + \ii \delta) \Z^2$, 
with the same notation
and conventions as in~
\cite{gheissari-hongler-park}), and then define
our function $G_{\meshsz}$ on $\dblcov{\dCmdl, 0}$.
While there are a few differences (e.g., the choice of the lattice),
our function is otherwise very similar.


Let us start by the extension of $G_{\dblcov{\dC, 0}}$
to the double cover $\dblcov{\dC, 0}$ (with the
notation and 45-degree rotated lattice of~
\cite{gheissari-hongler-park})
into an s-holomorphic function.

Define the 
constant~$\tilde{c}_{*}=\frac{1}{2 \sqrt{\pi}}$.
\begin{lem}
\label{lem:extension-spin-spin-sqrt}
With the notation and conventions of~%
\cite{gheissari-hongler-park},
the function $G_{\dblcov{\dC, 0}}$ defined on
$\mathcal{V}_{\dblcov{\dC, 0}}^{1}$ admits a unique
s-holomorphic extension with $-1$ monodromy to 
$\mathcal{V}_{\dblcov{\dC, 0}}^{\mathrm{\corn\mdl}}$,
which vanishes on $\mathcal{V}_{\dblcov{\dC, 0}}^{\ii}\cap \R_{+}$
and on $\mathcal{V}_{\dblcov{\dC, 0}}^{1} \cap \R_{-}$.
We have that 
$\frac{1}{\tilde{c}_{*}\sqrt{\meshsz}} G_{\dblcov{\dC, 0}}(z) 
    \to \sqrt{z}$
as $\meshsz \to 0$ uniformly on the compact subsets of $\dblcov{\dC, 0}$.
\end{lem}
\begin{proof}
This follows from~\cite{gheissari-hongler-park}. See Definition 3.24
in Section 3.2.4 (`Auxiliary Functions') for the construction, Remark
3.25 in Section 3.2.4 for the proof of s-holomorphicity, and Lemma
5.16 in Section 5.4.1 (`Convergence of the Full-plane Observable')
for the proof of convergence. The fact that the function vanishes
on $\mathcal{V}_{\dblcov{\dC, 0}}^{\ii} \cap \R_{+}$
and on $\mathcal{V}_{\dblcov{\dC, 0}}^{1} \cap \R_{-}$
comes directly from the construction. 
\end{proof}
We can now define the function $G_{\meshsz}$ appropriate for our setup,
as a 45-degree rotated version of $G_{\dblcov{\dC, 0}}$: 
\begin{defn}
We define $G_{\meshsz}  \colon  \dblcov{\dCmdl, 0} \to \C $
by
$G_{\meshsz}(z) :=
    e^{\frac{\pi \ii}{8}}
    G_{\dblcov{\dC, 0}} (\sqrt{2}\varrho z)$.\footnote{
The square lattice used in~\cite{gheissari-hongler-park}
is~$(1+\ii) \meshsz \, \Z^2$, and correspondingly the side-length of the 
squares is $\sqrt{2}\meshsz$ instead of~$\meshsz$, and the orientation differs 
by $45$~degrees.}
\end{defn}

The most important features of $G_{\meshsz}$ are summarized in the
following.
\begin{prop}
\label{prop:g-sqrt}The function $G_{\meshsz}$ is s-holomorphic on
$\dblcov{\dCmdl, 0}$, vanishes at $\frac{1-\ii}{4} \, \meshsz$,
and is ``discrete holomorphic at~$0$'', i.e.,
\[ G_{\meshsz} \Big(\frac{\meshsz}{2} e^{\ii 0} \Big)
- G_{\meshsz} \Big(\frac{\meshsz}{2} e^{\ii \pi} \Big)
+ \ii \, G_{\meshsz} \Big(\frac{\meshsz}{2} e^{\ii \pi / 2} \Big)
- \ii \, G_{\meshsz} \Big(\frac{\meshsz}{2} e^{\ii 3 \pi / 2} \Big)
       = 0 . \]
Moreover, we have that 
$\frac{1}{2^{1/4} \tilde{c}_{*} \sqrt{\meshsz}}G_{\meshsz}(z)  \to \sqrt{z}$
as $z \to \infty$ uniformly on the compact subsets of~$\dblcov{\C,0}$.
\end{prop}
\begin{proof}
The vanishing of $G_{\meshsz}$ at~$\frac{1-\ii}{4}\meshsz$ is direct from its 
construction based on~$G_{\dblcov{\dC, 0}}$.
The discrete holomorphicity at~$0$ follows from this: a failure of 
holomorphicity
at $0$ would come from an incompatibility between the opposite values
coming from the different sheets at a given corner; here, since one
of the corner values is zero here, this problem is absent. The convergence
follows from Lemma~\ref{lem:extension-spin-spin-sqrt}. 
\end{proof}

We are now in position to define $z_{\mdl}^{\left[\frac{1}{2}\right]}$: 
\begin{defn}
\label{def:discrete-square-root}We define $z_{\mdl}^{\left[\frac{1}{2}\right]}$
on $\dblcov{\dCmdl, 0} \to \C $ by 
\[
z^{\left[\frac{1}{2}\right]} 
:= \frac{1}{8 \, 2^{1/4} \, \tilde{c}_*}
    \sum_{k=0}^{7} e^{-\ii k \pi / 4} \,
    G_{\meshsz}\left(\ii^{k}z\right).
\]
\end{defn}

\begin{lem}
\label{lem:symmetries-of-disc-sqrt}
Suppose a function~$f \colon \dblcov{\dCmdl, 0} \to \C$ has the same 
90~degree rotational symmetry as $z \mapsto z^{m/2}$ for $m$ odd (and 
in particular $-1$ monodromy) and is ``discrete holomorphic at~$0$''
(in the sense of Proposition~\ref{prop:g-sqrt}). Then $f$ 
vanishes at each of the medial lattice points adjacent to~$0$, i.e., at
$\frac{\meshsz}{2} \, e^{\ii \frac{\pi}{2} k}$ for $k=0,1,\ldots,7$.
\end{lem}
\begin{proof}
The assumed symmetry implies that
$f \big( \frac{\meshsz}{2} \, e^{\ii \frac{\pi}{2} k} \big)
= e^{\ii \frac{\pi}{4} m k} \, f \big( \frac{\meshsz}{2} \big)$.
The vanishing of the discrete $\deebar$-derivative at~$0$ then amounts to 
\begin{align*}
0 = \; & \frac{1}{2} \left( 
	  f \big( \frac{\meshsz}{2} \big)
      - f\big( \frac{\meshsz}{2} \, e^{\ii \pi} \big) \right)
  + \frac{\ii}{2} \left(
      f\big( \frac{\meshsz}{2} \, e^{\ii \frac{\pi}{2}} \big)
      - f\big( \frac{\meshsz}{2} \, e^{\ii \frac{3\pi}{2}} \big) \right) \\
= \; & \frac{1}{2} \left( 
		1 + e^{\ii \frac{\pi}{4}(2m+4)}
		+ e^{\ii \frac{\pi}{4}(m+2)} + e^{\ii \frac{\pi}{4}(3m+6)} 
	\right) \; f \big( \frac{\meshsz}{2} \big) \,
= \,\frac{1}{1-e^{\ii \frac{\pi}{4}(m+2)}} \; f \big( \frac{\meshsz}{2} 
\big) .
\end{align*}
The constant in front of $f \big( \frac{\meshsz}{2} \big)$ above is non-zero, 
so we conclude that $f$ vanishes at $\frac{\meshsz}{2}$, and then also at all 
the points of the form
$\frac{\meshsz}{2} \, e^{\ii \frac{\pi}{2} k}$. 
\end{proof}
\begin{prop}
\label{prop:disc-sqrt}
The function $z_{\mdl}^{\left[\frac{1}{2}\right]}$
is discrete holomorphic on $\dblcov{\dCmdl, 0}$,
has the same 90~degree rotation symmetry as $z\mapsto\sqrt{z}$, and 
therefore vanishes
at $\frac{\meshsz}{2} \, e^{\ii \frac{\pi}{2} k}$ for $k=0,1,\ldots,7$. 
Moreover, we have
that $\frac{1}{\sqrt{\meshsz}}z_{\mdl}^{\left[\frac{1}{2}\right]} \to \sqrt{z}$
as $\meshsz \to 0$, uniformly on compact subsets of $\left[ \C,0\right]$.
\end{prop}
\begin{proof}
The symmetry is obvious by construction and the discrete holomorphicity
at~$0$ is inherited from that of~$G_{\dblcov{\dC, 0}}$.
The vanishing at the medial lattice points adjacent to~$0$ then follows from
Lemma~\ref{lem:symmetries-of-disc-sqrt}.
The convergence follows from Proposition \ref{prop:g-sqrt}.
\end{proof}

\subsection{Half-Integer Powers}

In the previous subsection, we defined the functions 
$z_{\mdl}^{\left[-\frac{1}{2}\right]}$
and $z_{\mdl}^{\left[\frac{1}{2}\right]}$. Provided these functions
can be differentiated and integrated in the space of functions with
a $-1$ monodromy around $0$, they uniquely determine the functions
$z_{\mdl}^{\left[k\right]}$ and $z_{\dmd}^{\left[k\right]}$ for
all $k\in\Z +\frac{1}{2}$: as usual, differentiating or integrating
shifts the power by $1$ and switches the lattice between 
$\dblcov{\dCmdl, 0}$
and $\dblcov{\dCdmd, 0}$. The relevant
90 degree symmetries are automatically inherited from those of 
$z_{\mdl}^{\left[-\frac{1}{2}\right]},z_{\mdl}^{\left[\frac{1}{2}\right]}$. 

\begin{figure}[tb]
\centering
\subfigure[$z \mapsto \dhimon{z}{-1/2}$] 
{
    \includegraphics[scale=.8]{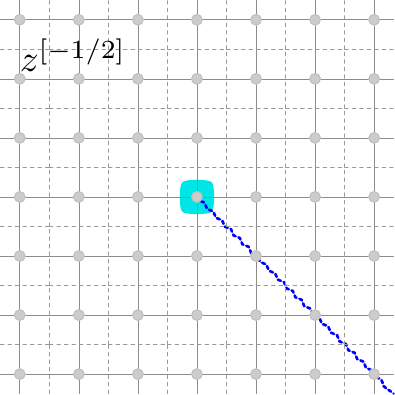}%
	\label{sfig: discrete z power minus half}
}
\hspace{0.6cm}
\subfigure[$z \mapsto \dhimon{z}{1/2}$] 
{
    \includegraphics[scale=.8]{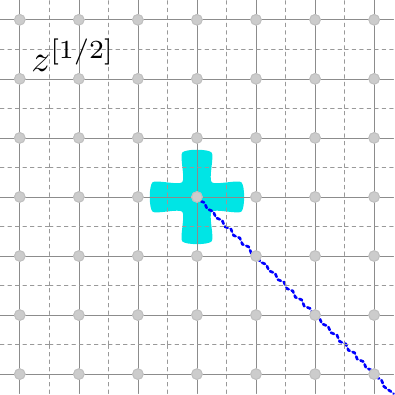}%
	\label{sfig: discrete z power minus half}
} \\
\vspace{0.4cm}
\subfigure[$z \mapsto \dhimon{z}{3/2}$] 
{
    \includegraphics[scale=.8]{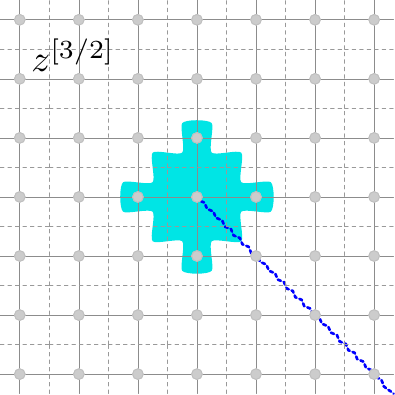}%
	\label{sfig: discrete z power minus half}
}
\hspace{0.6cm}
\subfigure[$z \mapsto \dhimon{z}{5/2}$] 
{
    \includegraphics[scale=.8]{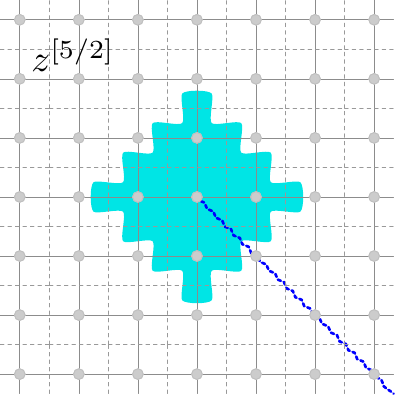}%
	\label{sfig: discrete z power minus half}
}
\caption{\label{fig:locus-vanishing}
The sets of vertices of~$\dblcov{\dCmdldmd,0}$ where discrete 
half-integer monomials vanish are illustrated by the shaded area.
}
\end{figure}

\begin{lem}
\label{lem:half-int-mon}There is a unique family of functions 
$\dhimon{z}{p} $
for $p\in\Z +\frac{1}{2}$ satisfying the conditions 1\textendash 6
and 8 of Proposition \ref{prop:half-integer-monomials}. 
\end{lem}
\begin{proof}
As explained above, the negative powers on 
$\dblcov{\dCmdldmd, 0}$
are obtained by differentiating $z_{\mdl}^{\left[-\frac{1}{2}\right]}$
and $z_{\mdl}^{\left[\frac{1}{2}\right]}$ and interpreting~$0^{[p]}=0$
for all $p$. The decay at infinity of $z_{\dmd}^{\left[-\frac{1}{2}\right]}$
follows from Harnack type estimates on the derivatives of discrete holomorphic
functions \cite{lawler-limic}, and for~$p<-\frac{1}{2}$, the 
same argument applies. The
rest of the properties for $p<0$ are straightforward to check.

The positive powers are obtained by integrating in the space of functions
with $-1$ monodromy. The discrete holomorphicity of the $\dhimon{z}{p} $
for all $p\geq\frac{1}{2}$ follows by induction from the next lemma. 
\end{proof}
\begin{lem}
\label{lem:pos-half-int-mon-vals}We have the following (see Figure
\ref{fig:locus-vanishing}): 
\begin{enumerate}
\item The values of the function $z_{\dmd}^{\left[\frac{1}{2}\right]}$
are given by
$\frac{C \sqrt{\sqrt{2}+1}}{2 }\sqrt{x/\meshsz}$
for $x$ living
above one of the four dual vertices
$\set{ \frac{\pm 1 \pm \ii}{2} \meshsz }$
and by $\frac{C}{2} \sqrt{x/\meshsz}$ 
for $x$ living above one of the four vertices
$\pm \meshsz, \pm \ii \meshsz $, where $C$ is the value of the function
$z_{\mdl}^{[-\frac{1}{2}]}$ at $\frac{\meshsz}{2}$. 
\item The functions 
$z_{\mdl}^{\left[\frac{1}{2}\right]},
z_{\mdl}^{\left[\frac{3}{2}\right]}$
vanish on the four medial vertices
$\set{ \pm \frac{\meshsz}{2} , \pm \frac{\ii \meshsz }{2} } $. 
\item The functions 
$z_{\dmd}^{\left[\frac{3}{2}\right]} , z_{\dmd}^{\left[\frac{5}{2}\right]}$
vanish on the nine diamond vertices
$\set{ 0, \frac{\pm 1 \pm \ii}{2} \meshsz , \pm \meshsz , \pm \ii \meshsz } $. 
\item In general, for half-integer $k\geq\frac{1}{2}$, if 
$z_{\dmd}^{\left[k\right]}$
vanishes on a neighborhood $\Lambda_{\dmd}$ of the origin, then
$z_{\mdl}^{\left[k+1\right]}$
vanishes on the set of medial vertices at distance $\frac{\meshsz}{2}$
from $\Lambda_{\dmd}$, and conversely, if $z_{\mdl}^{\left[k\right]}$
vanishes on a neighborhood $\Lambda_{\mdl}$ of the origin, then
$z_{\dmd}^{\left[k+1\right]}$
vanishes on the set of diamond vertices at distance $\frac{\meshsz}{2}$
from $\Lambda_{\mdl}$. 
\end{enumerate}
\end{lem}
\begin{proof}
Item~1 follows from explicit computation and the definition of 
$z_{\mdl}^{\left[-\frac{1}{2}\right]}$.
Then Lemma~\ref{lem:symmetries-of-disc-sqrt}
gives item~2, and the rest
from integration together with Lemma~\ref{lem:symmetries-of-disc-sqrt} again 
(also keeping in mind that the choice of the antiderivative
is made unique by the constraint that the monodromy is $-1$).
\end{proof}
%

\subsection{Contour Integrals}
\begin{lem}
\label{lem:gff-contour-integral-integer}For any $k,\ell\in\Z $
and any large enough contour $\contour$, we have 
\[
\frac{1}{2\pi \ii}
\dcint{\contour}
  z_{\mdl}^{\left[k\right]}
  z_{\dmd}^{\left[\ell\right]} \dd{z}
= \delta_{k+\ell+1}.
\]
\end{lem}

\begin{proof}
If $k+\ell\leq2$, we have the following three sub-cases: 
\begin{itemize}
\item If $k+\ell<-1$, then since 
$z_{\mdl}^{\left[k\right]}z_{\dmd}^{\left[\ell\right]}
=O\left(\left|z\right|^{k+\ell}\right)=O\left(\left|z\right|^{-2}\right)$
we can take the integration contour to be very large, containg 
$O\left(\left|z\right|\right)$
terms a show that the integral must vanish. 
\item If $k+\ell=-1$, we can use that 
$\delta^{k+\ell}z^{\left[k\right]}z^{\left[\ell\right]} \to  z^{k+\ell}$
as $\meshsz \to 0$ and multiply by $\meshsz$ to make the discrete contour
integral converge to a continuous integral, hence yielding the result:
the discrete contour integral has the `right scaling' to pass to the
limit without renormalization. 
\item If $k+\ell=0,1,2$, 
by contour deformation it suffices to consider a contour~$\contour$ 
which forms the boundary of a large square of the 
form~$\left[-R,R\right]^{2}$. For such contours, we can use symmetry to show 
that
the integral must vanish: if $k+\ell=0,2$, we have 
$z^{\left[k\right]}z^{\left[\ell\right]}=\left(-z\right)^{\left[k\right]}
\left(-z\right)^{\{\ell\}}$
and hence antipodal contributions will cancel each other (due to the
integration element), and if $k+\ell=1$, we have 
$\left(\ii z\right)^{\left[k\right]}\left(\ii z\right)^{\left[\ell\right]}
=\ii z^{\left[ k\right]}z^{\left[\ell\right]}$
and hence $\frac{\pi}{4}$ symmetric contributions will cancel each
other (due to the integration element being multiplied by $\ii$, too). 
\end{itemize}
Let us now look at the case $k+\ell\geq3$: 
\begin{itemize}
\item If $k,\ell\geq0$, the functions are discrete holomorphic everywhere
and we conclude readily. 
\item If $k=-1$, then $\ell\geq4$, and we can use
Stokes' formula and the fact that $z^{\left[\ell\right]}$
vanishes at the points where $z^{\left[k\right]}$ is not discrete
holomorphic. 
\item If $k<-1$, we can use integration by 
parts~\eqref{eq:discrete-contour-integration-by-parts} to raise the power of 
$z^{\left[k\right]}$
while decreasing the power of $z^{\left[\ell\right]}$, until we get
back to the previous case. 
\item The case $\ell\leq-1$ is symmetric to the previous two cases. 
\end{itemize}
\end{proof}
\begin{lem}
\label{lem:ising-contour-integral}If either $k,\ell\in\Z $
or $k,\ell\in\Z +\frac{1}{2}$, denoting 
$\dmonavg{z_{\mdl}}{\ell} 
= \frac{1}{4}\sum_{x\in\left\{ \pm1,\pm \ii \right\}} 
\dmon{ ( z-\frac{x\meshsz}{2} )}{\ell}$,
then for any large enough contour $\contour$, we have 
\[
\frac{1}{2\pi \ii}
\dcint{\contour}
  \dmon{z_{\dmd}}{k} \dmonavg{z_{\mdl}}{\ell} \, \dd{z}
= \delta_{k+\ell+1}.
\]
\end{lem}

\begin{proof}
The proofs of the cases $k+\ell\leq2$ are exactly the same as the
ones of the cases $k+\ell\leq2$ in the proof of Lemma 
\ref{lem:gff-contour-integral-integer},
even if $k,\ell$ are half-integer: all that is used is the decay
at infinity and the 90 degree symmetries. Let us now look at the cases
$k+\ell\geq3$, and show that the integral vanishes. We cover this in 
three separate cases: 
\begin{itemize}
\item If $k\geq1$ and $\ell\geq0$, to show that the integral vanishes,
we distinguish the integer and half-integer cases. 
\begin{itemize}
\item For integer $k,\ell$ this is straightforward: the functions 
$\dmon{z_{\dmd}}{k}$
and $\dmonavg{z_{\mdl}}{\ell}$ are discrete holomorphic everywhere,
and hence the contour integral vanishes. 
\item For half-integer $k,\ell$, we have that 
\begin{itemize}
\item $\dmon{z_{\dmd}}{k}$ vanishes (at least) on $\left\{ 
0,\frac{\pm1\pm \ii }{2}\meshsz,\pm\meshsz,\pm \ii \meshsz \right\} $
and is discrete holomorphic everywhere (since $k\geq\frac{3}{2}$) 
\item $\dmonavg{z_{\mdl}}{\ell}$ is (at least) discrete holomorphic
outside of the nine diamond vertices $\left\{ 0,\frac{\pm1\pm 
\ii}{2}\meshsz,\pm\meshsz,\pm \ii \meshsz \right\} $:
this is inherited from the properties of the function 
$\dmon{z_{\dmd}}{\ell}$,
shifted by a lattice spacing. 
\end{itemize}
We can hence deduce that 
\begin{align*}
\sum_{z_{\mdl} \in \dblcov{\dCmdl, 0}}
    \dmonavg{z_{\mdl}}{\ell} \,
    \ddeebar \dmon{z_{\dmd}}{k}
& = 0 \\
\sum_{z_{\dmd}\in \dblcov{\dCdmd, 0}}
    \dmon{z_{\dmd}}{k} \,
    \ddeebar \dmonavg{z_{\mdl}}{\ell}
& =0
\end{align*}
and conclude that the contour integral vanishes. 
\end{itemize}
\item If $k<1$, we again distinguish the integer and half-integer cases.
\begin{itemize}
\item For integer $k,\ell$ with $k+\ell \geq 3$ there are three cases:
\begin{itemize}
\item If $k,\ell \geq 0$, then the vanishing of the contour integral 
is clear by discrete holomorphicity of~$\dmon{z_{\dmd}}{k}$ 
and~$\dmonavg{z_{\mdl}}{\ell}$.
\item For $k=-1$ and $k=-2$ and any $\ell \geq 3-k$ we have that 
$\dmonavg{z_{\mdl}}{\ell}$ vanishes on each of the medial vertices 
where $\ddeebar \dmon{z_{\dmd}}{k}$ is non-zero, so the vanishing of the 
contour integral follows from Stokes' formula.
\item For $k<-2$ we can integrate by parts an even number of times to 
reduce to the previous case.
\end{itemize}
\item For half-integer $k,\ell$,
we can use two integrations by parts (on a large enough contour,
where both $z_{\dmd}^{\left[k\right]}$ at $z_{\mdl}^{\{\ell\} }$
are discrete holomorphic) to raise the power of $k$ by $2$ units
while decreasing the power of $\ell$ by $2$ units, keeping $k+\ell$
constant and $\ell\geq0$ until $k\geq1$, and reduce this case to
the previous one (the reason we use two integrations by parts is to
keep the functions on the same lattices: 
$\ddee \ddee z_{\dmd}^{\left[k+2\right]}
= (k+2) (k+1) z_{\dmd}^{\left[k\right]}$
and $\ddee \ddee z_{\mdl}^{\{\ell+2\}}
= (\ell+2) (\ell+1)
    z_{\mdl}^{\{\ell\} }$).
\end{itemize}
\item The $\ell<1$ case is symmetric to the previous one.
\end{itemize}
\end{proof}

\subsection{\label{subsec:proof-of-propositions-discrete-monomials}Proof of
Propositions \ref{prop:integer-monomials} and \ref{prop:half-integer-monomials}}

We are now in position to conclude the proofs of the two key propositions
about integer and half-integer monomials. 
\begin{proof}[Proof of Proposition \ref{prop:integer-monomials}]
Properties 1\textendash 8 were proven in Lemmas \ref{lem:pos-int-mon}
and \ref{lem:neg-int-mon}. Property 9 follows from Lemma 
\ref{lem:gff-contour-integral-integer}
and Property 10 follows from Lemma \ref{lem:ising-contour-integral}.
\end{proof}

\begin{proof}[Proof of Proposition \ref{prop:half-integer-monomials}]
Properties 1\textendash 6 and 8 follow from Lemma \ref{lem:half-int-mon}.
Property 7 follows from \ref{lem:pos-half-int-mon-vals}. Property
10 follows from Lemma \ref{lem:ising-contour-integral} and the proof of 
Property 9 is similar but easier. 
\end{proof}

\section{\label{sec:lattice-fermion-proofs}Lattice Fermion Proofs}

In this section, we give the proofs of the key properties of the fermions
stated in Section \ref{subsec:ising-model}. 

\subsection{Fermions as Parafermions}

In this subsection, we review the low-temperature expansion of the
Ising fermion, which yields to the family of parafermionic observables
(see \cite{smirnov-iv} for instance). This representation is useful
to reveal a number of symmetries (it is more canonical in some sense,
as it does not depend on a choice of defect line) and it is decisive
for the boundary value problem analysis of Ising correlations (see
\cite{chelkak-hongler-izyurov,hongler-smirnov,hongler} for instance). 
\begin{defn}
Let $\LTEsetin{\ddomain }$ denote the set of
even subgraphs\footnote{
Informally, an even subgraph $\LTEelem$ of~$\ddomaindual$ forms a collection of 
loops on the 
dual graph. Even subgraphs are also called \emph{contours} in the 
literature, but we have chosen to avoid this term because it is 
already used in the context of discrete integrals.}
of the dual graph~$\ddomaindual$,
i.e., subsets $\LTEelem$ of edges of~$\ddomaindual$ such that every 
dual vertex $p \in \ddomaindual$ is incident to an even number of edges 
of~$\LTEelem$.
For the Ising model with $+$ boundary conditions on $\ddomain $,
spin configurations $\spin \in \set{\pm 1}^{\ddomain}$ bijectively
correspond with even subgraphs $\LTEelem\in\LTEsetin{\ddomain }$ of the 
dual graph~$\ddomaindual$ by the condition that edges of~$\LTEelem$ separate 
spins with opposite values. For $V \subset \ddomain$,
we denote by $\spin_{\LTEelem}^{V}$ the value of the product
$\prod_{x\in V}\spin_{x}$ of spins in $V$ in the spin configuration~$\spin$ 
corresponding to~$\LTEelem$.
\end{defn}

\begin{rem}\label{rem: Boltzmann weights in LTE}
The Boltzmann weight $e^{-\invtemp \Hamiltonian[\spin]}$
of a spin configuration $\spin$ is 
proportional to~$e^{-2\invtemp\left|\LTEelem\right|}$, where $|\LTEelem|$ is 
the number of edges in the corresponding even subgraph~$\LTEelem$.
It follows that we have
\[ \correl{ \prod_{x\in V}\spin_{x} } =
\frac{1}{\mathcal{Z}}\sum_{\LTEelem\in\LTEsetin{\ddomain }}
e^{ -2\invtemp\left|\LTEelem\right|}\spin_{\LTEelem}^{V} , \]
where \[\mathcal{Z} :=
\sum_{\LTEelem\in\LTEsetin{\ddomain }}e^{-2\invtemp\left|\LTEelem\right|} . \]
\end{rem}

Let $\cordefline_{c_{1},c_{2}}$ be a corner defect line in $\ddomain $
and, denoting the symmetric difference by $\oplus$, set 
\[ \LTEsetptsin{c_{1},c_{2}}{\ddomain }
:= \set{\cordefline_{c_{1},c_{2}} \oplus \LTEelem \; \Big| \;
        \LTEelem \in \LTEsetin{\ddomain}} \]
which only depends on $c_{1}$ and $c_{2}$, and not on the defect 
line~$\cordefline_{c_{1},c_{2}}$.
An element $\LTEishelem \in \LTEsetptsin{c_{1},c_{2}}{\ddomain }$ 
will be called a $(c_1,c_2)$-subgraph: it has an even 
number of edges or corner ends adjacent to every dual vertex, but has one 
corner end adjacent to both of the the marked corners $c_{1}$ and $c_{2}$. 
For $\LTEishelem \in \LTEsetptsin{c_{1},c_{2}}{\ddomain }$,
we denote (by a slight abuse of notation) by 
$\mathbf{W}\left(\LTEishelem\right)\in 
\R/4\pi\Z $
the winding of the path in $\LTEishelem$ which goes from $c_{1}$ to $c_{2}$, as 
defined
in, e.g., \cite{hongler-smirnov}: if $\LTEishelem$ is a simple path, 
$\mathbf{W}\left(\LTEishelem\right)$
is defined as in Section \ref{subsec:corner-lattice-fermions}, and
if `ambiguities' arise (i.e. going from $c_{1}$ to $c_{2}$, one
has three choices about how to continue the path, left, right or straight)
one either turns right or left, but does not go straight, and 
$\mathbf{W}\left(\LTEishelem\right)$
modulo $4\pi$ is independent of the choice of the path in $\LTEishelem$,
as shown in Lemma 1.4 of \cite{hongler-smirnov}.
We use the convention that the length $|\LTEishelem|$ of
$\LTEishelem \in \LTEsetptsin{c_{1},c_{2}}{\ddomain }$ counts only the edges on 
the 
dual lattice $\ddomain^*$ and not the two corner ends connecting dual vertices 
to the corners $c_1 , c_2$.

\begin{figure}
\includegraphics[scale=0.55]{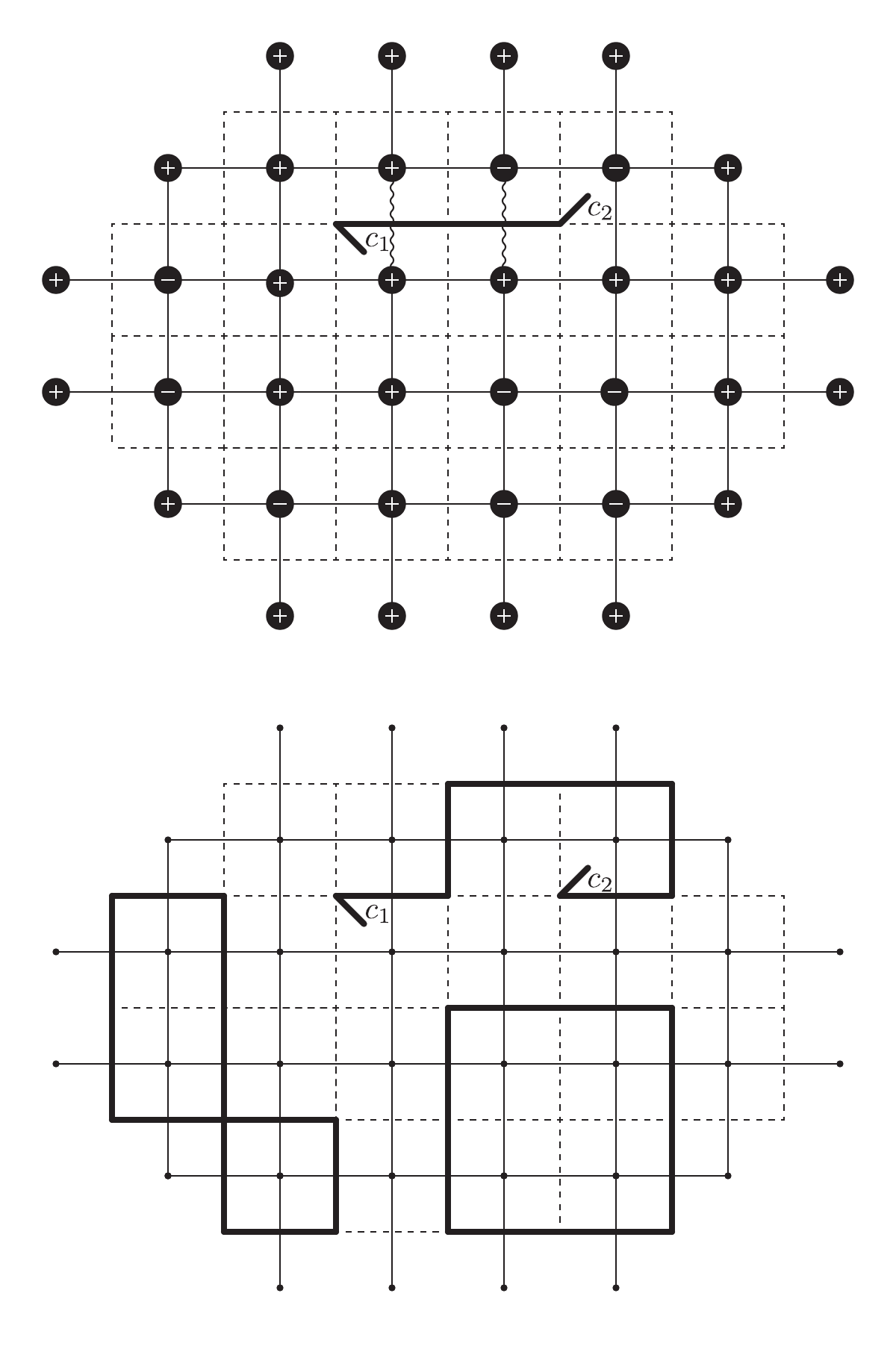}

\caption{\label{fig:low-t-fermion}Spin configuration with disorder (top picture)
and the associated $\LTEishelem$ configuration (bottom picture)}
\end{figure}

The following lemma connects the two-fermion correlations with the
observables of~\cite{hongler,hongler-smirnov,chelkak-hongler-izyurov}.
Recall that for a corner~$c \in \dCcor$ adjacent to vertex~$x \in 
\dC$ and dual vertex~$p \in \dCdual$ we denote $\corndir(c) := 
\frac{x-p}{|x-p|}$.
\begin{lem}
\label{lem:low-temperature-two-fermions-alone}Consider the Ising
model on a discrete domain $\ddomain $ with $+$~boundary conditions. For any 
corners $c_{1},c_{2}$, we have 
\[
\correl{ \ferm_{c_{1}}\ferm_{c_{2}} }
= - \overline{\corndir (c_{1})} \; \frac{1}{\mathcal{Z}}
  \sum_{\LTEishelem \in \LTEsetptsin{c_{1},c_{2}}{\ddomain }}
    e^{-\frac{\ii}{2}\mathbf{W}\left(\LTEishelem\right)}
    e^{-2\invtemp\left|\LTEishelem\right|}.
\]
\end{lem}
\begin{proof}
Pick a disorder line $\disline$ between the dual vertices $p_1$ and 
$p_2$ adjacent to corners $c_1$ and $c_2$, respectively.
Represent any Ising configuration $\spin \in \set{\pm1}^{\ddomain }$
by the following collection~$\LTEishelem$ of dual-edges: for any edge
$e=xy$ of $\ddomain $, its dual edge $e^{*}$ belongs to $\LTEishelem$
either if $e^{*}\notin\disline$ and $\spin_{x}\neq\spin_{y}$ or
if $e^{*}\in\disline$ and $\spin_{x}=\spin_{y}$ (see Figure 
\ref{fig:low-t-fermion}).%
\footnote{In other words, $\LTEishelem$ represents the `frustrated pairs' of 
adjacent
spins} It is elementary to check that the effective Boltzmann weight 
$e^{-\invtemp \Hamiltonian[\spin] - 2 \invtemp \disenergy{\disline} [\spin]}$
of a spin configuration~$\spin$ in the presence of the disorder 
line~$\disline$ is proportional to~$e^{-2\invtemp\left|\LTEishelem\right|}$ 
(with the same 
proportionality factor as in Remark~\ref{rem: Boltzmann weights in LTE}). 
Letting
$x_{1},x_{2} \in \ddomain $ be the vertices adjacent to $c_{1},c_{2}$,
it is again elementary to check that the value of $\spin_{x_{1}}\spin_{x_{2}}$
is determined by~$\LTEishelem$ and equals 
$e^{-\frac{\ii}{2}\mathbf{W}\left(\LTEishelem\right)}
  / e^{-\frac{\ii}{2}\mathbf{W} \left(\disline\right)}$.
The proof follows readily. 
\end{proof}

The above lemma generalizes to multiple spin insertions, if this time
we keep track of the defect line.
\begin{lem}
\label{lem:low-temperature-two-fermions-and-spins}Consider the Ising
model on a discrete domain $\ddomain $ with $+$~boundary conditions.
Let $V\subset\ddomain $. For any corners $c_{1},c_{2}$,
we have 
\begin{equation}
\correl{ \left(\prod_{x\in V}
  \spin_{x}\right) \left(\ferm_{c_{1}} 
    \ferm_{c_{2}}\right)_{\lambda} }
= - \overline{\corndir (c_{1})} \; \frac{1}{\mathcal{Z}}
  \sum_{\LTEishelem \in \LTEsetptsin{c_{1},c_{2}}{\ddomain}}
    e^{-\frac{\ii}{2}\mathbf{W}\left(\LTEishelem\right)}
    e^{-2\invtemp\left|\LTEishelem\right|}\spin_{\LTEishelem\oplus\lambda}^{V} .
\label{eq:low-temperature-exp-two-fermion-n-spin}
\end{equation}
\end{lem}
\begin{proof}
The proof is very similar to that of Lemma 
\ref{lem:low-temperature-two-fermions-alone}:
we define the same contour $\LTEishelem$ to describe the configuration.
The only additional thing to check is that, denoting as before $x_{1},x_{2}$
the vertices adjacent to $c_{1},c_{2}$, we have 
$\left(\prod_{v\in V}
  \spin_{x}\right) \spin_{x_{1}} \spin_{x_{2}}
= \frac{e^{-\frac{\ii}{2}
  \mathbf{W} 
    \left(\LTEishelem\right)}}{%
      e^{-\frac{\ii}{2}\mathbf{W}\left(\lambda\right)}}
\spin_{\LTEishelem\oplus\lambda}^{V}$.
This can be done by induction on the number of vertices in~$V$. 
\end{proof}

\subsection{\label{subsec:discrete-holomorphicity} Discrete Holomorphicity}

In this subsection, we give the proof of Proposition 
\ref{prop:discrete-holomorphicity-fermion-correlations},
which is central to the lattice Virasoro construction.

The proposition concerns the discrete holomorphicity and singularities of a 
certain function~$H$. We first prove intermediate results about two 
related functions, defined as follows. Consider the critical Ising model
on a domain $\ddomain $ with arbitrary boundary conditions.
The following will be fixed throughout below:
\begin{itemize}
\item Let $W \subset V \subset \ddomain$ be two subsets of vertices,
and assume that $V$ is connected.
\item Let $c_1 \in \dblcov{\ddomaincor , V}$ be a corner on the double cover 
of~$\ddomain$ ramified around~$V$, whose base point
$\dblcovproj{c}_1 \in \ddomaincor$ on the original lattice is not adjacent 
to~$V$. 
Denote the adjacent vertex and dual vertex to the base 
point corner~$\dblcovproj{c}_1$ by
$v_1 \in \ddomain$ and $p_1 \in \ddomaindual$.
We will also occasionally denote by
$c_1^* \in \dblcov{\ddomaincor , V}$ the corner with the same base 
point~$\dblcovproj{c}_1$ but on the opposite sheet of the 
double cover.
\end{itemize}
but 

We define a function
\[ H^\corn \colon \dblcov{\ddomaincor , V} \setminus \set{c_1, c_1^*} \to  \C 
\]
on the corners of the double cover graph by the following formula: for
$c_2 \in \dblcov{\ddomaincor , V} \setminus \set{c_1, c_1^*}$ set
\begin{align}\label{eq: def of H on corners}
H^\corn ( c_{2} ) = H^\corn ( c_{1} ; c_{2} )
:= \; & \correl{ \left(
    \prod_{u \in W} \spin_{u} \right)
    \ferm_{c_{1}} \ferm_{c_{2}} } .
\end{align}
The fact that this is well defined on the double cover is a consequence 
of~Proposition~\ref{prop:fermion-double-cover}. In concrete terms, if
$v_2 \in \ddomain$ and $p_2 \in \ddomaindual$ are the vertex and dual vertex 
adjacent to the base point ~$\dblcovproj{c}_2$, this definition can be 
unraveled to the form
\begin{align*}
H^\corn ( c_{2} ) = \; & - \cororibar{c_1} \; \correl{ \left(
    \prod_{u \in W} \spin_{u} \right) \, \spin_{v_1} \spin_{v_2} \,
    \dispair{p_1}{p_2}_{\disline}  } ,
\end{align*}
where the disorder line~$\disline$ is 
chosen so that its lift to the double cover connects~$c_1$ to~$c_2$
(when augmented by the two corner ends), and where 
as in Section~\ref{subsec:corner-lattice-fermions} we denote
$\corori{c_1} = \frac{v_1-p_1}{|v_1-p_1|}$.

We define a related function
\begin{align*}
H^\mdl \colon \dblcov{\ddomainmdl , V} \to  \C
\end{align*}
on the medial lattice by the following formula: for
$z \in \dblcov{\ddomainmdl , V}$ we set
\begin{align}\label{eq: def of H on medial lattice}
H^\mdl(z) = H^\mdl ( c_{1} ; z )
:= \frac{1}{2} \sum_{x\in\set{ \pm 1 \pm \ii } }
  \correl{ \left(\prod_{u \in W}\spin_{u}\right)
    \ferm_{c_{1}} \ferm_{z_{x}} } ,
\end{align}
where $z_{x} := z + \frac{\meshsz}{4} x \in \dblcov{\ddomaincor , V}$
denotes a corner adjacent to the medial vertex~$z$, in the direction 
specified by $x$, and in the case $z_{x}=c_{1}$ (resp. 
$z_{x}=c_{1}^*$),
we interpret
$\ferm_{c_{1}} \ferm_{z_{x}} = \frac{\cconj{z}-\cconj{w}}{\left|z-w\right|}$
(resp.
$\ferm_{c_{1}} \ferm_{z_{x}} = -\frac{\cconj{z}-\cconj{w}}{\left|z-w\right|}$)
where $w$ is the other medial edge besides $z$, which is adjacent to
$c_{1}$ (resp. $c_{1}^*$).

\begin{figure}
\includegraphics[scale=0.75]{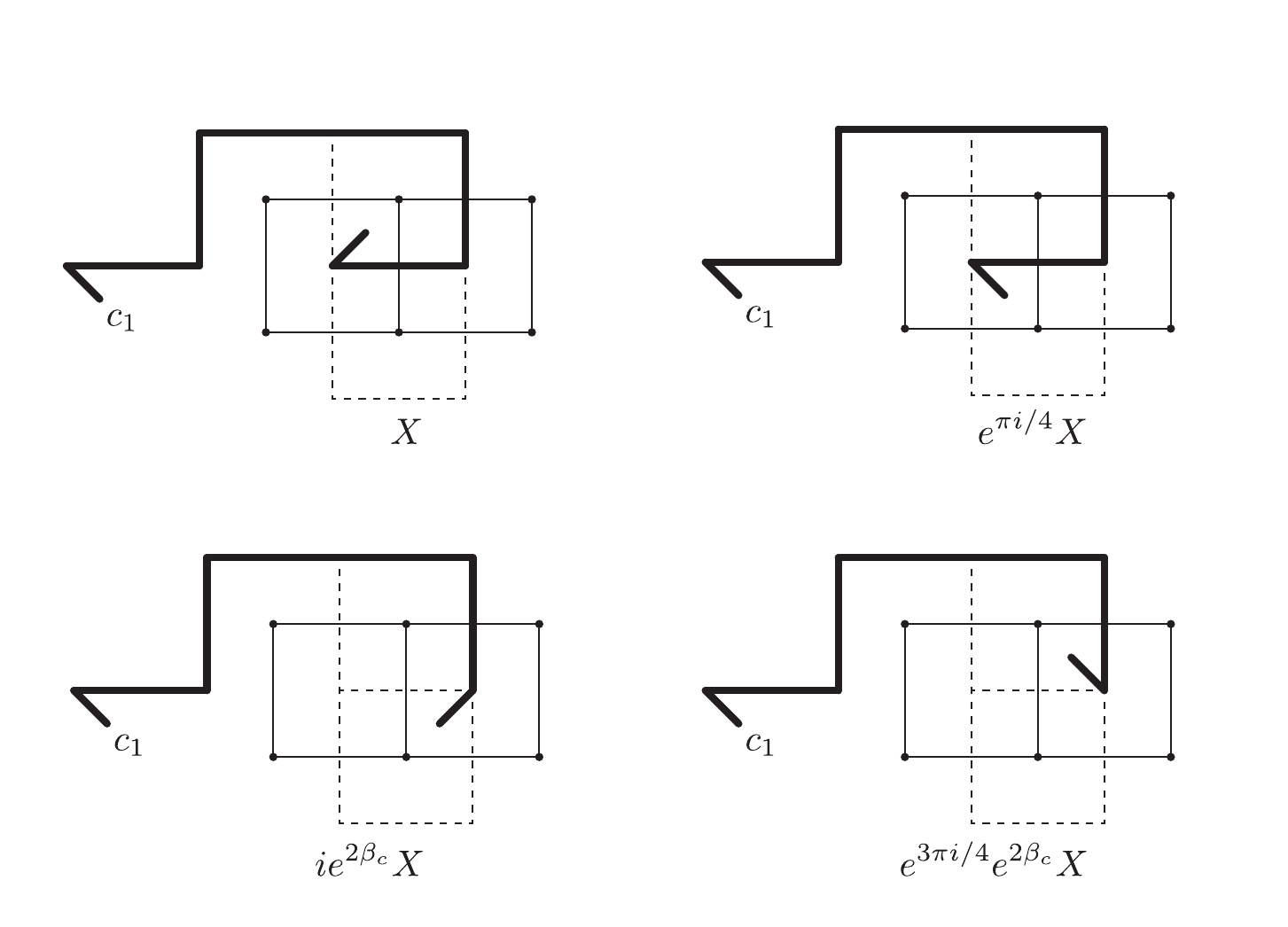}

\caption{\label{fig:proof-holomorphicity}Bijective correspondence between
the contour sets corresponding to the corners around a medial vertex,
and relative weights of the configuration.}
\end{figure}

\begin{figure}
\includegraphics[scale=0.75]{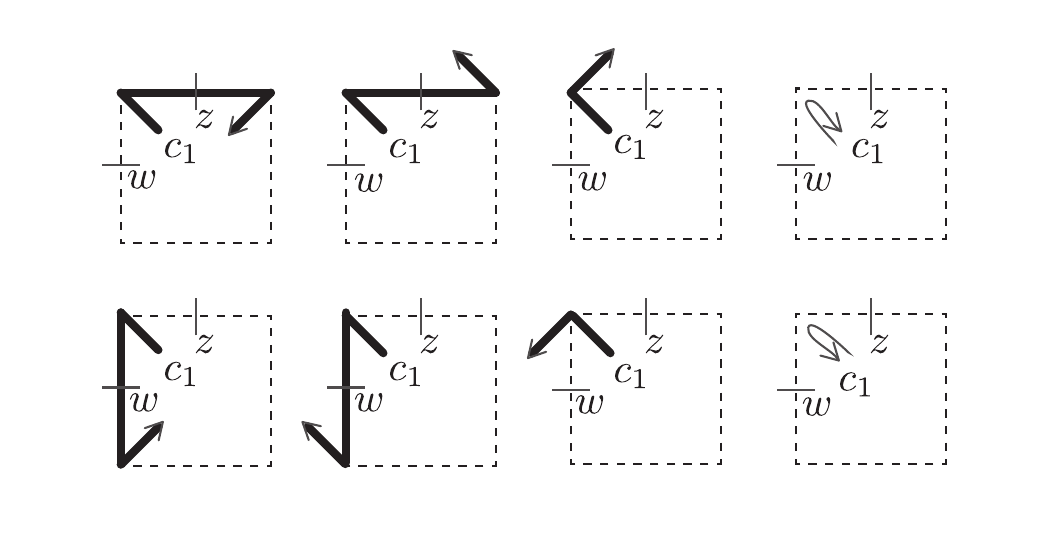}

\caption{\label{fig:discrete-singularity}Source of the discrete singularity
when $z$ is near the point $c_{1}$: performing the bijection as
before, we obtain two `conflicting' values of 
$\mathbf{W}\left(\LTEishelem\right)=-\pi$
(top) and $\mathbf{W}\left(\LTEishelem\right)=\pi$ (bottom) on the last
column.}
\end{figure}

\begin{lem}
\label{lem:corner-discrete-hol}
Let $z$ be a medial vertex of the double cover $\dblcov{\ddomainmdl , V}$.
If $z$ is not adjacent to the corners
$c_{1}, c_{1}^*$, i.e., $z_{x} \neq c_{1}, c_{1}^*$ for each
$x\in\set{ \pm1\pm  \ii }$, then we have 
\begin{equation}
H^\corn \left(z_{x}\right) + H^\corn \left(z_{-x}\right)
= H^\corn \left(z_{\ii x}\right) + H^\corn \left(z_{-\ii x}\right) .
\label{eq:corner-discrete-hol-relation}
\end{equation}
If $z_{x}=c_{1}$ for some $x\in\set{ \pm1\pm  \ii }$,
and $w \in \dblcov{\ddomainmdl , V}$ is the other medial edge adjacent
to~$c_{1}$, then we have
\begin{equation}
\frac{\cconj{z}-\cconj{w}}{\left|z-w\right|} \;
    \correl{ \prod_{x\in V} \spin_{x} }
 + H^\corn \left(z_{-x}\right)
= H^\corn \left(z_{\ii x}\right) + H^\corn \left(z_{-\ii x}\right) .
\label{eq:corner-discrete-sing-relation}
\end{equation}
\end{lem}
%
%
in 
\begin{proof}
Let us first assume that we have $+$ boundary conditions. Let $e$
be the dual edge whose midpoint is $z$ and for each 
$x\in\set{ \pm 1 \pm \ii } $,
let $e_{x}$ denote
the corner end between $z_x$ and a dual vertex adjacent to $z$.

Let us first assume that $z_{x} \neq c_{1}, c_{1}^*$
for all $x$.
Let us write
the values of $H^\corn \left(z_{x}\right)$ in terms of the low-temperature
expansion (Lemma \ref{lem:low-temperature-two-fermions-and-spins}),
choosing 
defect lines $\cordefline_{x}:c_{1}\leftrightarrow z_{x}$
which differ only locally,
and let $\LTEset_{x} := \LTEset_{c_{1},z_{x}}$ denote the relevant
sets of $(c_1 , z_x)$-subgraphs. Let $W_{x}$ denote the weight 
$e^{-\frac{\ii}{2}\mathbf{W}\left(\LTEishelem_{x}\right)}
  \alpha^{ \left| \LTEishelem_{x} \right| }
  \spin_{\LTEishelem_{x}\oplus\cordefline_{x}}^{V}$,
where $\alpha = e^{-2\invtempcrit }=\sqrt{2}-1$.

The $(c_1 , z_x)$-subgraphs $\LTEishelem_{x}\in\LTEset_{x}$
for different $x$ can be put in
bijective correspondence (see Figure~\ref{fig:proof-holomorphicity}):
the map $\LTEset_{x} \to \LTEset_{-x}$ can be defined by taking
the symmetric difference $\LTEishelem_{x}\mapsto\LTEishelem_{x}\oplus 
e_{x}\oplus 
e_{-x}\oplus e$,
for instance, and if $\LTEset_{x}$ and $\LTEset_{\ii x}$, the
bijection share the same dual vertex, 
$\LTEishelem_{x}\mapsto\LTEishelem_{x}\oplus 
e_{x}\oplus e_{\ii x}$.
It is elementary to check that for any 
$\left(\LTEishelem_{x},\LTEishelem_{\ii 
x},\LTEishelem_{-x},\LTEishelem_{-\ii x}\right)$
thus put in correspondence, we have 
\[
W_{x}+W_{-x}=W_{\ii x}+W_{-\ii x}.
\]
Summing over all contours, we obtain~\eqref{eq:corner-discrete-hol-relation},
as desired.

If $z_{x}=c_{1}$ for some $x \in\set{ \pm1 \pm \ii } $,
then $H^\corn \left(z_{x}\right)$ is not defined. However, if we would set
$H^\corn \left(z_{x}\right)$ equal to
$\frac{\cconj{z}-\cconj{w}}{|z-w|} \correl{ \prod_{x\in V} \spin_{x} }$
(this choice is one of the two ``conflicting values'' addressed below the 
proof),
and represent the latter in terms of even subgraphs in 
$\LTEset_{x} = \LTEset$
(there is no path in such even subgraphs, just loops), the bijective 
correspondence works in exactly the same way as above (see Figure 
\ref{fig:discrete-singularity}).
This hence yields~\eqref{eq:corner-discrete-sing-relation}.

For more general boundary conditions, simply notice that they can
be implemented by (a linear combination of) spin insertions on the
boundary, and hence can be included in $V$. 
\end{proof}
In the proof we temporarily used a certain value of the function~$H^\corn$ 
also at the special corner~$c_1$. However, the value depended on which of the 
two medial edges~$z$ adjacent to~$c_1$ was under consideration~--- the other 
adjacent medial edge would have required us to use the opposite value. These 
``conflicting values'' are the source of the singularity of our discrete 
holomorphic function.

The following lemma, combined with the previous one (Lemma 
\ref{lem:corner-discrete-hol}),
allows one to say that (at criticality) the values of the function~$H^\corn$ 
that appear in Equation~\eqref{eq:corner-discrete-hol-relation}
are actually the projection of an extension~$H^\mdl$ of~$H^\corn$ to the medial
vertices (as will Lemma \ref{lem:corner-medial-disc-hol} below show):
indeed the values given by opposite corners around a medial 
vertex
live on orthogonal lines of the complex plane (see \cite{smirnov-iii}
for a reference about such considerations). 

As usual, we set $\lambda := e^{\pi \ii /4}$. 

\begin{figure}

\includegraphics[scale=0.5]{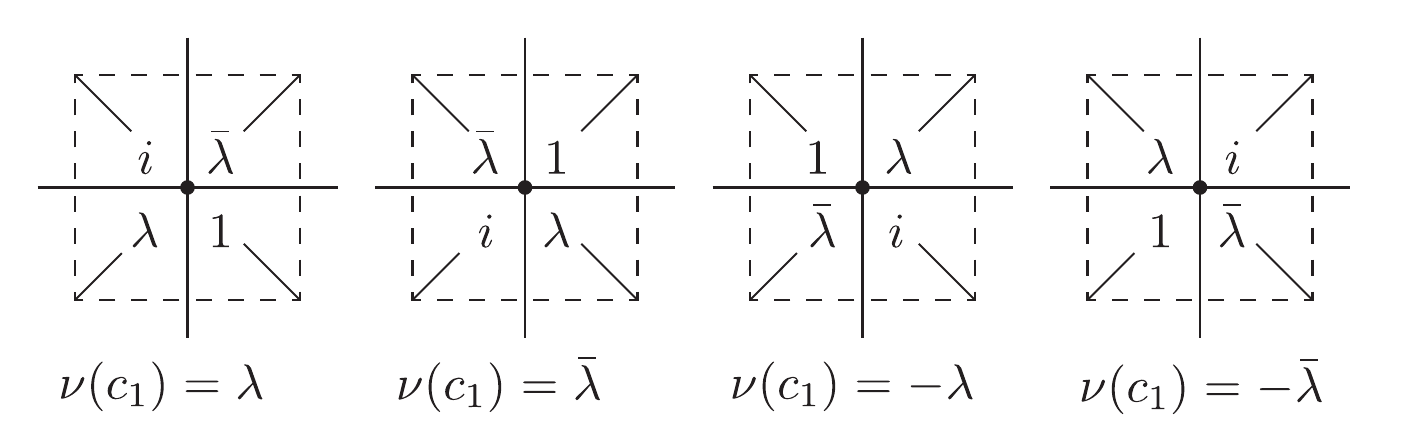}

\caption{\label{fig:lines-corner-fermion-projection}The directions of the
lines $\ell\left(c_{1},c_{2}\right)$ in $ \C $ on which 
$H^\corn\left(c_{2}\right)$
lives, for the four corners around a primal vertex.}

\end{figure}

\begin{lem}
\label{lem:phase-corner-fermion}
For any $c_2 \in
\dblcov{\ddomaincor , V} \setminus \set{c_1, c_1^*}$, we have that
$H^\corn \left(c_{2}\right) \in \ell\left(c_{1},c_{2}\right)$, where 
$\ell\left(c_{1},c_{2}\right) := \ii 
\sqrt{\bar{\corndir}(c_{1}) \bar{\corndir}(c_{2})} \R$
(see Figure \ref{fig:lines-corner-fermion-projection}). 
\end{lem}
\begin{proof}
This is elementary from the definition of~$H^\corn$ and of the corner fermion 
pair~$\fermpair{c_1}{c_2}$.
\end{proof}

We are now in position to state the following discrete holomorphicity
lemma for $H^\mdl$.
Set $\lambda := e^{\pi \ii /4}$ and $\eta := e^{\pi \ii /8}$ as usual. 
\begin{lem}
\label{lem:corner-medial-disc-hol}
If $\zeta \in \dblcov{\ddomaindmd , V}$ is away from
$V$, $\partial\ddomain $, and $c_{1}, c_{1}^{*}$ then we have
$\ddeebar H^\mdl (\zeta) =0$.
If $\zeta$~is the vertex or the dual vertex next to $c_{1}$, then
we have $\ddeebar  H^\mdl (\zeta) 
= \sqrt{2} \, \correl{ \prod_{x\in W} \spin_{x} } $.
\end{lem}
\begin{proof}
Consider a medial edge $z$ and the corners $z_x$ adjacent to it,
for $x \in \set{ \pm 1 \pm \ii } $. If $z$ is not adjacent to $c_1 , c_1^*$, 
then by Lemmas \ref{lem:corner-discrete-hol} and \ref{lem:phase-corner-fermion},
we have that the values of~$H^\corn (z_{x})$ at these corners
are given by the projections of the value of $H^\mdl (z) $ on
certain lines. The line $\ell(c_1, z_x) = \unitcnum_{c_1 , z_x} \, \R$ 
is specified by a complex number $\unitcnum_{c_1 , z_x}$ of unit modulus as in 
Figure~\ref{fig:lines-corner-fermion-projection}, and the projection relation 
explicitly reads
\begin{align}
\label{eq: projection relations for H}
H^\corn (z_{x}) = \projtoline{\ell(c_1, z_x)} \big[ H^\mdl (z) 
\big]
 = \frac{1}{2} \Big( H^\mdl (z)
     + \unitcnum_{c_1 , z_x}^2 \, \cconj{H^\mdl (z)} \Big) . 
\end{align}
When $z_{x} = c_{1}$, 
then the corresponding projection relation holds with one of the two
opposite ``conflicting values'' for $\ferm_{c_{1}} \ferm_{c_1}$.

Let us consider the case when $\corndir(c_{1}) = \lambda$
in detail; the other cases are symmetric. Set
$\Xi := \correl{ \prod_{x\in W} \spin_{x} } $.
We have the following four cases: 
\begin{enumerate}
\item If $\zeta$ is a primal vertex away from $c_{1}, c_{1}^{*}$, 
denoting
by $E,N,W,S$ the values of $H^\mdl$ at the four medial vertices next
to $\zeta$ in the directions $1,\ii ,-1,-\ii$, the projection relations
\eqref{eq: projection relations for H}
read: 
\begin{align*}
N-\ii \bar{N} & =E-\ii \bar{E}\\
S+\bar{S} & =E+\bar{E}\\
S+\ii \bar{S} & =W+\ii \bar{W}\\
N-\bar{N} & =W-\bar{W}.
\end{align*}
By taking an appropriate linear combination of the four equations
to eliminate the complex conjugate values,
one gets $E-W+\ii \left(N-S\right)=0$,
which shows $\ddeebar H^\mdl (\zeta) = 0$ as desired.
\item If $\zeta = v_1$ is the primal vertex adjacent to~$c_{1}$, 
plugging in the two ``conflicting'' values given by Equation 
\eqref{eq:corner-discrete-sing-relation}
of Lemma \ref{lem:corner-discrete-hol} (and reading them as projection
relations again), and using the same notation as in the previous item,
one gets
\begin{align*}
S + \ii \bar{S} & = 2 \lambda \, \Xi \\
- 2 \lambda \, \Xi & = W + \ii \bar{W}\\
S + \bar{S} & = E + \bar{E}\\
N - \ii \bar{N} & = E - \ii \bar{E}\\
N - \bar{N} & = W - \bar{W}.
\end{align*}%
By again taking an appropriate linear combination of these five equations,
one gets
$E - W + \ii \left(N-S\right) = 2 \sqrt{2} \, \Xi$
and hence
$\ddeebar H^\mdl (v_1) = \sqrt{2} \, \Xi$,
as desired. 
\item If $\zeta$ is a dual vertex away from $c_{1} ,c_{1}^{*}$, and if
we denote by $\mathcal{E},\mathcal{N},\mathcal{W},\mathcal{S}$ the
values of~$H^\mdl$ at the medial vertices around
$\zeta$ directions $1,\ii ,-1,-\ii$,
the projection relations~\eqref{eq: projection relations for H} read
\begin{align*}
\mathcal{N}+\bar{\mathcal{N}} & =\mathcal{W}+\bar{\mathcal{W}}\\
\mathcal{S}-\ii \bar{\mathcal{S}} & =\mathcal{W}-\ii \bar{\mathcal{W}}\\
\mathcal{S}-\bar{\mathcal{S}} & =\mathcal{E}-\bar{\mathcal{E}}\\
\mathcal{N}+\ii \bar{\mathcal{N}} & =\mathcal{E}+\ii \bar{\mathcal{E}}.
\end{align*}
As in the first item, an appropriate linear combination of these 
yields $\ddeebar H^\mdl(\zeta) = 0$.
\item Finally, if $\zeta=p_1$ is the dual vertex adjacent to 
$c_{1}$, using the same notation as in the previous item, one gets
\begin{align*}
2\lambda \, \Xi & =\mathcal{E} + \ii \bar{\mathcal{E}}\\
\mathcal{N} + \ii \bar{\mathcal{N}} & = - 2 \lambda \, \Xi\\
\mathcal{N} + \bar{\mathcal{N}} & = \mathcal{W} + \bar{\mathcal{W}}\\
\mathcal{S} - \ii \bar{\mathcal{S}} & = \mathcal{W} - \ii \bar{\mathcal{W}}\\
\mathcal{S} - \bar{\mathcal{S}} & = \mathcal{E} - \bar{\mathcal{E}},
\end{align*}
and one concludes $\ddeebar H^\mdl(p_1) = \sqrt{2} \, \Xi$
as in the second item. 
\end{enumerate}%
\end{proof}

The function
\[ H \colon \dblcov{\ddomainmdl , V} \to  \C\]
considered in
Proposition~\ref{prop:discrete-holomorphicity-fermion-correlations}
is straightforwardly reconstructed from~$H^\mdl$. For 
its definition, 
$c_1$ is no longer fixed, but we average over the four corners $c_1$ next to a 
fixed medial vertex $w \in \dblcov{\dCmdl , V}$. The function~$H$ is defined as
\begin{align*}
H(z) = H(w; z)
:= & \correl{ \left(\prod_{u\in W}\spin_{u}\right)
      \ferm(w) \ferm(z) } \\
  = & \frac{\pi}{8 \sqrt{2}} \sum_{x,y \in \set{\pm 1 \pm \ii}} 
      \correl{ \left(\prod_{u\in W}\spin_{u}\right)
          \ferm_{w_y} \ferm_{z_x} }
\end{align*}
Up to a multiplicative constant~$\frac{\pi}{4\sqrt{2}}$, this is just the sum
of~$H^\mdl(z)=H^\mdl(c_1;z)$ over the four corners~$c_1 = w_y$ adjacent to $w$.
We now recall and prove 
Proposition~\ref{prop:discrete-holomorphicity-fermion-correlations} concerning
the function~$H$.
\begin{prop*}[%
Proposition~\ref{prop:discrete-holomorphicity-fermion-correlations}]
For $\zeta \in \dblcov{\ddomaindmd , V}$ away from $\partial\ddomain $,
$V$, we have
$\ddeebar H(\zeta) =0$ if $\zeta \not \sim w, w^{*}$
and we have 
\[
\ddeebar  H(\zeta) 
= \frac{\pi}{2} \, \correl{ \prod_{x\in W} \spin_{x} }
\]
if $\zeta\sim w$.
\end{prop*}
\begin{proof}[Proof of Proposition 
\ref{prop:discrete-holomorphicity-fermion-correlations}]
This property of $H$ follows directly from 
Lemma~\ref{lem:corner-medial-disc-hol} about $H^\mdl$.
\end{proof}

\end{document}